\documentclass[11pt]{article}
\usepackage{epsfig}
\newcommand{\sect}[1]{ \section{#1} \setcounter{equation}{0} }

\newcommand{\third}{\mbox{\small{$\frac{1}{3}$}}} 
 
\newcommand{\pitwo}{\mbox{\small{$\frac{\pi}{2}$}}} 
 
\newcommand{\pisix}{\mbox{\small{$\frac{\pi}{6}$}}} 
\newcommand{\MSbar}{\overline{\mbox{MS}}} 
\newcommand{\MSbars}{\overline{\mbox{\footnotesize{MS}}}} 
\newcommand{\MOMg}{\mbox{MOMggg}}
\newcommand{\MOMgs}{\mbox{\footnotesize{MOMggg}}}
\newcommand{\MOMh}{\mbox{MOMh}}
\newcommand{\MOMhs}{\mbox{\footnotesize{MOMh}}}
\newcommand{\MOMq}{\mbox{MOMq}}
\newcommand{\MOMqs}{\mbox{\footnotesize{MOMq}}}
\newcommand{\mMOM}{\mbox{mMOM}}
\newcommand{\mMOMs}{\mbox{\footnotesize{mMOM}}}
\newcommand{\MOMi}{\mbox{MOMi}}
\newcommand{\MOMis}{\mbox{\footnotesize{MOMi}}}

\newcommand{\Nf}{N_{\!f}}
\newcommand{\NF}{N_{\!F}}
\newcommand{\NA}{N_{\!A}}

\begin{document}
\title{Momentum subtraction and the $R$-ratio}
\author{J.A. Gracey, \\ Theoretical Physics Division, \\ 
Department of Mathematical Sciences, \\ University of Liverpool, \\ P.O. Box 
147, \\ Liverpool, \\ L69 3BX, \\ United Kingdom.} 
\date{} 
\maketitle 

\vspace{5cm} 
\noindent 
{\bf Abstract.} We determine the $R$-ratio for massless quarks in several
renormalization schemes to various loop orders. These are the momentum
subtraction schemes of Celmaster and Gonsalves as well as the minimal momentum 
subtraction scheme. The dependence of the $R$-ratio on the schemes is analysed. 

\vspace{-15cm}
\hspace{13cm}
{\bf LTH 1025}

\newpage

\sect{Introduction.}

The electron-positron annihilation cross-section is a quantity which is of
immense interest experimentally and theoretically. It provides an avenue to
access the structure of hadrons via high energy particle beams. Known as the
$R$-ratio the $e^+ e^-$ hadronic cross-section can be computed in perturbation
theory. By this we mean that the corrections from Quantum Chromodynamics (QCD)
to the leading order parton prediction can in principle be determined order by
order in the strong coupling constant. Indeed as the parton model prediction
depends on the squares of the parton charges, experimental measurements were
used to confirm the fact that the partons themselves had fractional rather than
integer values. With the advent of higher energy colliders the QCD corrections 
were necessary for making precision measurements of the value of the strong 
coupling constant. Over several decades the $R$-ratio has been computed to 
four loop orders. The leading correction was determined in \cite{1,2}. 
Subsequently, several groups independently calculated the two loop graphs in 
\cite{3,4,5,6}. While the three, \cite{7,8,9}, and four loop, \cite{10,11,12}, 
results took longer to determine. In summarizing the perturbative side of our 
$R$-ratio knowledge we have omitted some of the technical issues. First, the 
computations of \cite{1,2,3,4,5,6,7,8,9,10,11,12} were for the unrealistic case
of massless quarks. In the real world quarks have a hierarchy of masses which 
have to be taken into account. In this respect there have been various analyses
of the $R$-ratio where quark mass effects have been included. See, for example,
\cite{13,14}. Another aspect of using a quantum field theory approach is that 
inevitably one has to include or estimate non-perturbative effects such as 
renormalons. Partly related to this is the estimation of the truncation errors 
on the series. This is important for making precision measurements. One way to 
quantify this is to use the last term of the perturbative expansion as a 
measure of the error bar. Though this would seem to be a limited use of the
considerable effort put into calculating the four or higher loop corrections in
the first place. Next in the early work of calculating the $R$-ratio it was 
noticed in \cite{5,6} that the choice of renormalization scheme used to handle 
the underlying divergences in the Feynman diagrams could lead to differing 
rates of the convergence of the series. This is not limited, of course, to the 
$R$-ratio but applies to any perturbatively computed quantity. In \cite{6} the 
observation was made in respect of comparing the minimal subtraction (MS) 
scheme with the modified minimal subtraction ($\MSbar$) scheme, \cite{15}. 
However, in \cite{5,6} the $R$-ratio was calculated at two loops in the then 
recently developed momentum subtraction (MOM) scheme of \cite{16,17}. The 
results in that scheme were numerically similar to the $\MSbar$ result to the 
same order in that the coefficients of the series appeared to be smaller than 
those for the MS scheme. Indeed for certain values of $\Nf$, which is the 
number of massless quark flavours, the MOM scheme seemed to have smaller 
numerical coefficients when compared to the $\MSbar$ ones. As a consequence it 
was suggested in \cite{5,6} that the MOM scheme might be considered as the 
preferred renormalization scheme.

There are several advantages and disadvantages to using a MOM scheme for
physical quantities instead of $\MSbar$. First, the $\MSbar$ scheme is widely
used since it is computationally easier to go to very high loop order. The
nature of the scheme is such that only poles with respect to the regularization
are removed from the quantity being computed. No finite parts are removed,
aside from the contribution of $\ln(4\pi e^{-\gamma})$, where $\gamma$ is the
Euler-Mascheroni constant. This factor in essence is the quantity which
differentiates between the MS and $\MSbar$ schemes. Hence, in light of the way
the $\MSbar$ scheme is defined one can essentially define the $\MSbar$
renormalization constants at any external momentum configuration for the
Green's function in question. For pragmatic reasons the canonical choice is a 
point at which it is straightforward to compute all the contributing Feynman 
integrals. This is assuming, of course, that such a momentum configuration does
not introduce infrared divergences. Although the infrared rearrangement 
technique of \cite{18,19} circumvents this in situations where such a problem 
arises it allows one to extract the unblemished renormalization constants. 
However, in this description of the $\MSbar$ scheme we have implicitly referred
to what could be regarded as a disadvantage of the scheme, which is that it is 
not physical. The $\MSbar$ renormalization of, say, the strong coupling 
constant which is derived from one of the vertices of the QCD Lagrangian 
carries no information about the external momentum configuration which one 
could imagine would be relevant to a physical measurement of such a decay 
process. Such information of the underlying vertex or process itself could 
never be quantified in the residues of the poles of the renormalization 
constant since such poles are dependent on the regulator which is lifted at the
end of the computation. Instead it would be included in the finite part of the 
renormalization constant and transmitted via the renormalization group 
evolution. This was the motivation behind the MOM schemes of \cite{16,17}. In 
\cite{17} the $3$-point vertices of the QCD Lagrangian were renormalized at the
completely symmetric point. By this we mean the values of the squared momenta 
of the external legs were all equal. This is not sufficient in itself to define
a MOM scheme as in this external momentum configuration the usual $\MSbar$ 
coupling constant renormalization constant emerges in the computation. Instead 
the MOM schemes are defined by choosing the coupling constant renormalization 
constant such that at the symmetric point after renormalization there are no 
$O(g^2)$ corrections where $g$ is the coupling constant. Thus the MOM schemes 
are physical and carry information about the vertex at a specific momentum 
configuration unlike $\MSbar$. As an aside this external momentum configuration
is non-exceptional and so there are no infrared issues. In introducing the MOM 
schemes of \cite{16,17} there are three different MOM schemes depending on 
whether it is defined relative to the triple gluon, ghost-gluon or quark-gluon 
vertices. They are denoted by $\MOMg$, $\MOMh$ and $\MOMq$ respectively. Though
it is a moot point as to whether $\MOMh$ can be regarded as a truly physical
scheme as it is defined with respect to a vertex which contains a fictitious 
field deriving from the gauge fixing condition. In \cite{16,17} the MOM schemes
were defined for QCD fixed in a linear covariant gauge. In other gauges the MOM
schemes will be different. In the context of the $R$-ratio of \cite{5,6} it was
the $\MOMq$ scheme which was analysed in detail since that scheme is most 
closely aligned with the nature of the underlying Feynman graphs. 

While the $\MSbar$ evaluation of the $R$-ratio has progressed to four loops
in recent years, \cite{7,8,9,10,11,12}, the MOM scheme renormalization of QCD
of \cite{16,17} has not been developed at the same rate and remained at the 
original two loop level until recently. In \cite{20} the two loop extension to 
\cite{17} was provided with the full analysis of all three $3$-point vertices 
of the QCD Lagrangian at the fully symmetric point. Via the renormalization 
group this means that the three loop $\beta$-functions for each of the three 
MOM schemes are now known for all linear covariant gauges, \cite{20}. For the 
Landau gauge, which will be the focus here for reasons which will be indicated
later, this means that the three loop coefficient differs from that of the 
$\MSbar$ $\beta$-function. This is the first place the scheme difference of the
$\beta$-functions becomes manifest. Now that the two loop extension of 
\cite{17} is available it is possible to examine the $\MOMq$ scheme $R$-ratio 
of \cite{6} to the next order and compare it with the $\MSbar$ expresssion. 
This is the main purpose of this article. However, as an exercise in analysing 
a physical quantity in various physical and unphysical schemes we will include 
the $\MOMg$ and $\MOMh$ schemes for completeness. This is partly for comparison
even though $\MOMq$ is the natural scheme for the $R$-ratio. One aim is to see 
if the coefficients of the series in the different schemes show improved 
convergence and if there are any differences between physical and unphysical 
schemes. In the analysis we will include another scheme which is termed 
minimal-MOM and denoted by $\mMOM$. It was introduced in \cite{21} with the 
four loop $\mMOM$ QCD $\beta$-function deduced from the results of \cite{22}. 
The renormalization group functions recorded in \cite{21} were verified by 
explicit computations in \cite{23}. The $\mMOM$ scheme is based on an extension
of a property of the ghost-gluon vertex. Specfically in the Landau gauge this 
vertex does not get renormalized when one leg has a nullified momentum and 
leads to a non-renormalization theorem, \cite{24}. In this sense the $\mMOM$ 
scheme could be regarded as being similar to $\MSbar$ in that the 
renormalization point is exceptional. Indeed examining the higher order $\mMOM$
$\beta$-function the mathematical structure involves rationals and the Riemann 
zeta functions like the $\MSbar$ one and not the polylogarithms of the MOM 
schemes of \cite{16,17}. One reason for including the $\mMOM$ scheme in our 
analysis is that its four loop $\beta$-function is known, \cite{21}, which will
allow us to compare its $R$-ratio with the $\MSbar$ one at the same order. 
Moreover it will complement a similar analysis carried out in \cite{21} for
other quantities such as the Adler function. The actual evaluation of the 
$R$-ratio in all of these schemes is straightforward as it entails using the 
mapping of the coupling constant variable from one scheme to the other. In some
sense one could regard the $\MSbar$ evaluation of a quantity as the 
foundational or bare scheme from which the physical scheme value is deduced by 
parameter mapping. So if there is some overall difference between the 
unphysical $\MSbar$ scheme value compared to a MOM one it should be 
quantifiable which is one of our secondary aims. Alternatively it could provide
a novel way of extracting theoretical errors on the strong coupling constant. 
Finally, as this is an exercise in comparing the behaviour of the $R$-ratio in 
different renormalization schemes we have to compare like quantities. Therefore
we do not include quark masses or non-perturbative renormalon features. The 
reason for the exclusion of the former is that the MOM schemes of \cite{16,17} 
were defined for massless quarks. Aside from \cite{14} as far as we are aware 
there have been no attempts to define quark mass dependent MOM schemes. 

The article is organized as follows. The relevant aspects of the
renormalization group functions for each scheme are reviewed in section $2$
where we also recall the coupling constant maps. Section $3$ contains the
$R$-ratio in each of the MOM schemes we consider and a discussion of the status
of the convergence. A more detailed analysis is given in section $4$ where the
effects of the running coupling constant are inlcuded. Issues to do with gauges
other than the Landau gauge are given there as well. Section $5$ summarizes our
conclusions.  

\sect{Background.}

We begin by recalling the various main features of the momentum subtraction
based schemes we will be concentrating on. First, we note that our reference
scheme is the $\MSbar$ scheme which is the modified minimal subtraction scheme
introduced in \cite{15}. It is an extension of the original minimal subtraction
(MS) scheme of \cite{25} where the only divergences with respect to the 
regularizing parameter are absorbed by the renormalization constants at the 
point chosen for the subtraction. As noted in \cite{15} the convergence of
perturbative series appears to be quicker if one also includes a specific 
finite part in the renormalization constants. This is $\ln(4\pi e^{-\gamma})$ 
and this quantity derives from the expansion of the $d$-dimensional factor in 
the measure of each loop integral corresponding to the volume of the 
$d$-dimensional unit sphere when the quantum field theory is dimensionally 
regularized. In including this extra finite piece the $\MSbar$ scheme still 
remains a mass independent scheme and the coefficients in the $\beta$-function 
remain gauge parameter independent unlike a mass dependent scheme such as the 
MOM ones, \cite{16,17}. To assist with making contact with conventions used in 
different articles we will use the following form for the $\MSbar$ 
$\beta$-function. Defining the $\beta$-function in the generic scheme 
${\cal S}$ by 
\begin{equation}
\beta^{\cal S}(a) ~=~ - \sum_{n=0}^\infty b^{\cal S}_n a^{n+1}
\end{equation} 
then the first four coefficients are, \cite{26,27,28,29,30,31,32},
\begin{eqnarray}
b_0^{\MSbars} &=& \frac{11}{3} C_A - \frac{4}{3} T_F \Nf 
\nonumber \\
b_1^{\MSbars} &=& \frac{34}{3} C_A^2 - 4 T_F C_F \Nf - \frac{20}{3} T_F \Nf C_A
\nonumber \\
b_2^{\MSbars} &=& \frac{2857}{54} C_A^3 + 2 C_F^2 T_F \Nf 
- \frac{205}{9} C_F C_A T_F \Nf - \frac{1415}{27} C_A^2 T_F \Nf \nonumber \\
&& +~ \frac{44}{9} C_F T_F^2 \Nf^2 + \frac{158}{27} C_A T_F^2 \Nf^2 
\nonumber \\ 
b_3^{\MSbars} &=&
\left[ \frac{150653}{486} - \frac{44}{9} \zeta(3) \right] C_A^4
+ \left[ - \frac{39143}{81} + \frac{136}{3} \zeta(3) \right] C_A^3 T_F \Nf
\nonumber \\
&& + \left[ \frac{7073}{243} - \frac{656}{9} \zeta(3) \right] C_A^2 C_F T_F \Nf
+ \left[ - \frac{4204}{27} + \frac{352}{9} \zeta(3) \right] C_A C_F^2 T_F \Nf
+ 46 C_F^3 T_F \Nf
\nonumber \\
&& + \left[ \frac{7930}{81} + \frac{224}{9} \zeta(3) \right] C_A^2 T_F^2 \Nf^2
+ \left[ \frac{1352}{27} - \frac{704}{9} \zeta(3) \right] C_F^2 T_F^2 \Nf^2
\nonumber \\
&& + \left[ \frac{17152}{243} + \frac{448}{9} \zeta(3) \right] 
C_A C_F T_F^2 \Nf^2
+ \frac{424}{243} C_A T_F^3 \Nf^3 + \frac{1232}{243} C_F T_F^3 \Nf^3
\nonumber \\
&& + \left[ \frac{704}{3} \zeta(3) - \frac{80}{9} \right] \!\! 
\frac{d_{AA}}{\NA}
+ \left[ \frac{512}{9} - \frac{1664}{3} \zeta(3) \right] \!\! 
\frac{\Nf d_{FA}}{\NA} + \left[ \frac{512}{3} \zeta(3) - \frac{704}{9} \right]
\!\! \frac{\Nf^2 d_{FF}}{\NA}
\end{eqnarray}
where $\zeta(z)$ is the Riemann zeta function, $C_F$, $C_A$ and $T_F$ are the 
usual colour group Casimirs and $\NA$ is the dimension of the adjoint 
representation of the colour group. At four loops several new rank four 
Casimirs arise, \cite{31,32}, which are defined by
\begin{equation}
d_{R_1 R_2} ~=~ d_{R_1}^{abcd} d_{R_2}^{abcd}
\end{equation}
where the completely symmetric tensor $d_{R_i}^{abcd}$ is defined by
\begin{equation} 
d_{R_i}^{abcd} ~=~ \frac{1}{6} \mbox{Tr} \left( T^a_{R_i} T^{(b}_{R_i}
T^c_{R_i} T^{d)}_{R_i} \right)
\end{equation}
with the group generators $T^a$ in representation $R_i$. Here $F$ and $A$
denote the fundamental and adjoint representations. The number of quarks is 
$\Nf$ and throughout we assume that the quark masses are zero. This is 
primarily because the MOM scheme renormalizations were established in the 
absence of quark masses. We note that we use $a$~$=$~$g^2/(16\pi^2)$ where $g$ 
is the gauge field coupling constant. The strong coupling constant $\alpha_s$ 
is related to $g$ by $\alpha_s$~$=$~$g^2/(4\pi)$. We will use $a$ throughout to
be consistent with previous work upon which this article is based, \cite{20}, 
but it is straightforward to convert to other conventions. 

The three main MOM schemes we consider here were introduced in \cite{16,17} and
are based on the three $3$-point vertices of the QCD Lagrangian when it is
fixed in a linear covariant gauge. In other words the triple gluon, ghost-gluon
and quark-gluon vertices which lead respectively to the $\MOMg$, $\MOMh$ and 
$\MOMq$ schemes. The definition of each scheme is in essence the same. The 
specific vertex is evaluated at the completely symmetric point where the value 
of the square of the momentum of each external leg is equal and set to 
$(-\mu^2)$ where $\mu$ is the scale introduced to ensure the coupling constant 
is dimensionless in $d$-dimensions when the theory is dimensionally 
regularized. Then at this particular external momentum configuration the 
coupling constant renormalization constant is defined by the condition that 
after renormalization there are no $O(a)$ corrections, \cite{16,17}. In 
addition the wave function renormalization constants of the fields associated 
with each vertex are also constructed in a similar way, \cite{16,17}. In other 
words for the gluon, ghost and quark $2$-point functions the wave function 
renormalization constants are determined by the condition that after 
renormalization there are no $O(a)$ contributions. The two loop extension of 
\cite{17} was carried out in \cite{20}. One interesting property is that once 
the $L$th loop computation has been accomplished it is possible to deduce the 
$(L+1)$th loop renormalization group functions provided that the $\MSbar$ 
renormalization group functions are available at that order. This is possible 
due to properties of the renormalization group equation and the fact that one 
can construct the relation between the coupling constant as defined with 
respect to the $\MSbar$ scheme and that in another scheme. Indeed this is the 
key to converting the $R$-ratio to the MOM schemes from the known $\MSbar$ 
result of \cite{1,2,3,4,5,6,7,8,9,10,11,12}. Also the three loop MOM 
$\beta$-functions will be required for our comparitive analysis later. The 
fourth momentum subtraction scheme we consider is $\mMOM$, \cite{21}. It 
derives from the Slavnov-Taylor identity for the ghost-gluon vertex in the 
Landau gauge. Specifically in that gauge the vertex undergoes no 
renormalization when an external leg momentum is set to zero. So the 
renormalization condition for the $\mMOM$ coupling constant renormalization is 
that the ghost-gluon {\em vertex} renormalization is the {\em same} as the 
$\MSbar$ case. In this respect the $\mMOM$ scheme differs from the MOM ones of 
\cite{16,17} in that the subtraction point where the scheme is defined is 
exceptional. However, the main reason for including it in the present analysis 
is that the $\mMOM$ $\beta$-function is the only other $\beta$-function 
available at four loops, \cite{22,23}.

Since the three loop MOM $\beta$-functions and coupling constant maps were
constructed in earlier work we merely record both sets of results numerically
for $\Nf$ flavours. Again this is partly due to space considerations but also
to make contact with conventions. The full expressions were provided in the
respective articles. However, for completeness here, we provide an attached 
data file where all the analytic expressions we have used are provided to 
assist an interested reader. For the MOM schemes at two loops special functions
evaluated at specific arguments occur and to assist the conversion to numerical
values we record the values we used were
\begin{eqnarray}
\zeta(3) &=& 1.20205690 ~~,~~
\zeta(5) ~=~ 1.03692776 ~~,~~
\zeta(7) ~=~ 1.00834928 \nonumber \\
\psi^\prime ( \third ) &=& 10.09559713 ~~,~~ 
\psi^{\prime\prime\prime} ( \third ) ~=~ 488.1838167 ~~,~~
s_2 ( \pitwo ) ~=~ 0.32225882 \nonumber \\
s_2 ( \pisix ) &=& 0.22459602 ~~,~~
s_3 ( \pitwo ) ~=~ 0.32948320 ~~,~~
s_3 ( \pisix ) ~=~ 0.19259341 
\end{eqnarray}
where $\psi(z)$ is the derivative of the logarithm of the Euler 
$\Gamma$-function and 
\begin{equation}
s_n(z) ~=~ \frac{1}{\sqrt{3}} \Im \left[ \mbox{Li}_n \left(
\frac{e^{iz}}{\sqrt{3}} \right) \right]
\end{equation}
with $\mbox{Li}_n(z)$ corresponding to the polylogarithm function. Equally we 
note that several other quantities can be expressed in terms of polylogarithms.
For example,
\begin{equation}
\psi^\prime ( \third ) ~=~ \frac{2\pi^2}{3} ~+~ 2 \sqrt{3} 
\mbox{Cl}_2 \left( \frac{\pi}{3} \right) 
\end{equation}
where $\mbox{Cl}_2(\theta)$ is the Clausen function which is related to the
dilogarithm by 
\begin{equation}
\mbox{Cl}_2(\theta) ~=~ \Im \left[ \mbox{Li}_2 \left( e^{i\theta} \right) 
\right] ~.
\end{equation}
Therefore, the numerical values of the various $\beta$-functions are, 
\cite{25,26,27,28,29,30,31,16,17,20,21}, 
\begin{eqnarray}
\beta^{\MSbars}(a) &=& 
\left[ 0.666667 \Nf - 11.000000 \right] a^2 
+ \left[ 12.666667 \Nf - 102.000000 \right] a^3
\nonumber \\
&& + \left[ - 6.018518 \Nf^2 + 279.611111 \Nf - 1428.500000 \right] a^4
\nonumber \\
&& + \left[ - 1.499314 \Nf^3 - 405.089040 \Nf^2 + 6946.289617 \Nf 
- 29242.964136 \right] a^5 \nonumber \\
&& +~ O(a^6) \nonumber \\
\beta^{\MOMgs}(a) &=& 
\left[ 0.666667 \Nf - 11.000000 \right] a^2 
+ \left[ 12.666667 \Nf - 102.000000 \right] a^3 
\nonumber \\
&& + \left[ - 2.658115 \Nf^3 + 67.089536 \Nf^2 - 0.565929 \Nf - 1570.984380 
\right] a^4 \nonumber \\
&& +~ O(a^5) \nonumber \\ 
\beta^{\MOMhs}(a) &=& 
\left[ 0.666667 \Nf - 11.000000 \right] a^2 
+ \left[ 12.666667 \Nf - 102.000000 \right] a^3 
\nonumber \\
&& + \left[ - 21.502818 \Nf^2 + 617.647154 \Nf - 2813.492948 
\right] a^4 ~+~ O(a^5) \nonumber \\ 
\beta^{\MOMqs}(a) &=& 
\left[ 0.666667 \Nf - 11.000000 \right] a^2 
\nonumber \\
&& + \left[ 12.666667 \Nf - 102.000000 \right] a^3 
\nonumber \\
&& + \left[ - 22.587812 \Nf^2 + 588.654845 \Nf - 1843.65273 
\right] a^4 ~+~ O(a^5) \nonumber \\ 
\beta^{\mMOMs}(a) &=& 
\left[ 0.666667 \Nf - 11.000000 \right] a^2 
+ \left[ 12.666667 \Nf - 102.000000 \right] a^3 
\nonumber \\
&& + \left[ - 19.383310 \Nf^2 + 625.386670 \Nf - 3040.482287 \right] a^4 
\nonumber \\
&& + \left[ 27.492640 \Nf^3 - 1625.402243 \Nf^2 + 24423.330550 \Nf 
\right. \nonumber \\
&& \left. ~~~
- 100541.058601 \right] a^5 ~+~ O(a^6) ~. 
\end{eqnarray} 
Throughout all our numerical expressions will be for the $SU(3)$ colour group.
For the $\mMOM$ scheme we have provided the four loop term as the coupling
constant map is known to the requisite order, \cite{22,23}. This is primarily 
because that scheme is defined with respect to a vertex function where there is
a nullified external momentum. However, in this scheme and the other three MOM 
schemes we should emphasise that as they are mass dependent renormalization 
schemes their $\beta$-functions are not only scheme dependent at {\em two} 
loops but they also gauge dependent. This is evident in the expressions given 
in \cite{16,17} and implies, moreover, that the coupling constant mappings are 
gauge dependent. However, we have presented the expressions here in the Landau 
gauge. One reason for concentrating on this gauge is that in the path integral 
formulation of the gauge fixing procedure the limit $\alpha$~$\rightarrow$~$0$ 
is included in order that the gauge condition $\partial^\mu A^a_\mu$~$=$~$0$ is
functionally implemented, where $A^a_\mu$ is the gluon and $\alpha$ is the 
gauge parameter of the linear covariant gauge. Ordinarily since one computes in 
$\MSbar$ and the $\beta$-function is gauge parameter independent in that
scheme, \cite{25}, one usually ignores this formal limit. Indeed one usually
chooses an alternative gauge such as the Feynman gauge where the actual Feynman
graphs are simpler and hence quicker to evaluate. For the mass dependent 
schemes one can no longer use this computational shortcut. Though we should 
note that in the Landau gauge the two loop term of each of the MOM 
$\beta$-functions precisely equate with the scheme independent term of the 
$\MSbar$ $\beta$-function as is numerically evident. 

In recording the numerical value of the $\MSbar$ scheme one can compare the 
values of the various corrections in different schemes. For example, 
considering the Yang-Mills case as a reference the three loop terms of the 
$\MOMg$ and $\MSbar$ schemes are of comparable size but smaller than the 
corresponding term in the other three schemes. Though for the $\mMOM$ and 
$\MSbar$ four loop terms the former has a smaller value and differing sign. One
basic lesson that one could learn from this is that a smaller coefficient in a 
scheme dependent term at a particular loop order in one scheme is not 
necessarily smaller in that scheme at the next loop order. These observations 
on the magnitude of the corrections in the terms in each of the schemes in 
question will be similar in the other quantities we consider here. For 
instance, the mappings of the coupling constants defined in the various schemes
numerically are, \cite{16,17,20,21},
\begin{eqnarray}
a_{\MOMgs} &=& a + \left[ - 3.416806 \Nf + 26.492489 \right] a^2
\nonumber \\
&& + \left[ 7.687393 \Nf^2 - 202.085011 \Nf + 960.462717 \right] a^3 
~+~ O(a^4) \nonumber \\
a_{\MOMhs} &=& a + \left[ - 1.1111111 \Nf + 18.5482754 \right] a^2 \nonumber \\
&& + \left[ 1.2345678 \Nf^2 - 85.5559502 \Nf + 641.9400674 \right] a^3 
~+~ O(a^4) \nonumber \\
a_{\MOMqs} &=& a + \left[ - 1.1111111 \Nf + 16.7157746 \right] a^2
\nonumber \\
&& + \left[ 1.2345678 \Nf^2 - 83.1112168 \Nf + 472.1590953 \right] a^3 
~+~ O(a^4) \nonumber \\
a_{\mMOMs} &=& a + \left[ - 1.111111 \Nf + 14.083333 \right] a^2 \nonumber \\
&& + \left[ 1.2345678 \Nf^2 - 72.454594 \Nf + 475.475031 \right] a^3 
\nonumber \\
&& + \left[ - 1.371742 \Nf^3 + 209.401255 \Nf^2 - 4109.724062 \Nf 
+ 18652.393278 \right] a^4 \nonumber \\
&& +~ O(a^5) 
\end{eqnarray} 
where we use $\MSbar$ are the reference scheme on the right hand side and omit
the label as our convention. We do not include the mappings of the gauge 
parameter in the various schemes since we have chosen the Landau gauge. All the
gauge parameter mappings from the MOM schemes to the $\MSbar$ scheme are 
proportional to the gauge parameter. So choosing the Landau gauge in one scheme
implies that one has to use the Landau gauge in the other scheme. Related to 
the coupling constant scheme mappings are the relations between the $\Lambda$ 
parameters in the various schemes. This parameter in essence is the constant of
integration when one solves the first order differential equation defining the 
running of the coupling constant in terms of the $\beta$-function. It is 
different in different schemes but the ratio of $\Lambda$ in two schemes can be
determined by a simple one loop calculation. As it will play a role in our 
analysis later we record the relations we will need for the various schemes and
values of $\Nf$ in Table $1$ for the Landau gauge for completeness where 
$\MOMi$ indicates one of the four MOM schemes of interest. Those for $\MOMg$, 
$\MOMh$ and $\MOMq$ were given in \cite{17} and the $\mMOM$ values were 
provided in \cite{21}.

{\begin{table}[ht]
\begin{center}
\begin{tabular}{|c||c|c|c|c|c|}
\hline
$\Nf$ & $\MOMg$ & $\MOMh$ & $\MOMq$ & $\mMOM$ \\
\hline
3 & 2.4654 & 2.3286 & 2.1032 & 1.8171 \\
4 & 2.1587 & 2.3308 & 2.0881 & 1.7831 \\
5 & 1.8471 & 2.3335 & 2.0706 & 1.7440 \\
6 & 1.5341 & 2.3366 & 2.0499 & 1.6985 \\
\hline
\end{tabular}
\end{center}
\begin{center}
{Table $1$. Values of $\Lambda^{\MOMis}/\Lambda^{\MSbars}$ for $SU(3)$ in the 
Landau gauge.}
\end{center}
\end{table}}

\sect{MOM $R$-ratios.}

Having discussed the various renormalization schemes and the relation between
them we can construct the expressions for the $R$-ratio of the MOM schemes. The
starting point is the $\MSbar$ $R$-ratio, \cite{1,2,3,4,5,6,7,8,9,10,11,12}. To
reproduce the two loop $\MOMq$ correction of \cite{5,6} the method is the same 
as that of the previous section for the renormalization group functions. One 
maps the $\MSbar$ coupling constant to the coupling constant of the MOM scheme 
of interest. Therefore to repeat the exercise of \cite{5,6} to the next order 
we begin with the three and four loop $\MSbar$ $R$-ratios of 
\cite{7,8,9,10,11,12}. The latter is required for the conversion to the $\mMOM$
scheme since the coupling constant map is known to the right order. For 
reference, we note that the three loop $R$-ratio is
\begin{eqnarray}
R^{\MSbars}(s) &=& \NF \sum_f Q_f^2 \left[ \frac{}{}
1 + 3 C_F a \right. \nonumber \\
&& \left. ~~~~~~~~~~~~~~
+ \left[ \left[ \frac{123}{2} - 44 \zeta(3) \right] C_F C_A - \frac{3}{2} C_F^2 
+ [ - 22 + 16 \zeta(3) ] C_F T_F \Nf \right] a^2 \right. \nonumber \\
&& \left. ~~~~~~~~~~~~~~ + \left[ - \frac{69}{2} C_F^3 
+ [ - 127 - 572 \zeta(3) + 880 \zeta(5) ] C_F^2 C_A
\right. \right. \nonumber \\
&& \left. \left. ~~~~~~~~~~~~~~~~~~
+ \left[ \frac{90445}{54} - \frac{10948}{9} \zeta(3) - \frac{440}{3} \zeta(5) 
\right] C_F C_A^2
\right. \right. \nonumber \\
&& \left. \left. ~~~~~~~~~~~~~~~~~~
+ \left[ - 29 + 304 \zeta(3) - 320 \zeta(5) \right] C_F^2 T_F \Nf 
\right. \right. \nonumber \\
&& \left. \left. ~~~~~~~~~~~~~~~~~~
+ \left[ - \frac{31040}{27} + \frac{7168}{9} \zeta(3) + \frac{160}{3} \zeta(5)
\right] C_F C_A T_F \Nf
\right. \right. \nonumber \\
&& \left. \left. ~~~~~~~~~~~~~~~~~~
+ \left[ \frac{4832}{27} - \frac{1216}{9} \zeta(3) \right] C_F T_F^2 \Nf^2 
- \pi^2 \left[ \frac{11}{3} C_A - \frac{4}{3} T_F \Nf \right]^2 C_F 
\right] a^3 \right] \nonumber \\
&& + \left( \sum_f Q_f \right)^2 d^{abc} d^{abc} \left[ \frac{11}{3} 
- 8 \zeta(3) \right] a^3 ~+~ O(a^4) 
\label{Rms}
\end{eqnarray}
where $s$ is the centre of mass energy, $\NF$ is the dimension of the 
fundamental representation, $Q_f$ is the charge of the $\Nf$ active quarks and 
$d^{abc}$ is the totally symmetric rank $3$ tensor. It is related to the trace 
of three colour group generators and originates from three loop diagrams where 
there are two separate quark loops. The appearance of $\pi^2$ derives from the 
imaginary part of the mapping of the momentum from the Euclidean to the 
physical region as discussed, for instance, in \cite{7}. The convention is that
the scheme label on the quantity on the left side indicates the scheme the 
coupling constant is in on the right hand side. In order to save space we have 
not included the four loop term which is already available in \cite{10,11,12}.

We can now take the analytic form for the coupling constant map and convert
(\ref{Rms}) to each of the MOM schemes. For $\MOMg$ we have
\begin{eqnarray}
R^{\MOMgs}(s) &=& \NF \sum_f Q_f^2 \left[ \frac{}{}
1 + 3 C_F a \right. \nonumber \\ 
&& \left. ~~~~~~~~~~~~~
+ \left[
\left[
16 \zeta(3) - 14 - \frac{64}{27} \pi^2
+ \frac{32}{9} \psi^\prime \left( \third \right)
\right] \Nf T_F C_F 
\right. \right. \nonumber \\
&& \left. \left. ~~~~~~~~~~~~~~~~~
+ \left[ 
\frac{79}{2}
- 44 \zeta(3)
+ \frac{23}{27} \pi^2
- \frac{23}{18} \psi^\prime \left( \third \right)
\right]
- \frac{3}{2} C_F^2 \right] a^2
\right. \nonumber \\
&& \left. ~~~~~~~~~~~~~
+ \left[ 
\left[
\frac{6368}{81} \zeta(3) \pi^2
- \frac{29317}{54}
+ \frac{3520}{9} \zeta(3)
+ \frac{160}{3} \zeta(5)
- \frac{2866}{81} \pi^2
\right. \right. \right. \nonumber \\
&& \left. \left. \left. ~~~~~~~~~~~~~~~~~~
- \frac{2560}{2187} \pi^4
- 240 s_2 \left( \pisix \right)
+ 480 s_2 \left( \pitwo \right)
+ 400 s_3 \left( \pisix \right)
- 320 s_3 \left( \pitwo \right)
\right. \right. \right. \nonumber \\
&& \left. \left. \left. ~~~~~~~~~~~~~~~~~~
+ \frac{1829}{27} \psi^\prime \left( \third \right)
- \frac{3184}{27} \psi^\prime \left( \third \right) \zeta(3)
+ \frac{5152}{729} \psi^\prime \left( \third \right) \pi^2
\right. \right. \right. \nonumber \\
&& \left. \left. \left. ~~~~~~~~~~~~~~~~~~
- \frac{1288}{243} \left( \psi^\prime \left( \third \right) \right)^2 
- \frac{4}{9} \psi^{\prime\prime\prime} \left( \third \right)
- \frac{5}{9} \ln^2(3) \sqrt{3} \pi
+ \frac{20}{3} \ln(3) \sqrt{3} \pi
\right. \right. \right. \nonumber \\
&& \left. \left. \left. ~~~~~~~~~~~~~~~~~~
+ \frac{145}{243} \sqrt{3} \pi^3
\right] \Nf T_F C_F C_A 
\right. \right. \nonumber \\
&& \left. \left. ~~~~~~~~~~~~~
+ \left[ 
14
+ 256 \zeta(3)
- 320 \zeta(5)
- \frac{128}{27} \pi^2
+ \frac{64}{9} \psi^\prime \left( \third \right)
\right] \Nf T_F C_F^2 
\right. \right. \nonumber \\
&& \left. \left. ~~~~~~~~~~~~~
+ \left[ 
\frac{2252}{27}
- \frac{2048}{81} \zeta(3) \pi^2
- \frac{448}{9} \zeta(3)
+ \frac{1520}{81} \pi^2
+ \frac{7168}{2187} \pi^4
\right. \right. \right. \nonumber \\
&& \left. \left. \left. ~~~~~~~~~~~~~~~~~~
- \frac{832}{27} \psi^\prime \left( \third \right)
+ \frac{1024}{27} \psi^\prime \left( \third \right) \zeta(3)
- \frac{7168}{729} \psi^\prime \left( \third \right) \pi^2
\right. \right. \right. \nonumber \\
&& \left. \left. \left. ~~~~~~~~~~~~~~~~~~
+ \frac{1792}{243} \left( \psi^\prime \left( \third \right) \right)^2
\right] \Nf^2 T_F^2 C_F 
\right. \right. \nonumber \\
&& \left. \left. ~~~~~~~~~~~~~
+ \left[
\frac{1397759}{1728}
- \frac{2024}{81} \zeta(3) \pi^2
- \frac{90781}{144} \zeta(3)
- \frac{440}{3} \zeta(5)
- \frac{329}{9} \pi^2
\right. \right. \right. \nonumber \\
&& \left. \left. \left. ~~~~~~~~~~~~~~~~~~
- \frac{27181}{17496} \pi^4
+ 519 s_2 \left( \pisix \right)
- 1038 s_2 \left( \pitwo \right)
- 865 s_3 \left( \pisix \right)
\right. \right. \right. \nonumber \\
&& \left. \left. \left. ~~~~~~~~~~~~~~~~~~
+ 692 s_3 \left( \pitwo \right)
+ \frac{104}{3} \psi^\prime \left( \third \right)
+ \frac{1012}{27} \psi^\prime \left( \third \right) \zeta(3)
\right. \right. \right. \nonumber \\
&& \left. \left. \left. ~~~~~~~~~~~~~~~~~~
- \frac{3703}{2916} \psi^\prime \left( \third \right) \pi^2
+ \frac{3703}{3888} \left( \psi^\prime \left( \third \right) \right)^2
+ \frac{427}{576} \psi^{\prime\prime\prime} \left( \third \right)
\right. \right. \right. \nonumber \\
&& \left. \left. \left. ~~~~~~~~~~~~~~~~~~
+ \frac{173}{144} \ln^2(3) \sqrt{3} \pi
- \frac{173}{12} \ln(3) \sqrt{3} \pi
- \frac{5017}{3888} \sqrt{3} \pi^3
\right] C_F C_A^2
\right. \right. \nonumber \\
&& \left. \left. ~~~~~~~~~~~~~
+ \left[ 
880 \zeta(5)
- 105
- 572 \zeta(3)
- \frac{23}{27} \pi^2
+ \frac{23}{18} \psi^\prime \left( \third \right)
\right] C_F^2 C_A 
\right. \right. \nonumber \\
&& \left. \left. ~~~~~~~~~~~~~
- \frac{69}{2} C_F^3 \right] a^3 \right]
+ \left( \sum_f Q_f \right)^2 d^{abc} d^{abc} \left[ \frac{11}{3} 
- 8 \zeta(3) \right] a^3 \nonumber \\
&& +~ O(a^4) 
\end{eqnarray}
at three loops. For all the MOM schemes at this order the $d^{abc} d^{abc}$ 
term will be formally the same as the $\MSbar$ case. However, the numbers 
associated with the definition of the symmetric point renormalization of the 
underlying $3$-point vertices naturally appear in the full expression. Equally 
for the $\MOMq$ case we have
\begin{eqnarray}
R^{\MOMqs}(s) &=& \NF \sum_f Q_f^2 \left[ \frac{}{}
1 + 3 C_F a \right. \nonumber \\ 
&& \left. ~~~~~~~~~~~~~
+ \left[
\left[ 
- \frac{46}{3}
+ 16 \zeta(3)
\right] \Nf T_F C_F 
\right. \right. \nonumber \\
&& \left. \left. ~~~~~~~~~~~~~~~~~
+ \left[ 
\frac{407}{12}
- 44 \zeta(3)
- \frac{13}{9} \pi^2
+ \frac{13}{6} \psi^\prime \left( \third \right)
\right] C_F C_A 
\right. \right. \nonumber \\
&& \left. \left. ~~~~~~~~~~~~~~~~~
+ \left[ 
\frac{21}{2}
+ \frac{8}{9} \pi^2
- \frac{4}{3} \psi^\prime \left( \third \right)
\right] C_F^2 
\right] a^2 \right. \nonumber \\ 
&& \left. ~~~~~~~~~~~~~
+ \left[
\left[ 
- \frac{1553}{3}
- \frac{416}{27} \zeta(3) \pi^2
+ \frac{1004}{3} \zeta(3)
+ \frac{160}{3} \zeta(5)
+ \frac{2840}{81} \pi^2
- \frac{8}{27} \pi^4
\right. \right. \right. \nonumber \\
&& \left. \left. \left. ~~~~~~~~~~~~~~~~~~
- 24 s_2 \left( \pisix \right) 
+ 48 s_2 \left( \pitwo \right) 
+ 40 s_3 \left( \pisix \right) 
- 32 s_3 \left( \pitwo \right) 
- \frac{1024}{27} \psi^\prime \left( \third \right)
\right. \right. \right. \nonumber \\
&& \left. \left. \left. ~~~~~~~~~~~~~~~~~~
+ \frac{208}{9} \psi^\prime \left( \third \right) \zeta(3)
+ \frac{1}{9} \psi^{\prime\prime\prime} \left( \third \right)
- \frac{1}{18} \ln^2(3) \sqrt{3} \pi
+ \frac{2}{3} \ln(3) \sqrt{3} \pi
\right. \right. \right. \nonumber \\
&& \left. \left. \left. ~~~~~~~~~~~~~~~~~~
+ \frac{29}{486} \sqrt{3} \pi^3
\right] \Nf T_F C_F C_A 
\right. \right. \nonumber \\
&& \left. \left. ~~~~~~~~~~~~~~~~~
+ \left[ 
- \frac{346}{3}
+ \frac{256}{27} \zeta(3) \pi^2
+ 384 \zeta(3)
- 320 \zeta(5)
- \frac{880}{81} \pi^2
\right. \right. \right. \nonumber \\
&& \left. \left. \left. ~~~~~~~~~~~~~~~~~~~~~
+ \frac{440}{27} \psi^\prime \left( \third \right)
- \frac{128}{9} \psi^\prime \left( \third \right) \zeta(3)
\right] \Nf T_F C_F^2 
\right. \right. \nonumber \\
&& \left. \left. ~~~~~~~~~~~~~~~~~
+ \left[
96
- 64 \zeta(3)
- \frac{16}{9} \pi^2
\right] \Nf^2 T_F^2 C_F 
\right. \right. \nonumber \\
&& \left. \left. ~~~~~~~~~~~~~~~~~
+ \left[
\frac{7555}{12}
+ \frac{1144}{27} \zeta(3) \pi^2
- \frac{18535}{48} \zeta(3)
- \frac{440}{3} \zeta(5)
- \frac{32159}{648} \pi^2
\right. \right. \right. \nonumber \\
&& \left. \left. \left. ~~~~~~~~~~~~~~~~~~~~~
+ \frac{983}{486} \pi^4
- \frac{171}{2} s_2 \left( \pisix \right) 
+ 171 s_2 \left( \pitwo \right) 
+ \frac{285}{2} s_3 \left( \pisix \right) 
- 114 s_3 \left( \pitwo \right) 
\right. \right. \right. \nonumber \\
&& \left. \left. \left. ~~~~~~~~~~~~~~~~~~~~~
+ \frac{23447}{432} \psi^\prime \left( \third \right)
- \frac{572}{9} \psi^\prime \left( \third \right) \zeta(3)
- \frac{1759}{324} \psi^\prime \left( \third \right) \pi^2
\right. \right. \right. \nonumber \\
&& \left. \left. \left. ~~~~~~~~~~~~~~~~~~~~~
+ \frac{1759}{432} \left( \psi^\prime \left( \third \right) \right)^2
- \frac{23}{288} \psi^{\prime\prime\prime} \left( \third \right)
- \frac{19}{96} \ln^2(3) \sqrt{3} \pi
\right. \right. \right. \nonumber \\
&& \left. \left. \left. ~~~~~~~~~~~~~~~~~~~~~
+ \frac{19}{8} \ln(3) \sqrt{3} \pi
+ \frac{551}{2592} \sqrt{3} \pi^3
\right] C_F C_A^2 
\right. \right. \nonumber \\
&& \left. \left. ~~~~~~~~~~~~~~~~~
+ \left[ 
\frac{2441}{12}
- \frac{704}{27} \zeta(3) \pi^2
- 930 \zeta(3)
+ 880 \zeta(5)
- \frac{2665}{81} \pi^2
- \frac{688}{243} \pi^4
\right. \right. \right. \nonumber \\
&& \left. \left. \left. ~~~~~~~~~~~~~~~~~~~~~
+ 252 s_2 \left( \pisix \right) 
- 504 s_2 \left( \pitwo \right)
- 420 s_3 \left( \pisix \right) 
+ 336 s_3 \left( \pitwo \right)
\right. \right. \right. \nonumber \\
&& \left. \left. \left. ~~~~~~~~~~~~~~~~~~~~~
+ \frac{2665}{54} \psi^\prime \left( \third \right)
+ \frac{352}{9} \psi^\prime \left( \third \right) \zeta(3)
+ \frac{796}{81} \psi^\prime \left( \third \right) \pi^2
\right. \right. \right. \nonumber \\
&& \left. \left. \left. ~~~~~~~~~~~~~~~~~~~~~
- \frac{199}{27} \left( \psi^\prime \left( \third \right) \right)^2
- \frac{1}{6} \psi^{\prime\prime\prime} \left( \third \right)
+ \frac{7}{12} \ln^2(3) \sqrt{3} \pi
- 7 \ln(3) \sqrt{3} \pi
\right. \right. \right. \nonumber \\
&& \left. \left. \left. ~~~~~~~~~~~~~~~~~~~~~
- \frac{203}{324} \sqrt{3} \pi^3
\right] C_F^2 C_A 
\right. \right. \nonumber \\
&& \left. \left. ~~~~~~~~~~~~~~~~~
+ \left[ 
\frac{33}{2}
- 56 \zeta(3)
+ \frac{116}{3} \pi^2
- \frac{176}{243} \pi^4
+ 48 s_2 \left( \pisix \right) 
- 96 s_2 \left( \pitwo \right)
\right. \right. \right. \nonumber \\
&& \left. \left. \left. ~~~~~~~~~~~~~~~~~~~~~
- 80 s_3 \left( \pisix \right) 
+ 64 s_3 \left( \pitwo \right) 
- 58 \psi^\prime \left( \third \right)
- \frac{400}{81} \psi^\prime \left( \third \right) \pi^2
\right. \right. \right. \nonumber \\
&& \left. \left. \left. ~~~~~~~~~~~~~~~~~~~~~
+ \frac{100}{27} \left( \psi^\prime \left( \third \right) \right)^2
+ \frac{8}{9} \psi^{\prime\prime\prime} \left( \third \right)
+ \frac{1}{9} \ln^2(3) \sqrt{3} \pi
- \frac{4}{3} \ln(3) \sqrt{3} \pi
\right. \right. \right. \nonumber \\
&& \left. \left. \left. ~~~~~~~~~~~~~~~~~~~~~
- \frac{29}{243} \sqrt{3} \pi^3
\right] C_F^3 
\right. \right. \nonumber \\
&& \left. \left. ~~~~~~~~~~~~~~~~
- \frac{69}{2} C_F^3 \right] a^3 \right]
+ \left( \sum_f Q_f \right)^2 d^{abc} d^{abc} \left[ \frac{11}{3} 
- 8 \zeta(3) \right] a^3 \nonumber \\
&& +~ O(a^4) ~. 
\end{eqnarray}
As this $\MOMq$ scheme was the scheme which Celmaster and Gonsalves used for
their discussion of the magnitude of the higher order terms given the
underlying quark nature of the $R$-ratio, we have checked that the $O(a^2)$
term is in precise agreement with the corresponding term of \cite{5,6}.

Rather than repeat similar expressions for the other schemes as they add no
further enlightenment and are included in the associated date file anyway, we 
present the results in numerical form. This is more practical for relative 
comparison. To facillitate this we use a similar notation to \cite{10,11,12} 
and define $r^{\cal S}(s)$ by
\begin{equation}
R^{\cal S}(s) ~=~ \NF \left( \sum_f Q_f^2 \right) r^{\cal S}(s) ~.
\end{equation}
For arbitrary $\Nf$ we have for $SU(3)$ 
\begin{eqnarray}
r^{\MSbars}(s) &=& 
1 + 4.000000 a + \left[ - 1.844726 \Nf + 31.771318 \right] a^2 \nonumber \\
&& + \left[ - 0.331415 \Nf^2 - 76.808579 \Nf - 424.763877 - 26.443505 \eta^Q 
\right] a^3 \nonumber \\
&& + \left[ 5.508123 \Nf^3 - 204.143191 \Nf^2 + 4806.339848 \Nf 
+ 49.0568463 \Nf \eta^Q 
\right. \nonumber \\
&& \left. ~~~
- 1521.214892 \eta^Q - 40091.676394 \right] a^4 ~+~ O(a^5) \nonumber \\
r^{\MOMgs}(s) &=& 
1 + 4.000000 a + \left[ 11.822499 \Nf - 74.198637 \right] a^2 \nonumber \\
&& + \left[ 49.709397 \Nf^2 - 401.928165 \Nf - 335.201605 \right. 
\nonumber \\
&& \left. ~~~ - 26.443505 \eta^Q 
\right] a^3 ~+~ O(a^4) \nonumber \\
r^{\MOMhs}(s) &=& 
1 + 4.000000 a + \left[ 2.599718 \Nf - 42.421783 \right] a^2 \nonumber \\
&& + \left[ 0.507465 \Nf^2 + 74.704019 \Nf - 1418.822320 \right. 
\nonumber \\
&& \left. ~~~ - 26.443505 \eta^Q
\right] a^3 ~+~ O(a^4) \nonumber \\
r^{\MOMqs}(s) &=& 
1 + 4.000000 a + \left[ 2.599718 \Nf - 35.091780 \right] a^2 \nonumber \\
&& + \left[ 0.507465 \Nf^2 + 90.741952 \Nf - 1140.227694 \right. 
\nonumber \\
&& \left. ~~~ - 26.443505 \eta^Q
\right] a^3 ~+~ O(a^4) \nonumber \\
r^{\mMOMs}(s) &=& 
1 + 4.000000 a + \left[ 2.599718 \Nf - 24.562015 \right] a^2 \nonumber \\
&& + \left[ 0.507465 \Nf^2 + 85.202150 \Nf - 1634.833914 - 26.443505 \eta^Q 
\right] a^3 \nonumber \\
&& + \left[ 3.058056 \Nf^3 - 230.126428 \Nf^2 + 4880.206236 \Nf - 17400.630112 
\right. \nonumber \\
&& \left. ~~~ - 39.088169 \Nf \eta^Q - 403.976819 \eta^Q \right] a^4 ~+~ O(a^5)
\label{Rnf}
\end{eqnarray} 
where we have introduced
\begin{equation}
\eta^Q ~=~ \frac{\left( \sum_f Q_f \right)^2}{\left( \sum_f Q_f^2 \right)} ~.
\end{equation}
In effect (\ref{Rnf}) represents the main results of the article. With these
expressions one sees a similar feature to the $\beta$-functions in the sense
that comparing with the $\MSbar$ expression as the reference there is no clear 
pattern to the magnitude or signs of the coefficients. This is borne out if 
they are evaluated explicitly for various values of $\Nf$. For the non-singlet 
(NS) case, defined by $\eta^Q$~$=$~$0$, in ascending order of $\Nf$ we have 
\begin{eqnarray}
\left. r^{\MSbars}(s) \right|^{\mbox{\footnotesize{NS}}}_{\Nf=3} &=& 
1 + 4.000000 a + 26.237139 a^2 - 658.172348 a^3 \nonumber \\
&& -~ 27361.226258 a^4 ~+~ O(a^5) \nonumber \\
\left. r^{\MOMgs}(s) \right|^{\mbox{\footnotesize{NS}}}_{\Nf=3} &=& 
1 + 4.000000 a - 38.731139 a^2 - 1039.601525 a^3 ~+~ O(a^4) \nonumber \\
\left. r^{\MOMhs}(s) \right|^{\mbox{\footnotesize{NS}}}_{\Nf=3} &=& 
1 + 4.000000 a - 34.622629 a^2 - 1190.143080 a^3 ~+~ O(a^4) \nonumber \\
\left. r^{\MOMqs}(s) \right|^{\mbox{\footnotesize{NS}}}_{\Nf=3} &=& 
1 + 4.000000 a - 27.292626 a^2 - 863.434654 a^3 ~+~ O(a^4) \nonumber \\
\left. r^{\mMOMs}(s) \right|^{\mbox{\footnotesize{NS}}}_{\Nf=3} &=& 
1 + 4.000000 a - 16.762861 a^2 - 1374.660279 a^3 \nonumber \\
&& -~ 4748.581755 a^4 ~+~ O(a^5)
\label{rnf3}
\end{eqnarray} 
\begin{eqnarray}
\left. r^{\MSbars}(s) \right|^{\mbox{\footnotesize{NS}}}_{\Nf=4} &=& 
1 + 4.000000 a + 24.392413 a^2 - 737.300831 a^3 \nonumber \\
&& -~ 23780.088207 a^4 ~+~ O(a^5) \nonumber \\
\left. r^{\MOMgs}(s) \right|^{\mbox{\footnotesize{NS}}}_{\Nf=4} &=& 
1 + 4.000000 a - 26.908640 a^2 - 1147.563910 a^3 ~+~ O(a^4) \nonumber \\
\left. r^{\MOMhs}(s) \right|^{\mbox{\footnotesize{NS}}}_{\Nf=4} &=& 
1 + 4.000000 a - 32.022911 a^2 - 1111.886807 a^3 ~+~ O(a^4) \nonumber \\
\left. r^{\MOMqs}(s) \right|^{\mbox{\footnotesize{NS}}}_{\Nf=4} &=& 
1 + 4.000000 a - 24.692908 a^2 - 769.140448 a^3 ~+~ O(a^4) \nonumber \\
\left. r^{\mMOMs}(s) \right|^{\mbox{\footnotesize{NS}}}_{\Nf=4} &=& 
1 + 4.000000 a - 14.163143 a^2 - 1285.905875 a^3 \nonumber \\
&& -~ 1366.112459 a^4 ~+~ O(a^5)
\label{rnf4}
\end{eqnarray} 
\begin{eqnarray}
\left. r^{\MSbars}(s) \right|^{\mbox{\footnotesize{NS}}}_{\Nf=5} &=& 
1 + 4.000000 a + 22.547686 a^2 - 817.092143 a^3 \nonumber \\
&& -~ 20475.041592 a^4 ~+~ O(a^5) \nonumber \\
\left. r^{\MOMgs}(s) \right|^{\mbox{\footnotesize{NS}}}_{\Nf=5} &=& 
1 + 4.000000 a - 15.086140 a^2 - 1102.107500 a^3 ~+~ O(a^4) \nonumber \\
\left. r^{\MOMhs}(s) \right|^{\mbox{\footnotesize{NS}}}_{\Nf=5} &=& 
1 + 4.000000 a - 29.423193 a^2 - 1032.615604 a^3 ~+~ O(a^4) \nonumber \\
\left. r^{\MOMqs}(s) \right|^{\mbox{\footnotesize{NS}}}_{\Nf=5} &=& 
1 + 4.000000 a - 22.093189 a^2 - 673.831312 a^3 ~+~ O(a^4) \nonumber \\
\left. r^{\mMOMs}(s) \right|^{\mbox{\footnotesize{NS}}}_{\Nf=5} &=& 
1 + 4.000000 a - 11.563425 a^2 - 1196.136540 a^3 \nonumber \\
&& +~ 1629.497315 a^4 ~+~ O(a^5)
\label{rnf5}
\end{eqnarray} 
and
\begin{eqnarray}
\left. r^{\MSbars}(s) \right|^{\mbox{\footnotesize{NS}}}_{\Nf=6} &=& 
1 + 4.000000 a + 20.702961 a^2 - 897.546285 a^3 \nonumber \\
&& -~ 17413.037676 a^4 ~+~ O(a^5) \nonumber \\
\left. r^{\MOMgs}(s) \right|^{\mbox{\footnotesize{NS}}}_{\Nf=6} &=& 
1 + 4.000000 a - 3.263641 a^2 - 957.232297 a^3 ~+~ O(a^4) \nonumber \\
\left. r^{\MOMhs}(s) \right|^{\mbox{\footnotesize{NS}}}_{\Nf=6} &=& 
1 + 4.000000 a - 26.823475 a^2 - 952.329471 a^3 ~+~ O(a^4) \nonumber \\
\left. r^{\MOMqs}(s) \right|^{\mbox{\footnotesize{NS}}}_{\Nf=6} &=& 
1 + 4.000000 a - 19.493471 a^2 - 577.507245 a^3 ~+~ O(a^4) \nonumber \\
\left. r^{\mMOMs}(s) \right|^{\mbox{\footnotesize{NS}}}_{\Nf=6} &=& 
1 + 4.000000 a - 8.963706 a^2 - 1105.352275 a^3 \nonumber \\
&& +~ 4256.595899 a^4 ~+~ O(a^5) ~.
\label{rnf6}
\end{eqnarray} 
Interestingly the four loop term of the $\mMOM$ expression for each $\Nf$ value
is an order of magnitude smaller than the $\MSbar$ term, while the three loop
term is consistently higher but not by the same amount. For the larger values
of $\Nf$ the two loop term of the $\MOMq$ scheme is marginally smaller (in our
conventions) than the $\MSbar$ partner which would support the original 
observation of \cite{5,6}. This is also the case for the $\MOMg$ scheme but 
not for $\MOMh$. Both these schemes were not examined in \cite{5,6}. However,
it should not be the case that the large variation in the corrections lead to 
significantly different interpretations of the $R$-ratio especially in the 
context of comparing with experimental data. Therefore we devote the next 
section to a more detailed analysis of these results. 

\sect{Analysis.}

For the first part of our analysis we note that we have reproduced the results
of Table III of \cite{6}. In that table the two loop correction to
$r^{\MOMqs}(s)$ was evaluated for each $\Nf$ at a particular value of the 
momentum and representative value of $\Lambda^{\MOMqs}$. It was compared to the
same evaluation for the $\MSbar$ scheme. The observation was that the higher 
order corrections appeared to improve the convergence with both schemes giving 
approximately the same values. In repeating this exercise we need to first 
recall the method used and within this append the necessary formalism to extend
it to higher loop orders. First, we denote the running coupling constant at the
$L$th loop order by $a_L^{\cal S}(Q,\Lambda)$ and define each of the ones we 
require by
\begin{eqnarray}
a_2^{\cal S}(Q,\Lambda^{\cal S}) &=&
\frac{1}{b_0^{\cal S} L^{\cal S}}
\left[ 1 - \frac{b_1^{\cal S} \ln (L^{\cal S})}{{b_0^{\cal S}}^2 L^{\cal S}}
\right] \nonumber \\
a_3^{\cal S}(Q,\Lambda^{\cal S}) &=&
\frac{1}{b_0^{\cal S} L^{\cal S}}
\left[ 1 - \frac{b_1^{\cal S} \ln (L^{\cal S})}{{b_0^{\cal S}}^2 L^{\cal S}}
+ \left[ {b_1^{\cal S}}^2 \left[ \ln^2 (L^{\cal S}) - \ln (L^{\cal S}) - 1 
\right]
+ b_0^{\cal S} b_2^{\cal S} \right] \frac{1}{{b_0^{\cal S}}^4 {L^{\cal S}}^2}
\right] \nonumber \\
a_4^{\cal S}(Q,\Lambda^{\cal S}) &=&
\frac{1}{b_0^{\cal S} L^{\cal S}}
\left[ 1 - \frac{b_1^{\cal S} \ln (L^{\cal S})}{{b_0^{\cal S}}^2 L^{\cal S}}
+ \left[ {b_1^{\cal S}}^2 \left[ \ln^2 (L^{\cal S}) - \ln (L^{\cal S}) - 1 
\right]
+ b_0^{\cal S} b_2^{\cal S} \right] \frac{1}{{b_0^{\cal S}}^4 {L^{\cal S}}^2}
\right. \nonumber \\
&& \left. ~~~~~~~~~
- \left[ {b_1^{\cal S}}^3 \left[ \ln^3 (L^{\cal S}) - \frac{5}{2} \ln^2 
(L^{\cal S})
- 2 \ln (L^{\cal S}) + \frac{1}{2} \right]
+ 3 b_0^{\cal S} b_1^{\cal S} b_2^{\cal S} \ln (L^{\cal S}) 
\right. \right. \nonumber \\
&& \left. \left. ~~~~~~~~~~~~~
- \frac{1}{2} {b_0^{\cal S}}^2 b_3^{\cal S} \right]
\frac{1}{{b_0^{\cal S}}^6 {L^{\cal S}}^3} \right] 
\end{eqnarray}
for each scheme ${\cal S}$ where 
\begin{equation}
L^{\cal S} ~=~ \ln \left( \frac{Q^2}{{\Lambda^{\cal S}}^2} \right) ~.
\end{equation} 
We note that there are different choices for the higher order expressions but 
we have chosen to take the definitions recorded in the review in \cite{35}. 
Next if we define the perturbative coefficients of $r^{\cal S}(s)$ via
\begin{equation}
r^{\cal S}(s) ~=~ \sum_{n=0}^\infty r_{n}^{\cal S}(s) {a^{\cal S}}^n
\end{equation}
where $r_0^{\cal S}$~$=$~$1$ for all schemes then we can define the partial
sums of the series by
\begin{equation}
a_{pq}^{\cal S} \left( \frac{Q^2}{{\Lambda^{\cal S}}^2} \right) ~=~ 
\sum_{n=1}^p r_n^{\cal S}(s) 
\left( a_q^{\cal S}(Q,\Lambda^{\cal S}) \right)^n ~.
\end{equation}
We note that the series starts from the $O(a)$ term and hence one has in effect
defined an effective coupling constant whose root is in the $R$-ratio. Indeed
as a side comment we note that this is a starting point for the method of
effective charges discussed in \cite{36,37,38,39}. In \cite{7} the three loop 
$\beta$-function for this effective charge was constructed based on the 
formalism given in \cite{38}. We have repeated this exercise as part of the 
check on our partial sum construction here in each of the MOM schemes of 
interest and found that they are all formally equivalent. This is as it should 
be since the $\beta$-function of this effective charge method is a 
renormalization group invariant. 

{\begin{table}[ht]
\begin{center}
\begin{tabular}{|c|c||c|c|c|c|c|}
\hline
$p$ & $q$ & $\MSbar$ & $\MOMg$ & $\MOMh$ & $\MOMq$ & $\mMOM$ \\
\hline
1 & 1 & 0.0707 & 0.0848 & 0.0918 & 0.0881 & 0.0833 \\
1 & 2 & 0.0581 & 0.0683 & 0.0733 & 0.0707 & 0.0672 \\
1 & 3 & 0.0592 & 0.0699 & 0.0753 & 0.0681 & 0.0695 \\
1 & 4 & 0.0571 &        &        &        & 0.0698 \\
\hline
2 & 2 & 0.0629 & 0.0639 & 0.0635 & 0.0638 & 0.0640 \\
2 & 3 & 0.0641 & 0.0653 & 0.0649 & 0.0617 & 0.0661 \\
2 & 4 & 0.0643 &        &        &        & 0.0663 \\
\hline
3 & 3 & 0.0615 & 0.0594 & 0.0580 & 0.0584 & 0.0598 \\
3 & 4 & 0.0616 &        &        &        & 0.0599 \\
\hline
4 & 4 & 0.0606 &        &        &        & 0.0601 \\
\hline
\end{tabular}
\end{center}
\begin{center}
{Table $2$. Values of $a_{pq}^{\cal S} \left( 
\frac{Q^2}{{\Lambda^{\cal S}}^2} \right)$ for $\Nf$~$=$~$5$ with 
$\Lambda^{\MSbars}$~$=$~$500$MeV and $Q$~$=$~$20$GeV.}
\end{center}
\end{table}}

Returning to the comparison with Table III of \cite{6} in the present notation 
the results which were presented in \cite{6} were $a^{\cal S}_{12}$ and 
$a^{\cal S}_{22}$ for $\MSbar$ and $\MOMq$ with $Q$~$=$~$3$, $5$, $20$ and 
$40$GeV respectively for $\Nf$~$=$~$3$, $4$, $5$ and $6$. The choices for the 
values of $\Lambda$ were $\Lambda^{\MSbars}$~$=$~$500$MeV and 
$\Lambda^{\MOMqs}$~$=$~$850$MeV. From the formalism given here with these 
specific values it is straightforward to recover Table III of \cite{6}. 
However, we have extended it for each of the MOM schemes as well as for the 
$\MSbar$ scheme. The results for $\Nf$~$=$~$5$ are given in Table $2$ with 
Table $3$ corresponding to the $\Nf$~$=$~$6$ case. Several general comments are
in order. First, for $\MOMg$, $\MOMh$ and $\MOMq$ the absent entries are due to
the fact that the four loop $\beta$-functions for these schemes are not known. 
{\begin{table}[ht]
\begin{center}
\begin{tabular}{|c|c||c|c|c|c|c|}
\hline
$p$ & $q$ & $\MSbar$ & $\MOMg$ & $\MOMh$ & $\MOMq$ & $\mMOM$ \\
\hline
1 & 1 & 0.0652 & 0.0723 & 0.0809 & 0.0780 & 0.0742 \\
1 & 2 & 0.0566 & 0.0622 & 0.0690 & 0.0667 & 0.0637 \\
1 & 3 & 0.0569 & 0.0617 & 0.0688 & 0.0634 & 0.0641 \\
1 & 4 & 0.0571 &        &        &        & 0.0645 \\
\hline
2 & 2 & 0.0608 & 0.0614 & 0.0610 & 0.0613 & 0.0615 \\
2 & 3 & 0.0611 & 0.0609 & 0.0609 & 0.0585 & 0.0618 \\
2 & 4 & 0.0613 &        &        &        & 0.0621 \\
\hline
3 & 3 & 0.0585 & 0.0574 & 0.0560 & 0.0562 & 0.0572 \\
3 & 4 & 0.0587 &        &        &        & 0.0575 \\
\hline
4 & 4 & 0.0580 &        &        &        & 0.0580 \\
\hline
\end{tabular}
\end{center}
\begin{center}
{Table $3$. Values of $a_{pq}^{\cal S} \left( 
\frac{Q^2}{{\Lambda^{\cal S}}^2} \right)$ for $\Nf$~$=$~$6$ with 
$\Lambda^{\MSbars}$~$=$~$500$MeV and $Q$~$=$~$40$GeV.}
\end{center}
\end{table}}
Next, in each Table we have used $\Lambda^{\MSbars}$~$=$~$500$MeV for each of 
the values of $\Nf$. This is merely to gain a general appreciation of the 
effect of including higher order corrections but a different choice could be 
made. Rather than take $\Lambda^{\MOMqs}$ to be $850$MeV we use the values of 
the $\Lambda$ parameter ratios for the Landau gauge given in Table $1$. The 
small variation in the ratio for $\MOMq$ appears to be the justification behind
the choice of $850$MeV used in \cite{6} for all values of $\Nf$ considered 
there. However, we have chosen to apply the more precise ratios here mainly 
because the variation over the range in $\Nf$ is more significant for several 
of the other schemes. Therefore, the values given in Tables $2$ and $3$ which 
are to be compared to those in \cite{6} will not be precisely the same for this
reason. Though we emphasise that taking the same values of the momentum and 
parameters used in \cite{6} we do find exact agreement. We have not provided 
tables for $\Nf$~$=$~$3$ and $4$ because the values of $Q$ used in \cite{6} are
too low. At three loops for these values of $Q$ the contributions from the 
logarithm are large and affect the analysis. 

Examining the entries in Tables $2$ and $3$ perhaps a reasonable guide to the 
rate of convergence can be seen from looking at the values for 
$a_{LL}^{\cal S}$ at the $L$th loop. Though in this discussion we exclude 
$L$~$=$~$1$ as it has no true contact with loop corrections. For $\MSbar$ and 
$\mMOM$ there are three such terms and for both $\Nf$~$=$~$5$ and $6$ it 
appears that not only both converge but they seem to be converging to the same 
value. In most cases the convergence is not monotonic decreasing unlike for 
$\Nf$~$=$~$5$ for the $\MSbar$ scheme. For the other three schemes there are 
only two values to comment on and while each appear to be roughly the same 
value for $L$~$=$~$3$ a conclusion cannot really be drawn as to the final 
value. Although they are not dissimilar to the corresponding number for the 
$\mMOM$ scheme. More instructive in understanding the situation is to examine 
plots of these partial sums for each of the schemes. 
In Figures $1$ and $2$ we have plotted $a_{LL}^{\cal S}(x)$ for $L$~$=$~$3$ and
$4$ respectively for each of the four values of $\Nf$ where we use the 
dimensionless variable $x$ as shorthand for $x$~$=$~$s/\Lambda$. Although we 
only have two schemes to compare in Figure $2$. Unlike the data in Tables $2$ 
and $3$, for the plots we provide we do not fix the value of 
$\Lambda^{\MSbars}$ but use it as the reference $\Lambda$ to which the 
parameter in the other schemes are related to it by the ratios in Table $1$. We
have not provided the graphs for the two loop case as the lines for all the 
schemes lie on top of each other. This is because at this order the scheme 
dependence cancels out from the $\Lambda$ parameter relation and the coupling 
constant map as well as the fact that in the Landau gauge the one and two loop 
coefficients of each $\beta$-function are the same. Only at three loops does 
the scheme dependence become evident which is due in the main to the scheme 
differences in $b_3^{\cal S}$. Indeed it is for this reason that scheme issues 
could not be examined in the $R$-ratio until the three loop MOM QCD
$\beta$-functions were known, \cite{20}. However for larger values of $Q$ the 
variation in values between each scheme diminishes which is not unexpected. 
{\begin{figure}[ht]
\includegraphics[width=7.6cm,height=8cm]{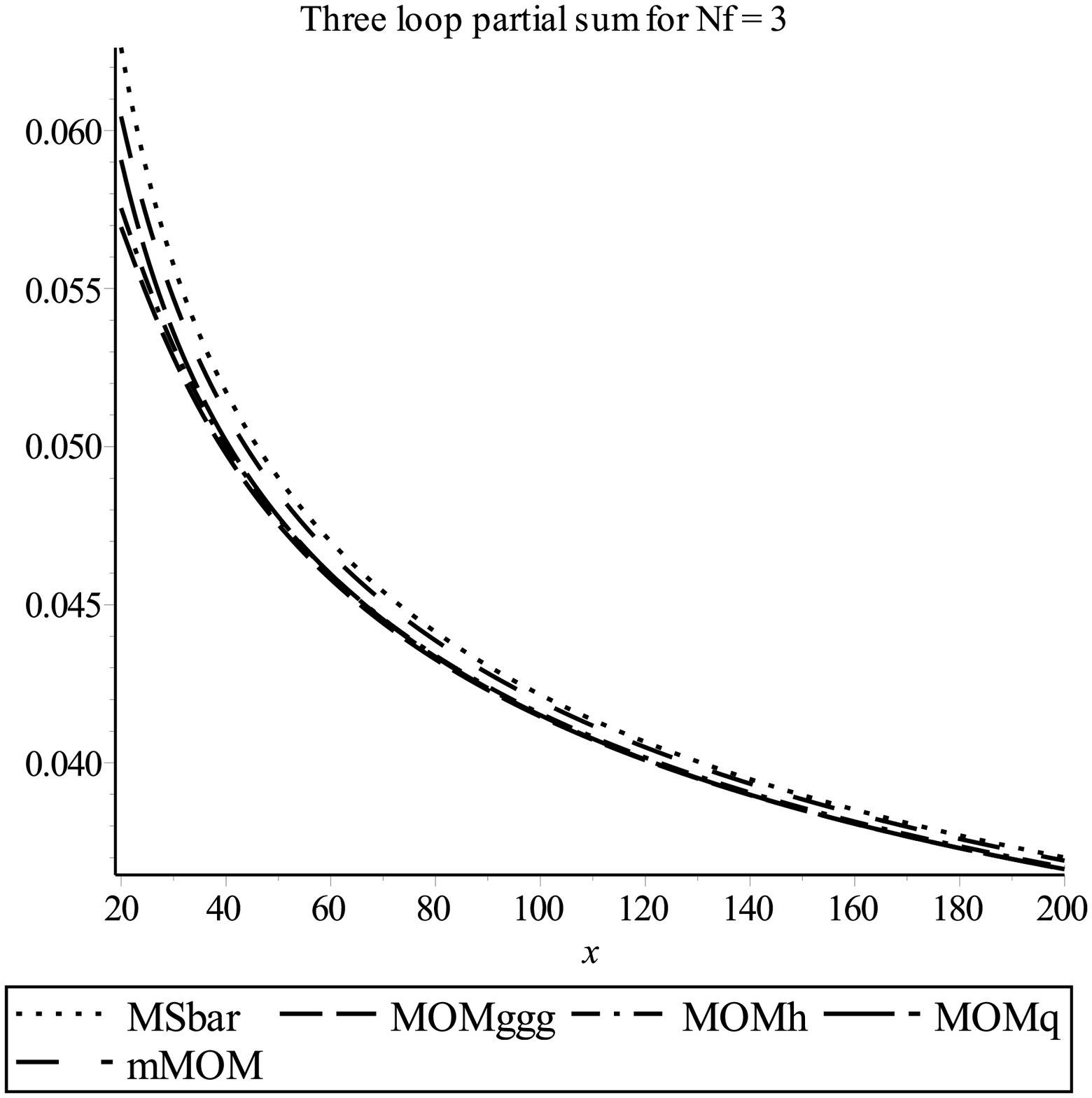}
\quad
\includegraphics[width=7.6cm,height=8cm]{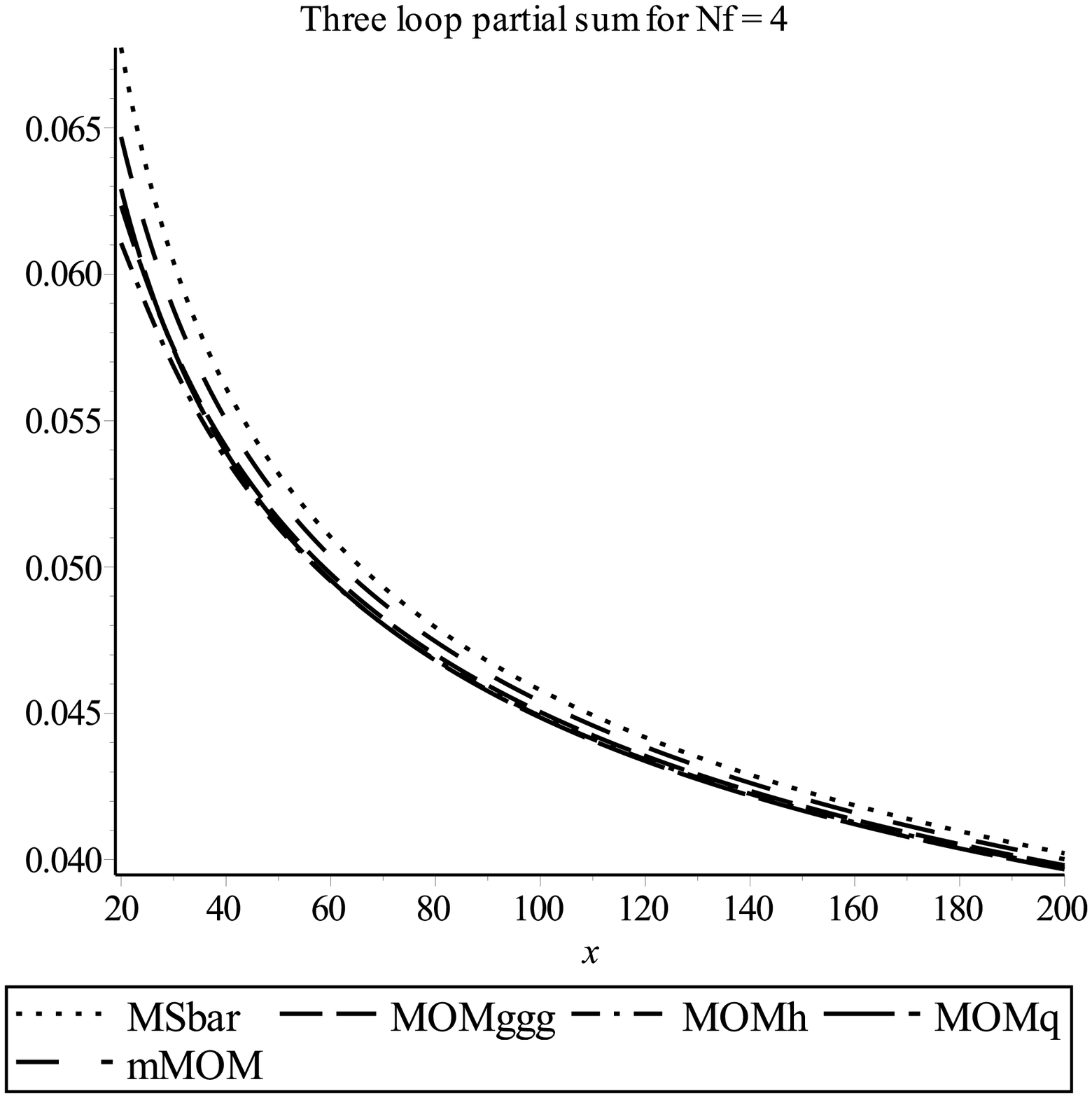}

\vspace{0.8cm}
\includegraphics[width=7.6cm,height=8cm]{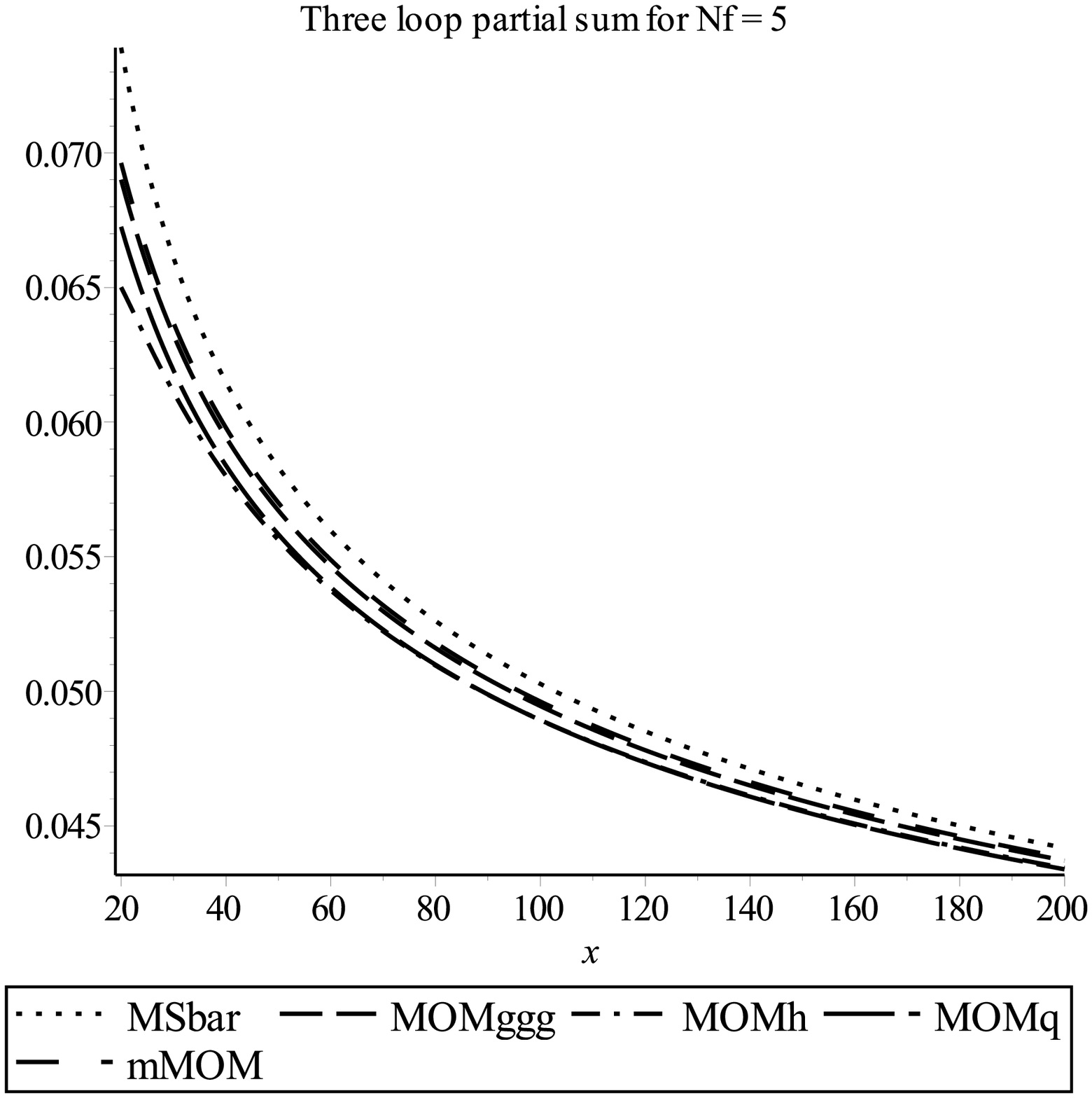}
\quad
\includegraphics[width=7.6cm,height=8cm]{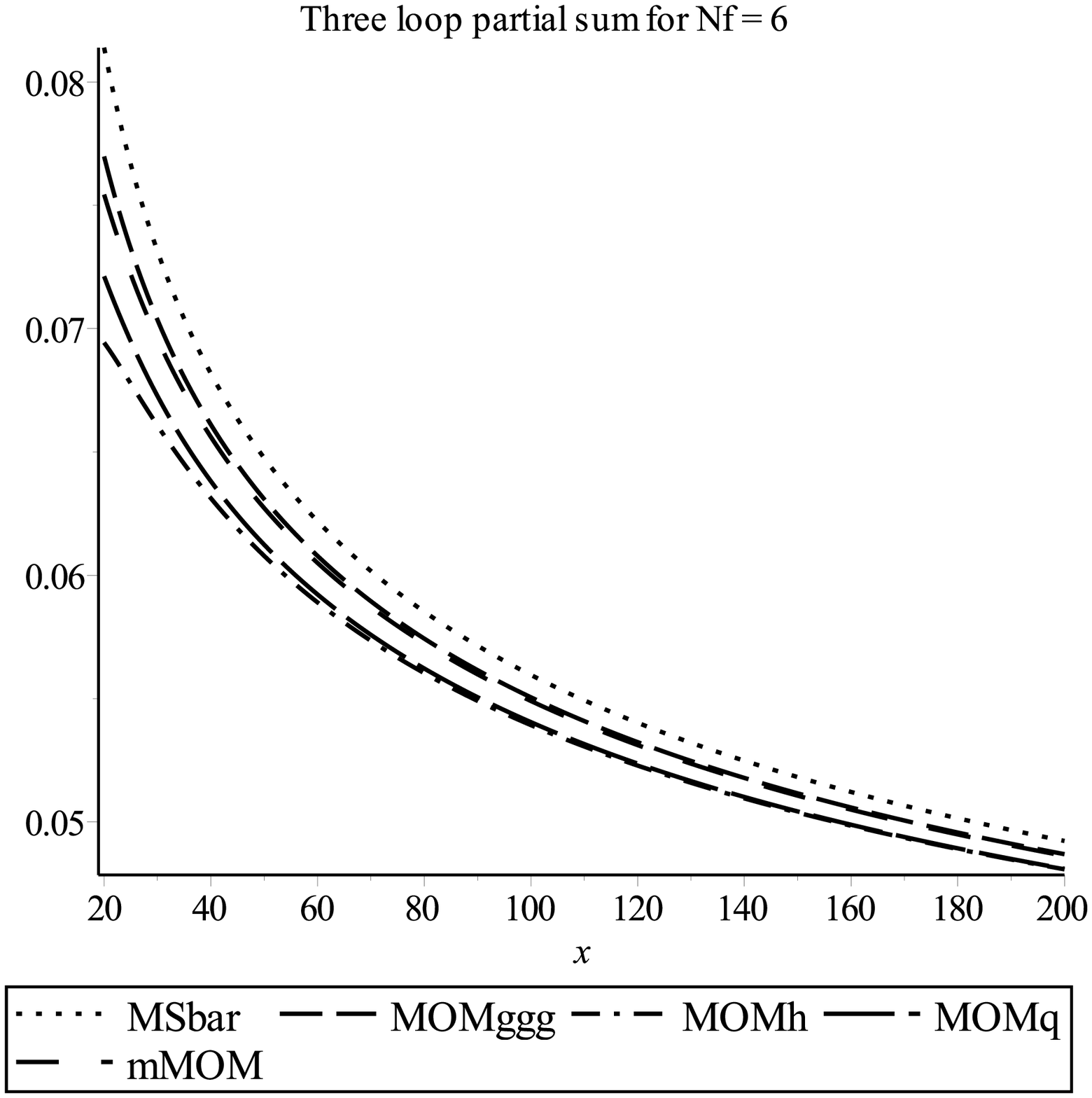}
\caption{Comparison of $a_{33}^{\cal S}(x)$ for the various schemes for 
$\Nf$~$=$~$3$, $4$, $5$ and $6$.}
\end{figure}}
At three loops the variation in values is not huge for $x$~$=$~$20$ except for 
$\Nf$~$=$~$6$. With these plots one can examine the values in Tables $2$ and 
$3$. Recalling that for these if we have $\Lambda^{\MSbars}$~$=$~$500$MeV then 
at $x$~$=$~$80$ in the $\Nf$~$=$~$5$ and $6$ plots of Figure $1$ we can see 
that the $\MOMg$ and $\mMOM$ scheme values are on a par with each other. 
Additionally the $\MOMh$ and $\MOMq$ lines are virtually the same with the 
$\MSbar$ line appearing to be at odds with both. Although at four loops we have
only two schemes to compare with there is very little to distinguish the curves
for all values of $\Nf$. As discussed earlier this may be due to these schemes 
being of a similar nature. 

While this comparison between schemes is instructive in observing over what 
ranges the schemes give similar values for the partial sums, it is also useful 
to compare the convergence within each scheme for $a_{LL}^{\cal S}(x)$. We have
provided these plots in Figures $3$, $4$, $5$, $6$ and $7$ for the $\MSbar$, 
$\MOMg$, $\MOMh$, $\MOMq$ and $\mMOM$ schemes respectively for $L$~$=$~$2$, $3$
and $4$. 
{\begin{figure}[ht]
\includegraphics[width=7.6cm,height=8cm]{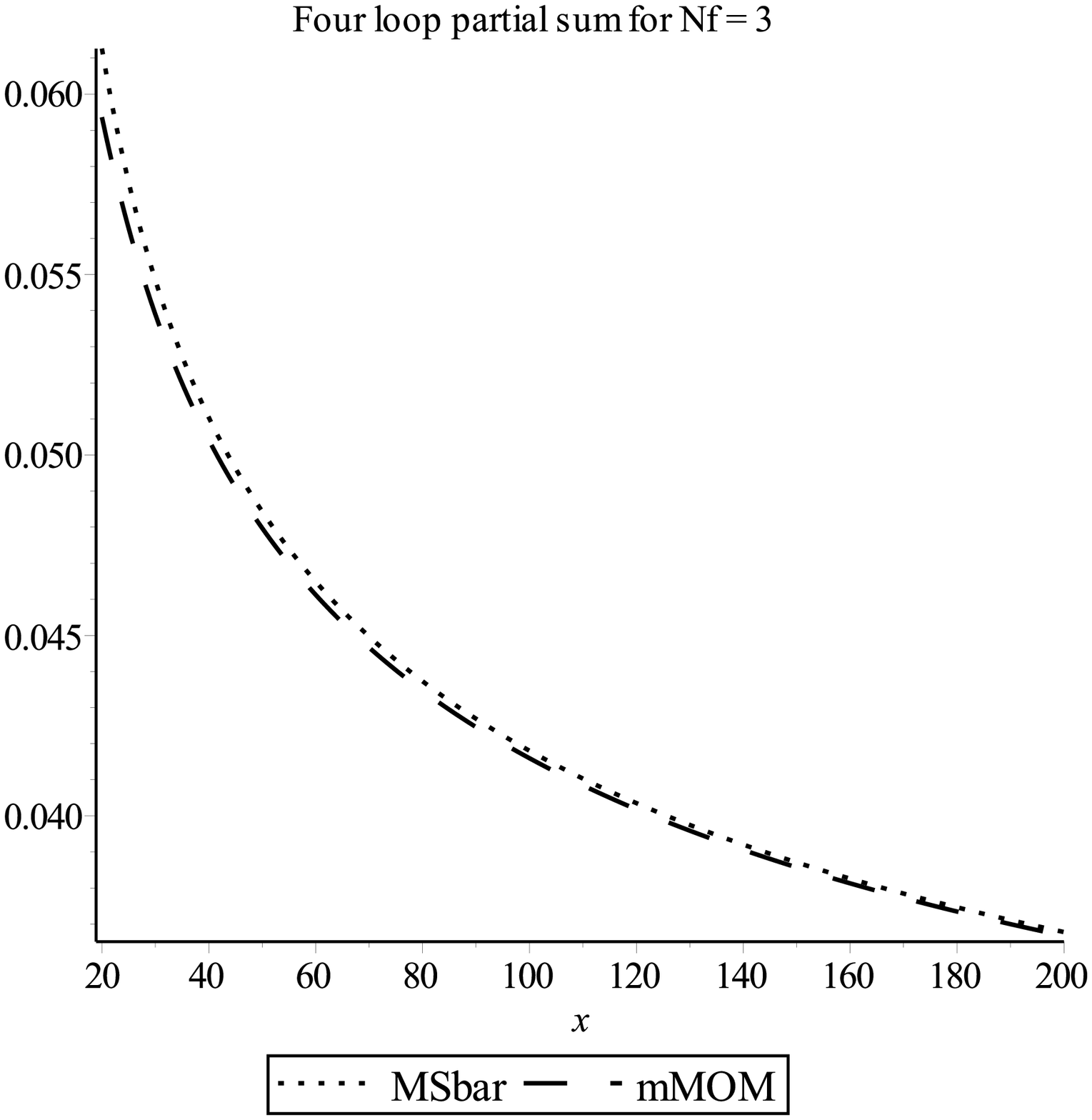}
\quad
\includegraphics[width=7.6cm,height=8cm]{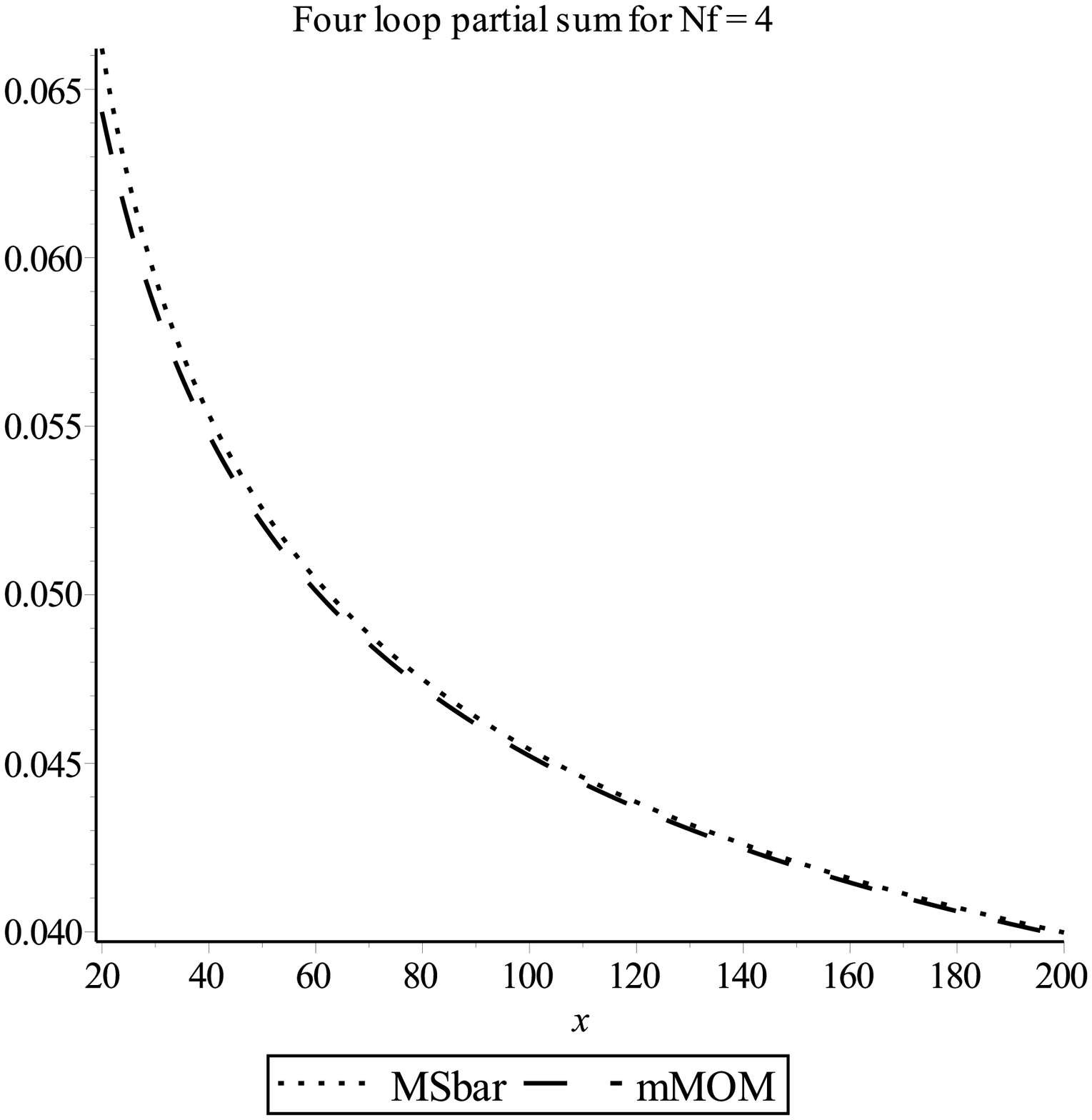}

\vspace{0.8cm}
\includegraphics[width=7.6cm,height=8cm]{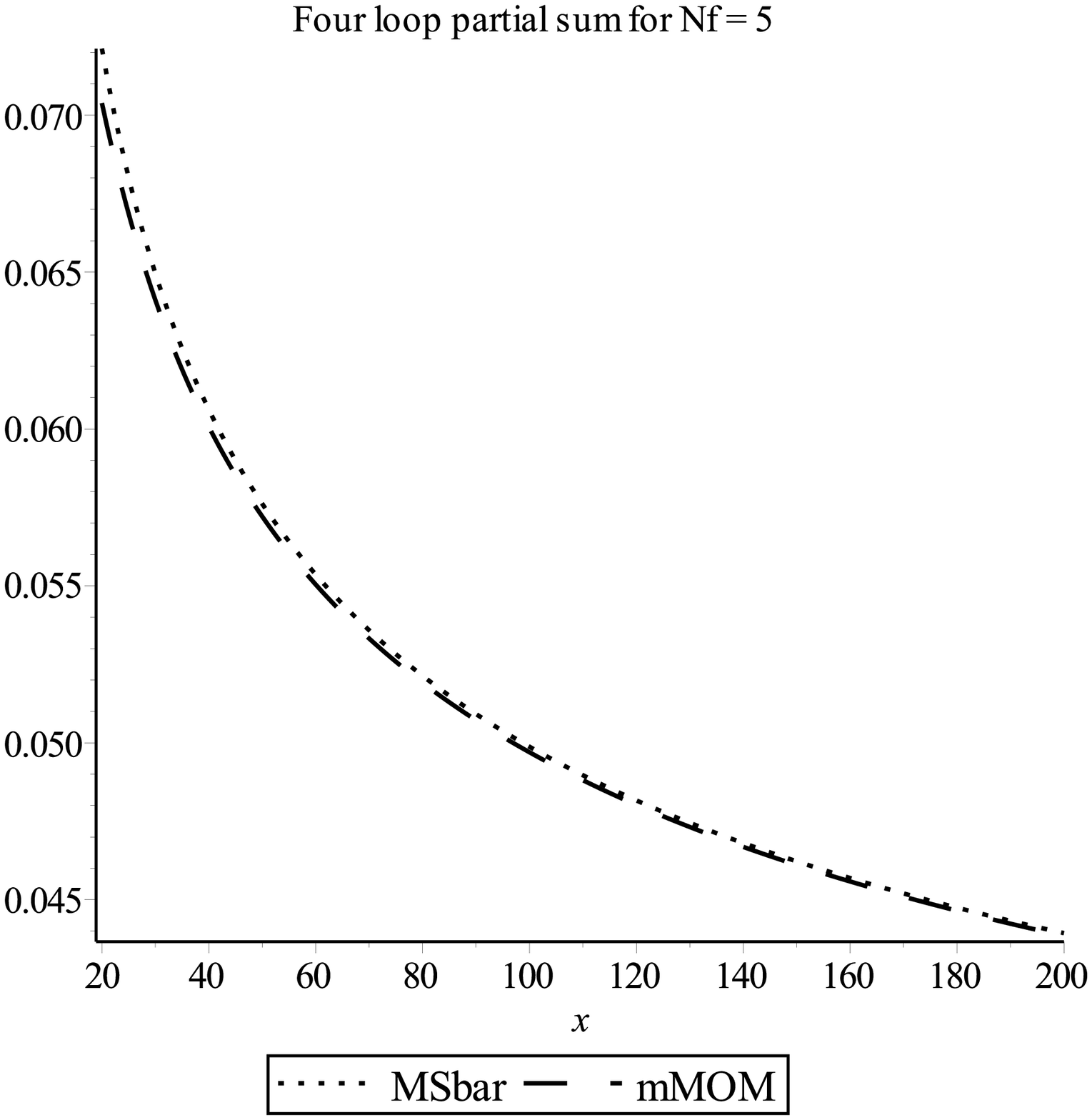}
\quad
\includegraphics[width=7.6cm,height=8cm]{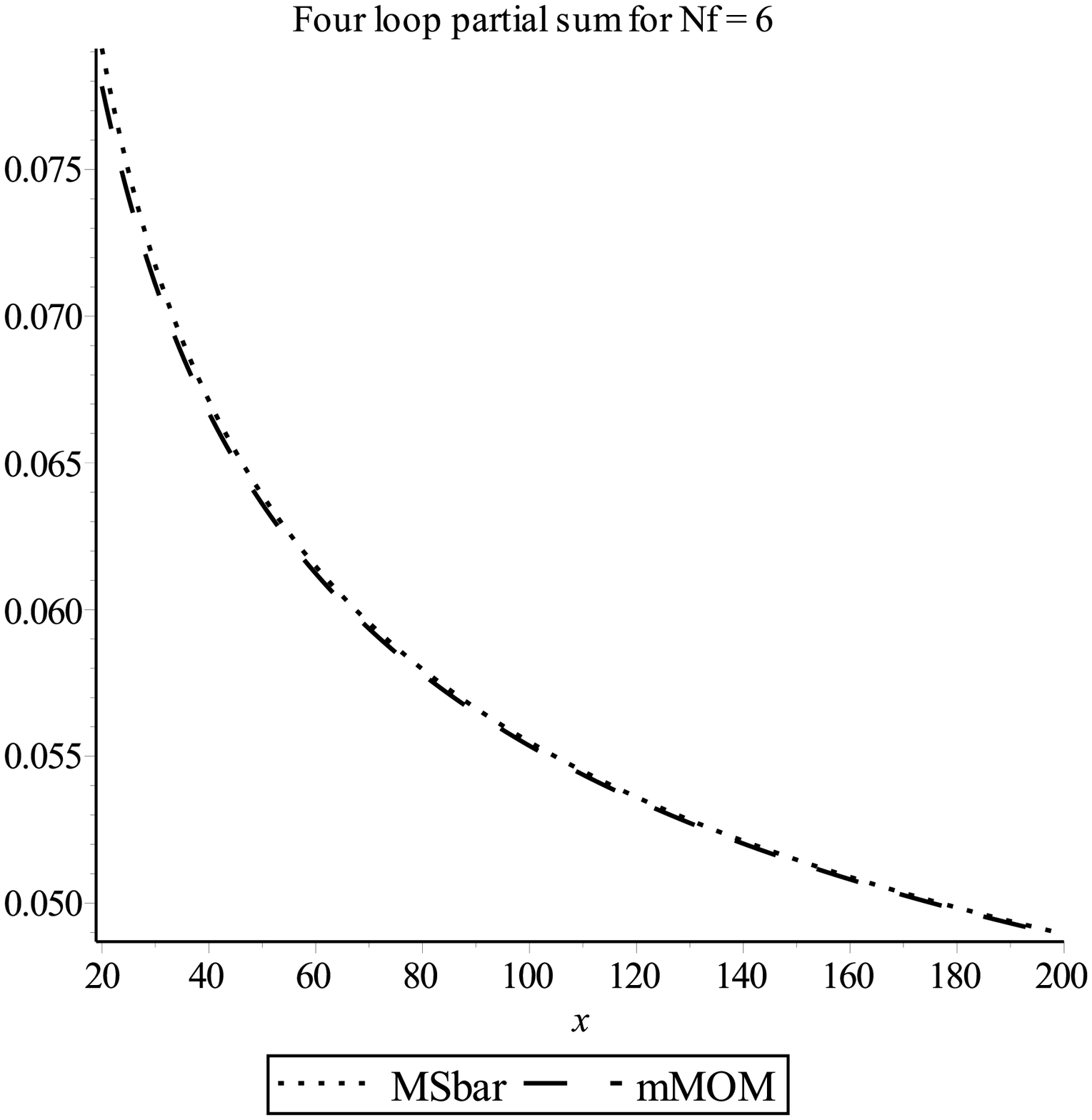}
\caption{Comparison of $a_{44}^{\cal S}(x)$ for $\MSbar$ and $\mMOM$ for
$\Nf$~$=$~$3$, $4$, $5$ and $6$.}
\end{figure}}
Overall the partial sums decrease in value as $L$ increases, with the 
largest reduction being for larger values of $\Nf$. For the two cases where 
four loop information is available the four loop plots are not significantly 
different from the three loop ones. This would be consistent with the 
observation that an increase in loop order may not improve precision by very 
much. However, the situation with the other three schemes is not conclusive. 
Each share the same property that there is a relatively large drop in the value
from $L$~$=$~$2$ to $3$. In light of a similar observation for the $\MSbar$ and
$\mMOM$ cases this would suggest that a determination of the four loop 
$\beta$-function would appear necessary in order to see whether that correction
was significantly different from the three loop one. Once that was resolved the
main issue would be if each of the schemes gave a similar precision. The curves
in Figure $2$ would be encouraging in this respect.
{\begin{figure}[hb]
\includegraphics[width=7.6cm,height=8cm]{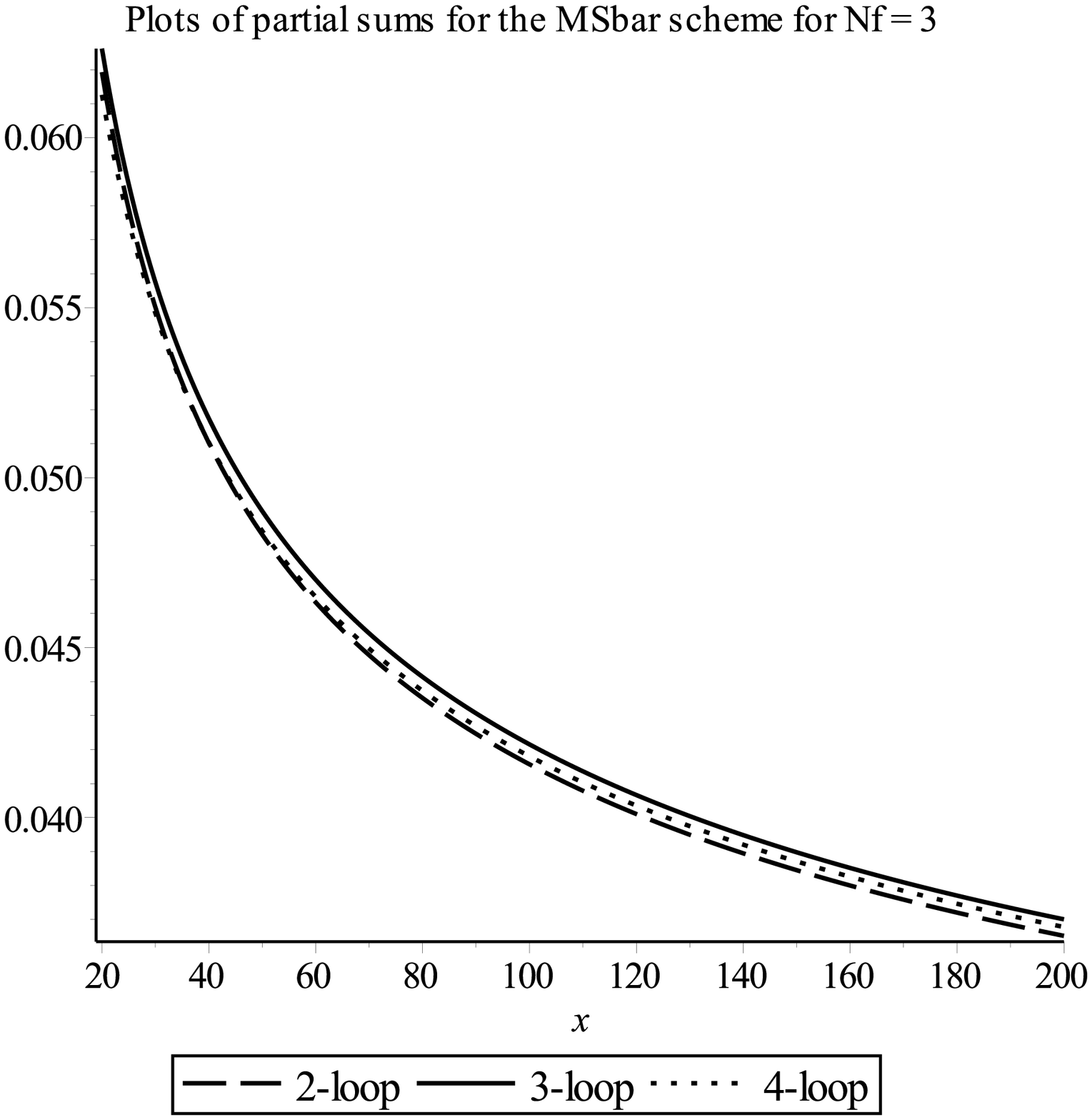}
\quad
\includegraphics[width=7.6cm,height=8cm]{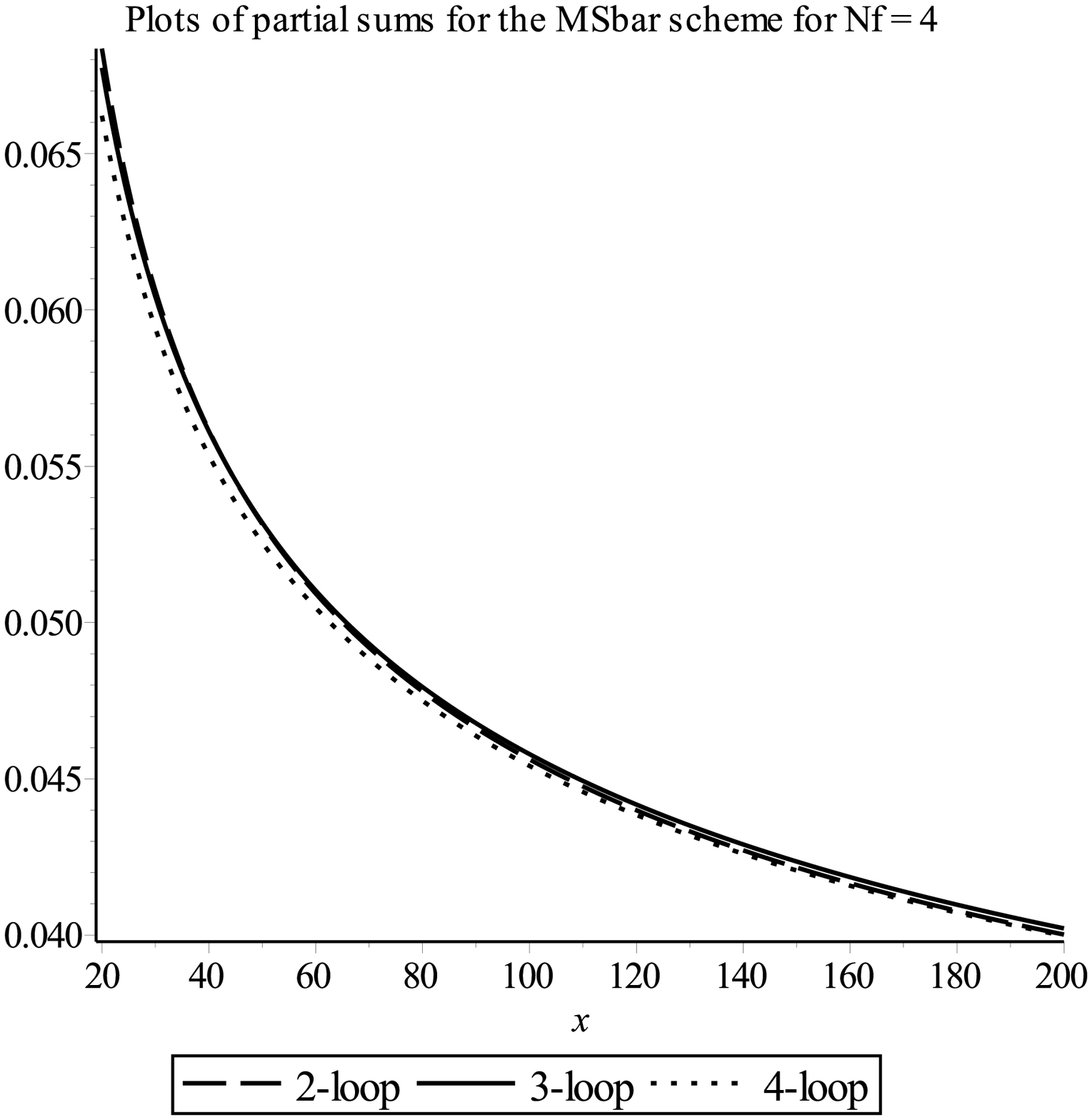}

\vspace{0.8cm}
\includegraphics[width=7.6cm,height=8cm]{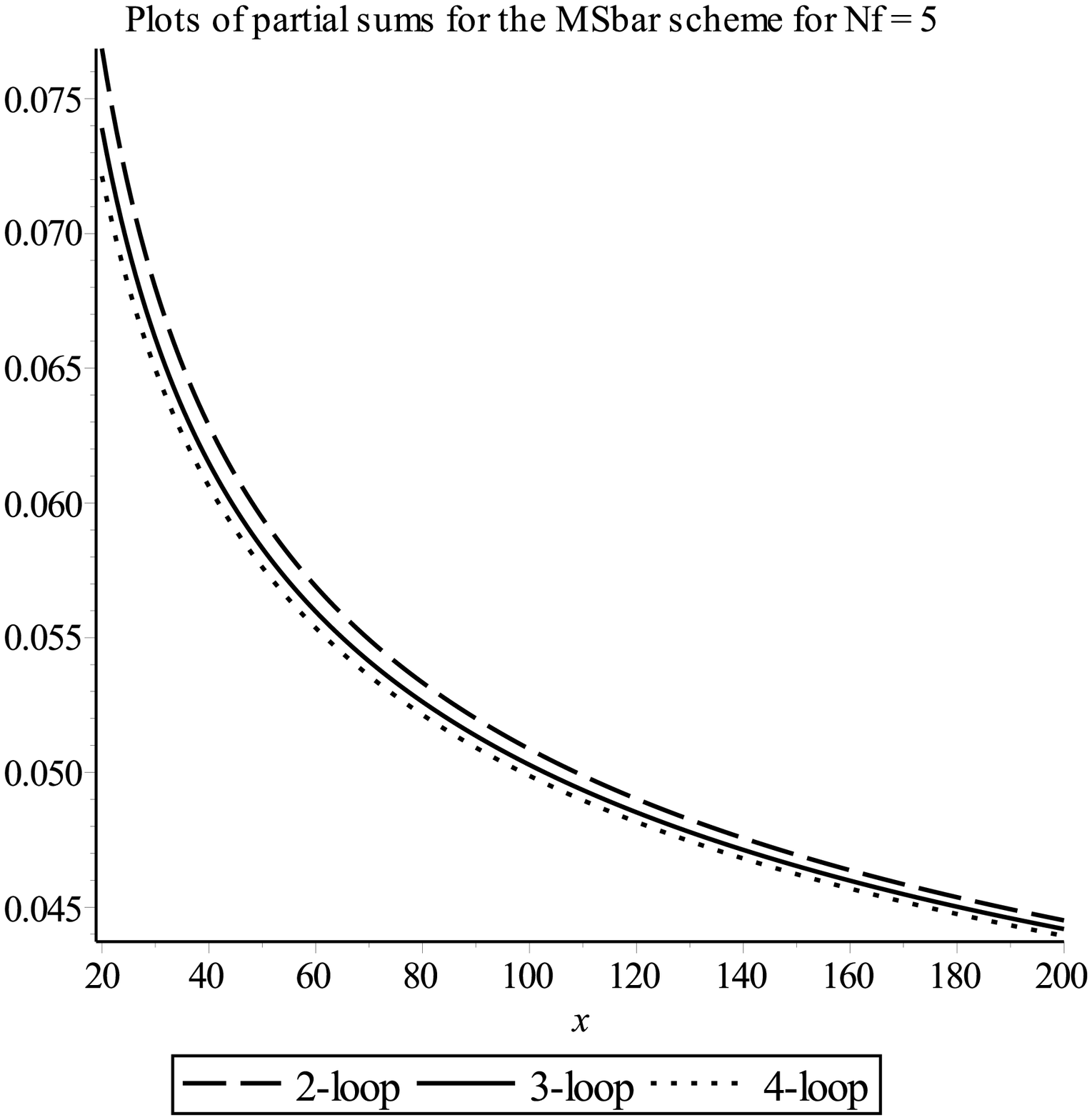}
\quad
\includegraphics[width=7.6cm,height=8cm]{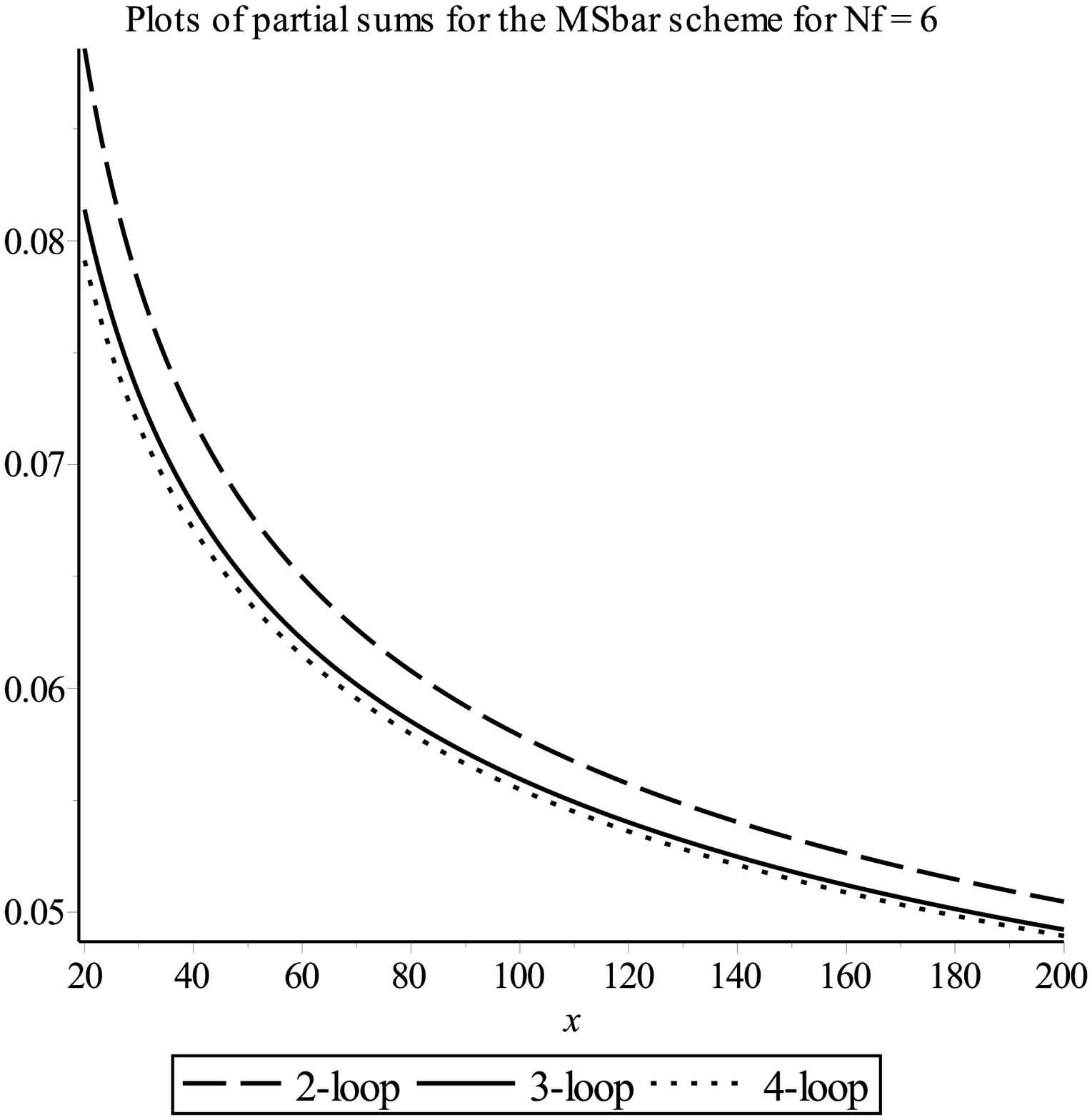}
\caption{Plots of $a_{LL}^{\MSbars}(x)$ for $L$~$=$~$2$, $3$ and $4$.}
\end{figure}}
However, this has to be tempered with the fact that both the $\MSbar$ and 
$\mMOM$ schemes have a similar aspect in their definition. In some sense they 
are related to an external momentum configuration of the vertex function where 
the subtraction is defined at a point where one external leg is nullified. This
is evident for the $\mMOM$ scheme. For the $\MSbar$ scheme it is less apparent.
To determine the renormalization constants for the coupling constant in QCD in 
$\MSbar$ there are several computational approaches. For reasons of 
calculational ease for each of the three $3$-point vertices in the QCD 
Lagrangian one can set the external momentum of a specific leg to zero. This 
reduces all the Feynman graphs in effect to $2$-point ones and so the 
extraction of the poles in the regularizing parameter are no more difficult 
than in computing the poles for the wave function renormalization. Moreover, 
this nullification procedure, known as infrared rearrangement \cite{18,19}, 
does not introduce any infrared problems. 
{\begin{figure}[ht]
\includegraphics[width=7.6cm,height=8cm]{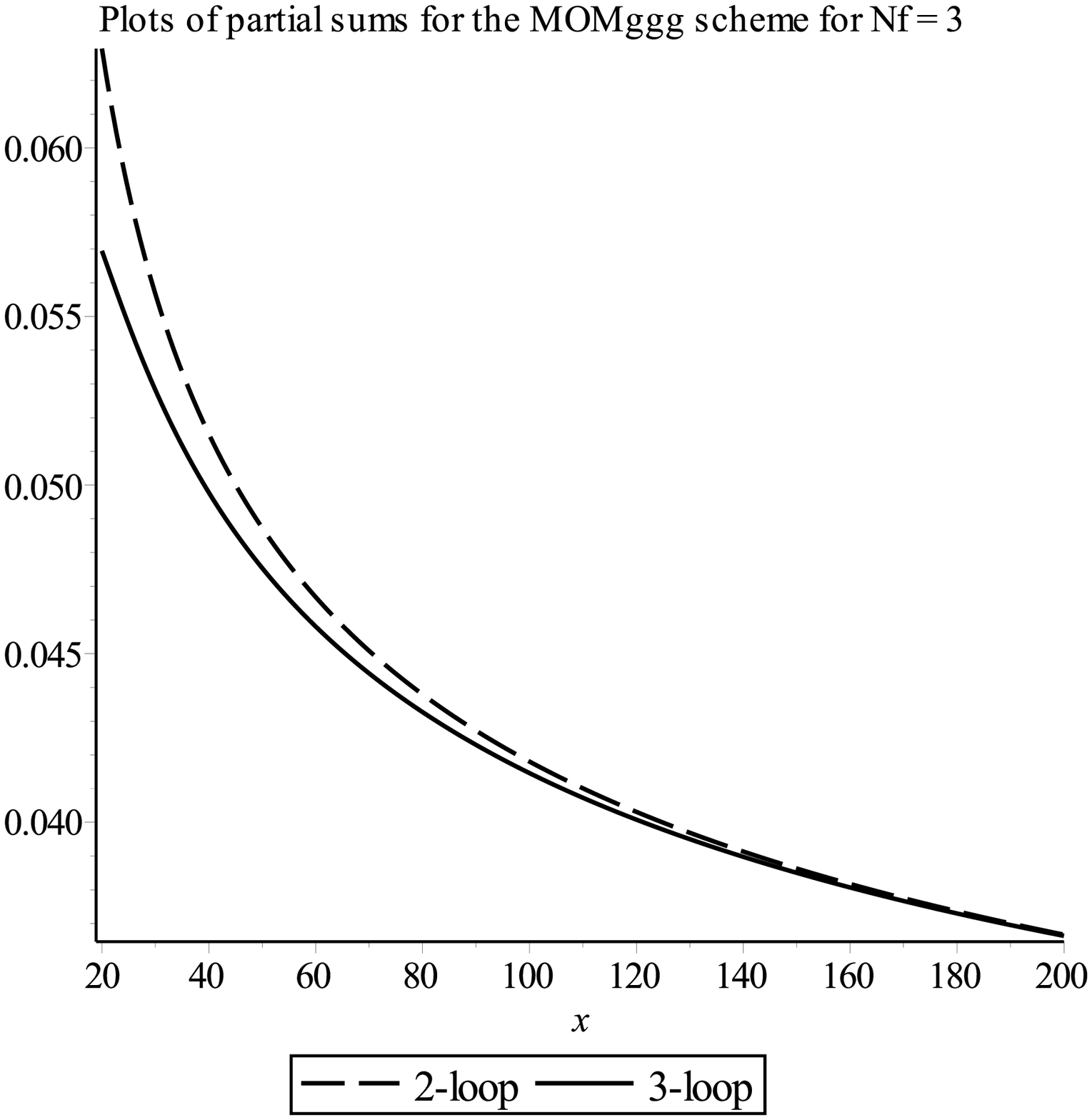}
\quad
\includegraphics[width=7.6cm,height=8cm]{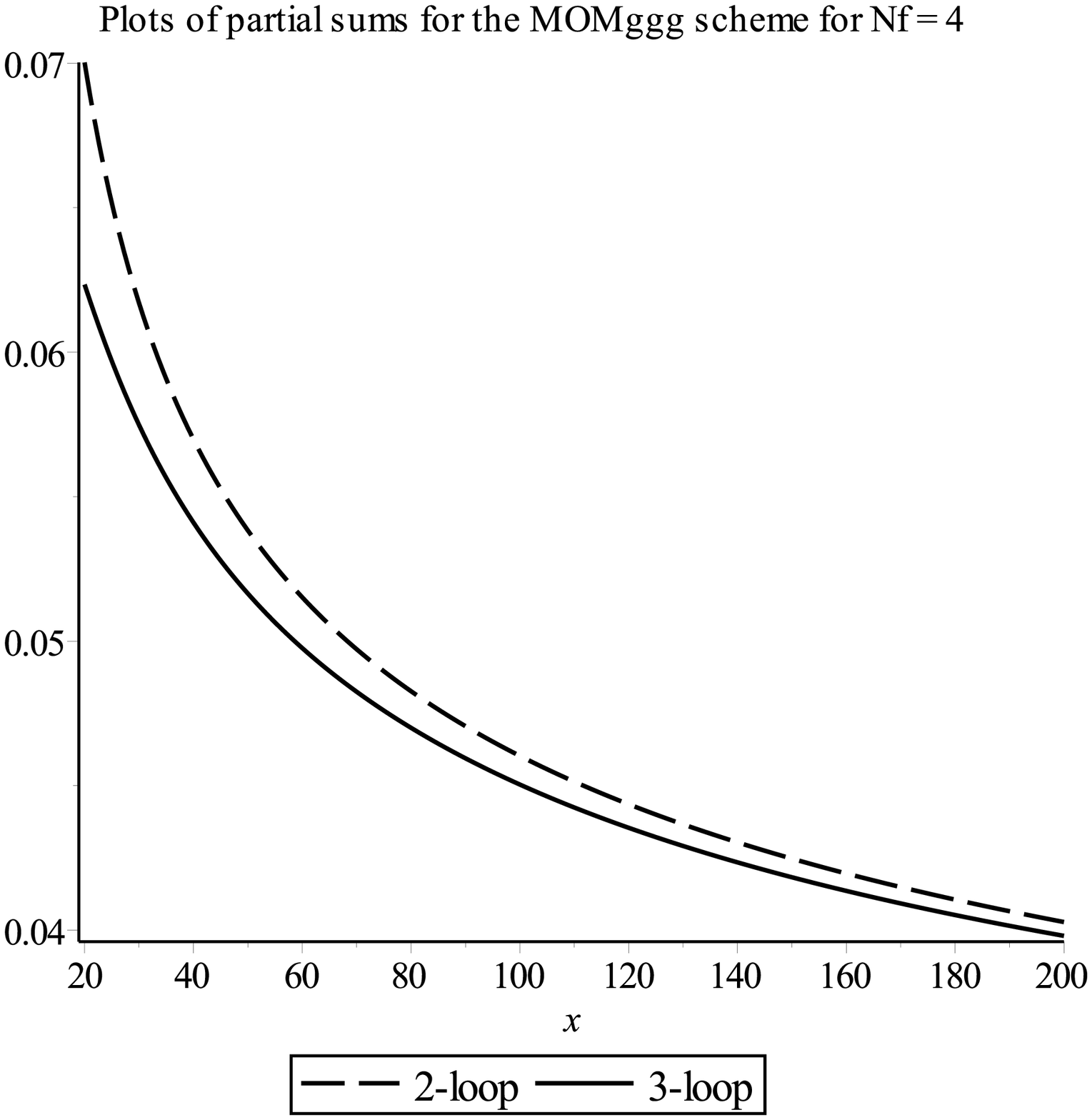}

\vspace{0.8cm}
\includegraphics[width=7.6cm,height=8cm]{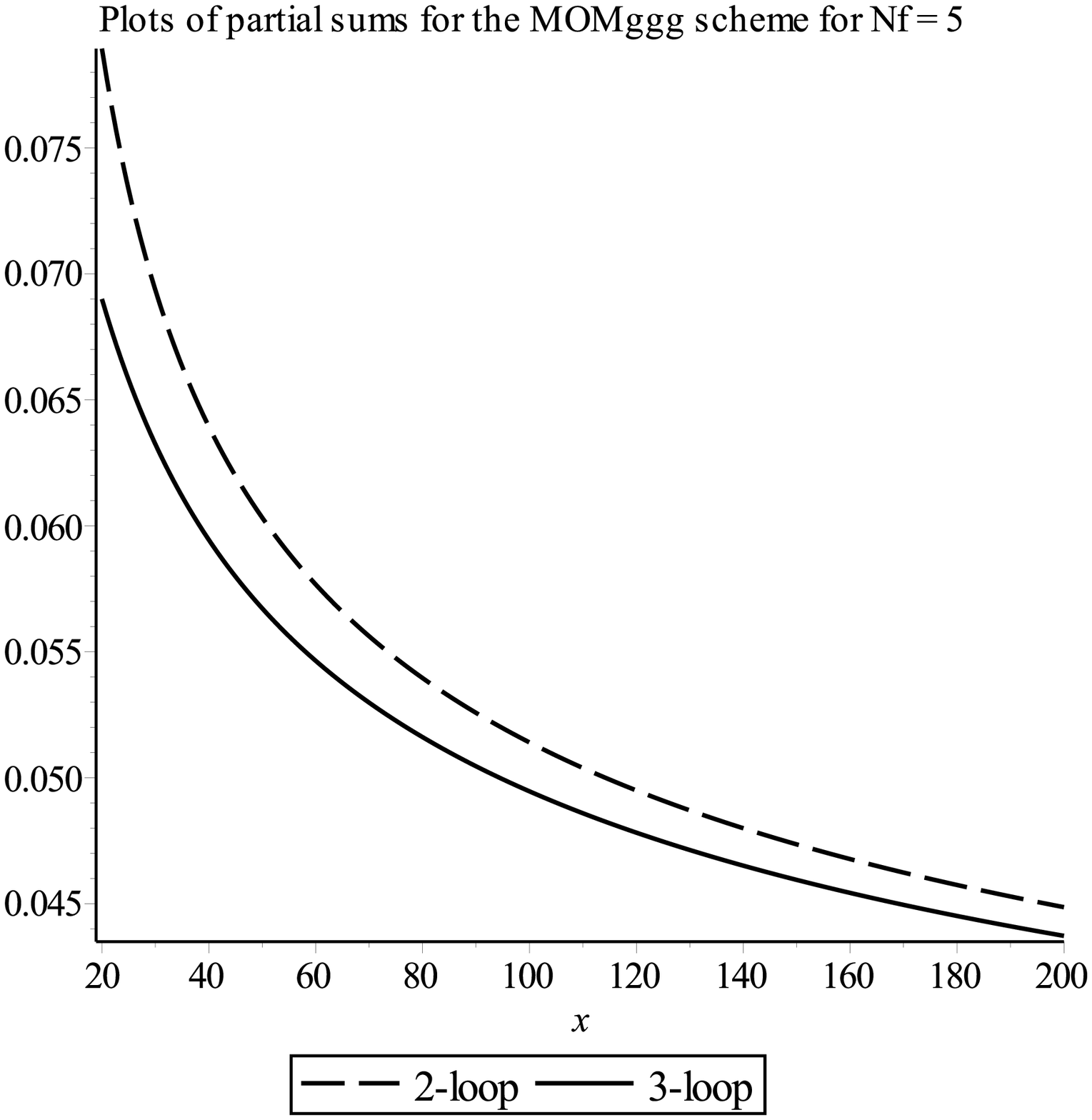}
\quad
\includegraphics[width=7.6cm,height=8cm]{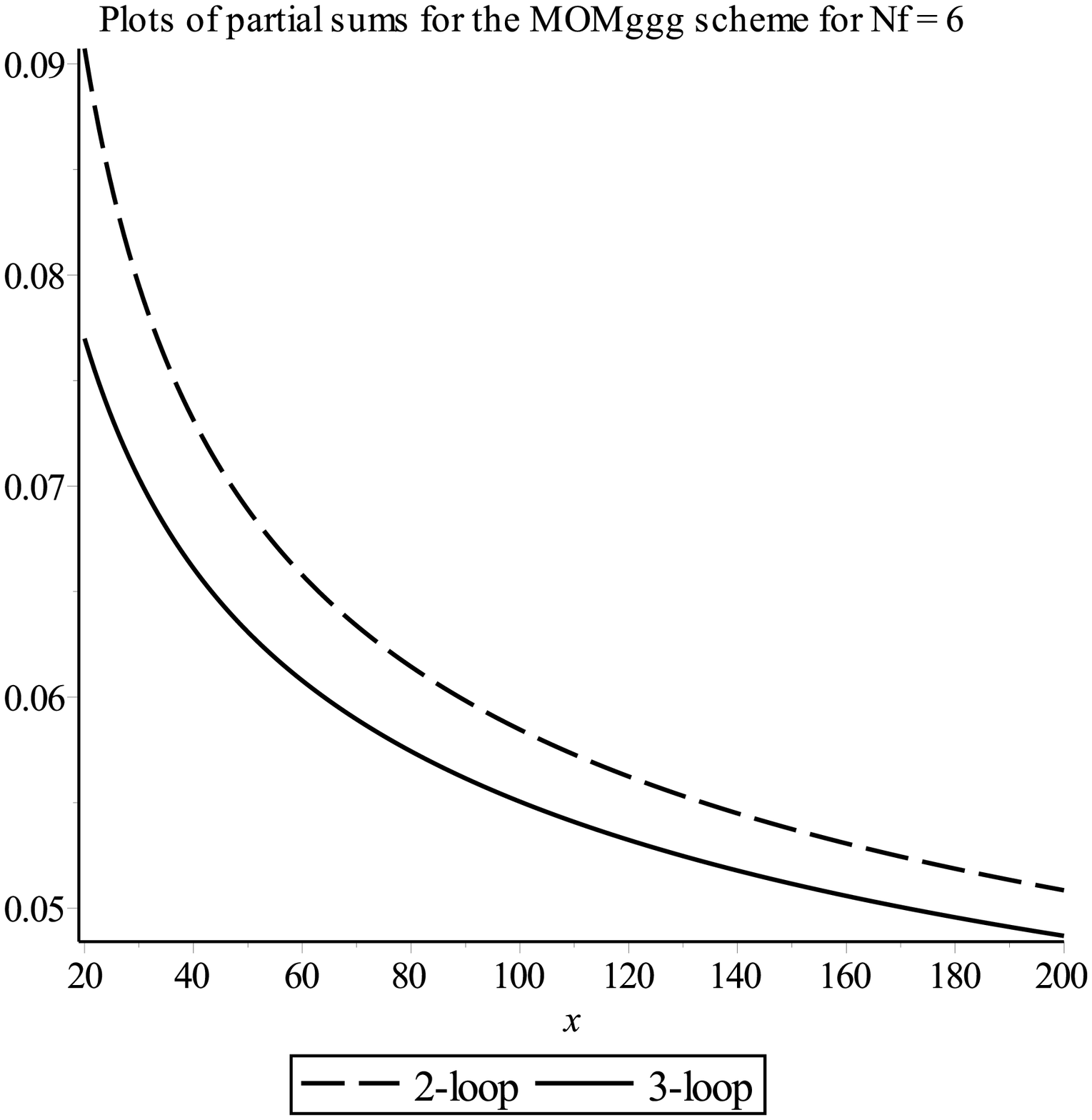}
\caption{Plots of $a_{LL}^{\MOMgs}(x)$ for $L$~$=$~$2$ and $3$.}
\end{figure}}
A nullified external momentum means that either the quark or gluon is on-shell.
While this is an efficient method to determine the $\MSbar$ coupling constant 
renormalization which has allowed for the construction of the $\beta$-function 
at very high loop order, \cite{30}, one does not have to apply a nullification. 
Instead it is possible to extract the $\MSbar$ coupling constant 
renormalization constant at more general momentum configurations. Indeed in the
original MOM approach of \cite{16,17} the $\MSbar$ renormalization was carried 
out at the fully symmetric point at one loop. Moreover, in \cite{20} this was 
used as a computational check on the extension to two loops as well as for the 
general off-shell case, \cite{40}. However, in the context of $\MSbar$ and 
$\mMOM$ being similar this independence of the definition of the $\MSbar$ 
scheme to the subtraction point is what makes it akin to the $\mMOM$ scheme 
where a nullification is inherent in the construction. The $\MSbar$ scheme can 
be defined at a nullified point. In the context of perturbative quantum field 
theory this is somewhat unnatural. 
{\begin{figure}[hb]
\includegraphics[width=7.6cm,height=8cm]{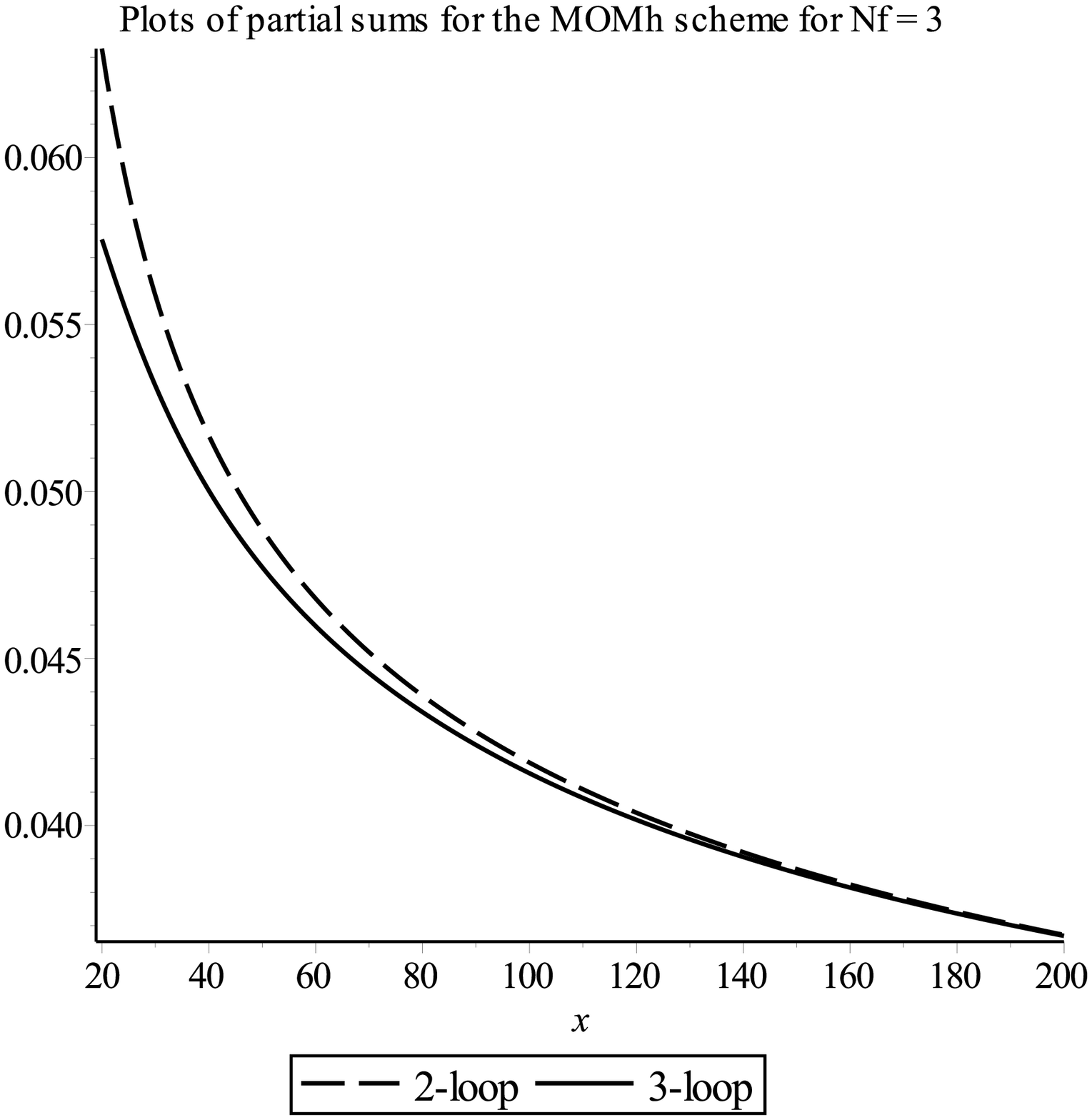}
\quad
\includegraphics[width=7.6cm,height=8cm]{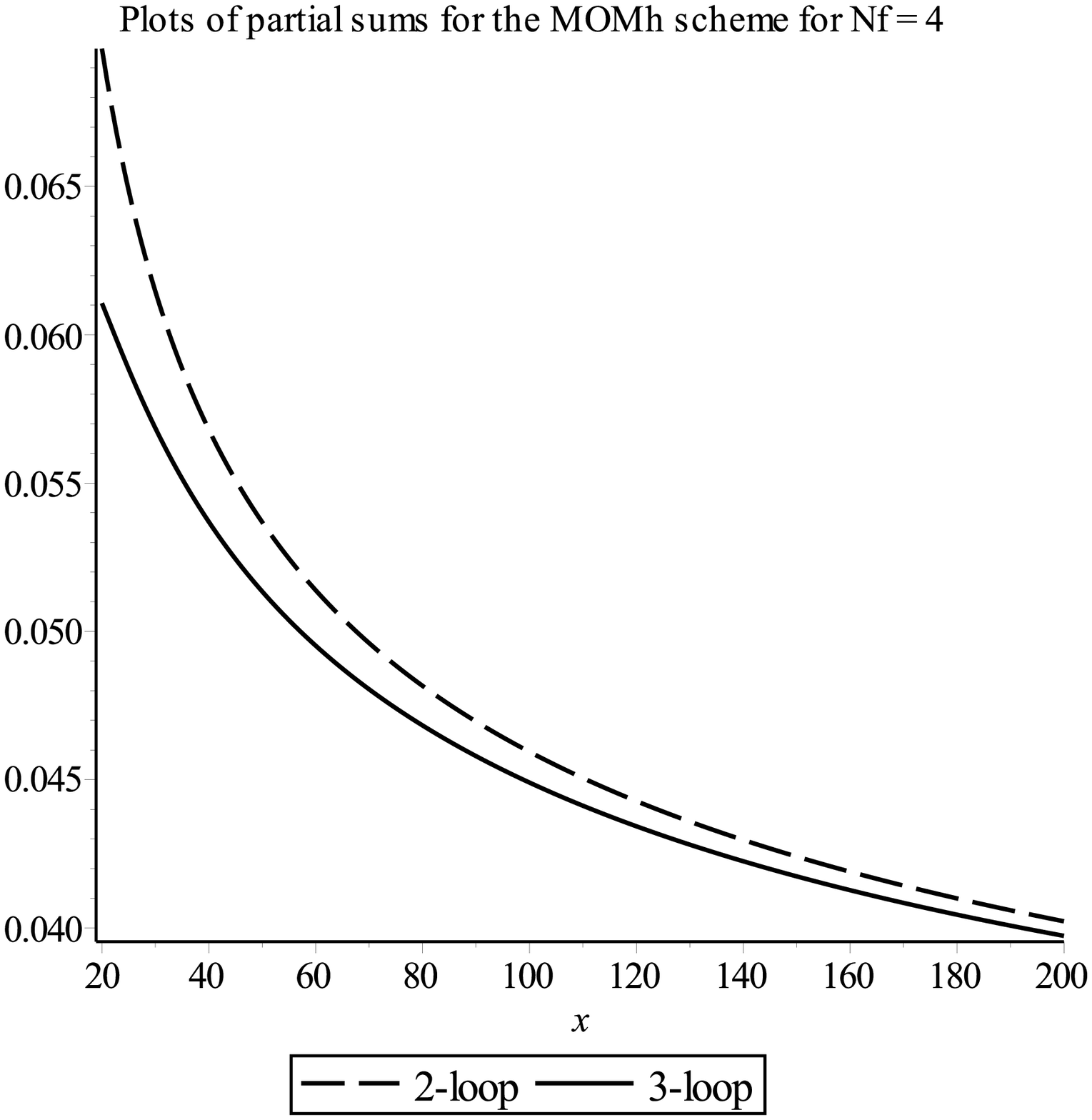}

\vspace{0.8cm}
\includegraphics[width=7.6cm,height=8cm]{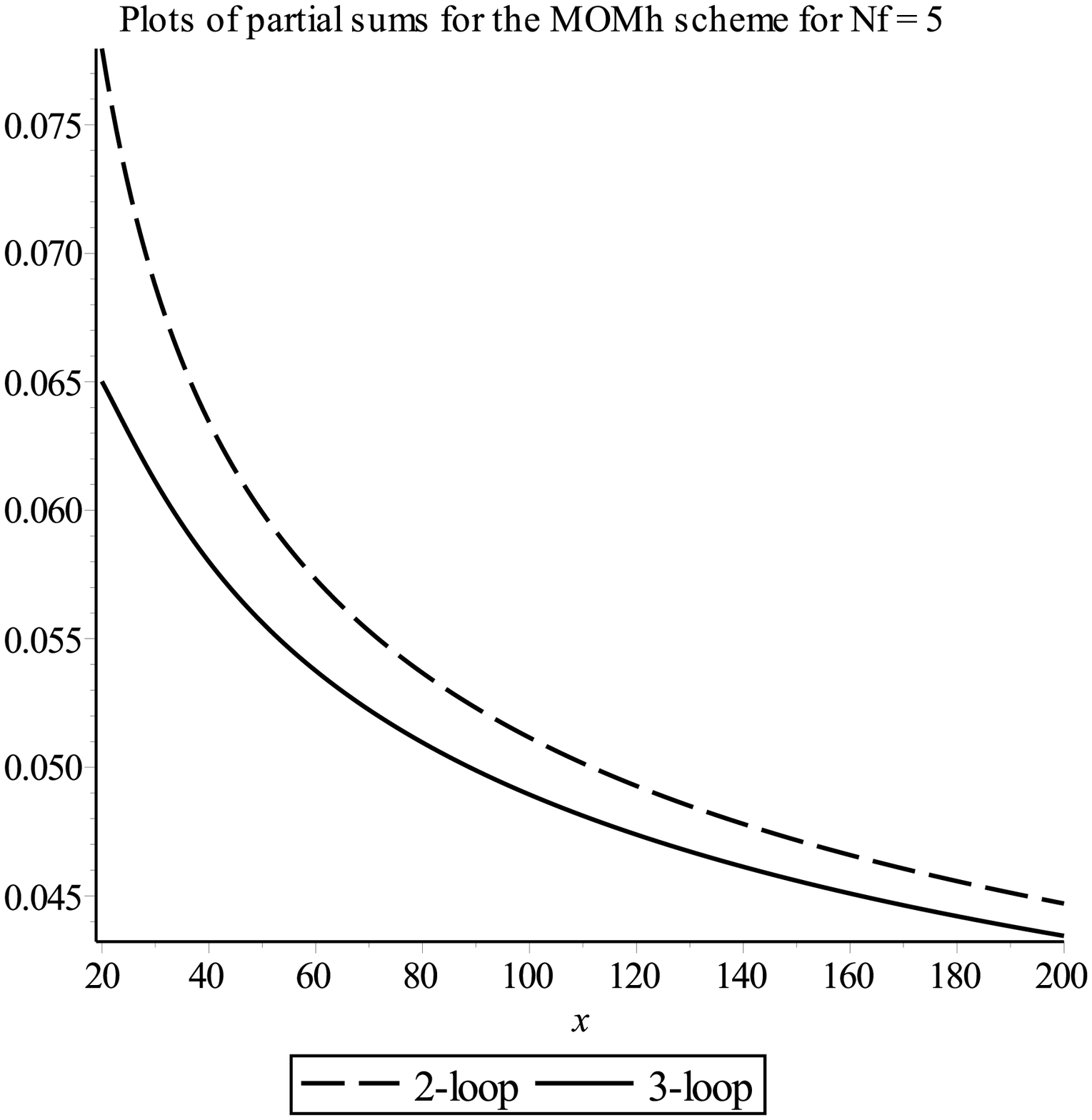}
\quad
\includegraphics[width=7.6cm,height=8cm]{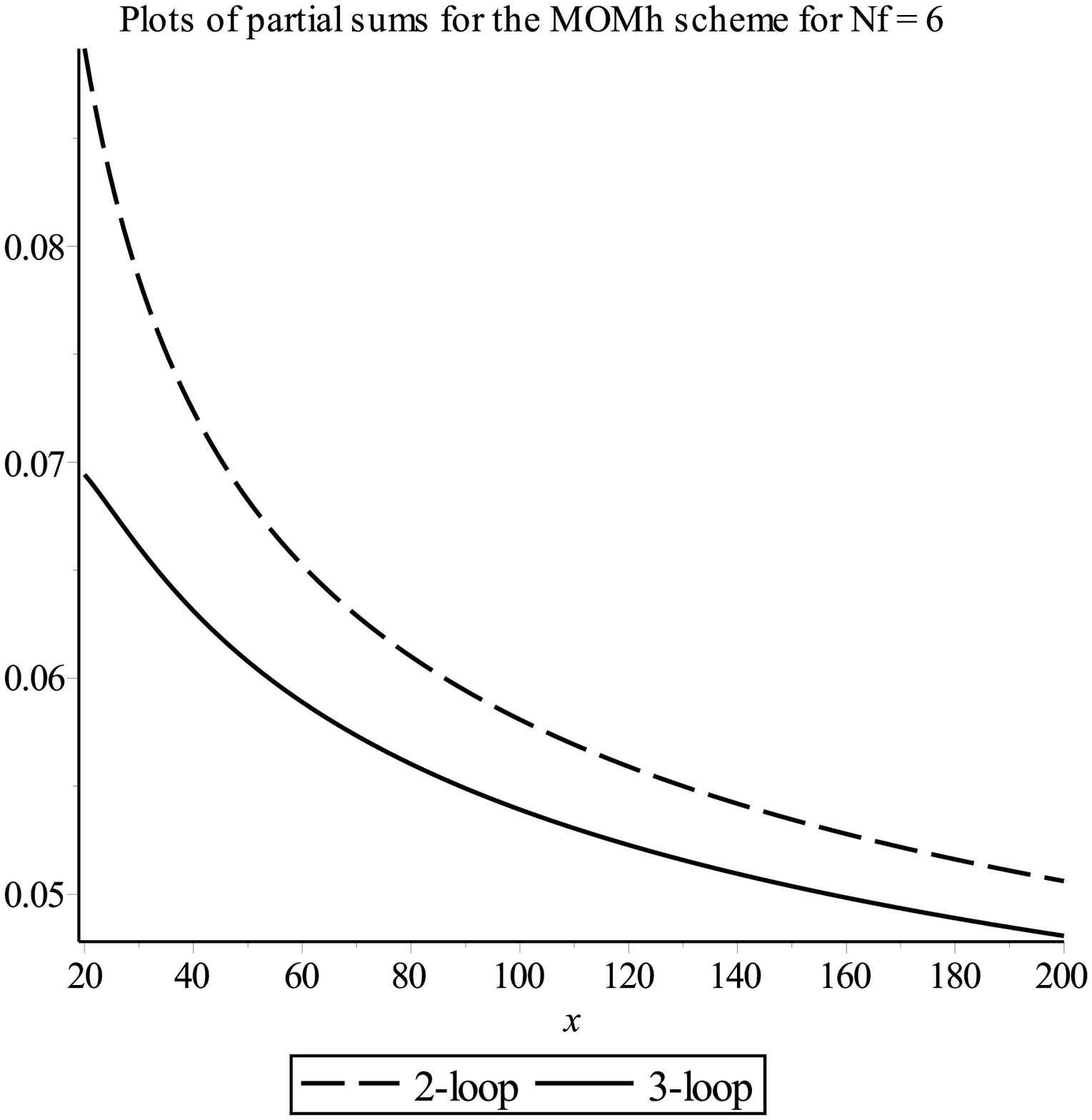}
\caption{Plots of $a_{LL}^{\MOMhs}(x)$ for $L$~$=$~$2$ and $3$.}
\end{figure}}
(We recall that we are using chiral quarks here.) Therefore, the coupling 
constant renormalization is being examined at a point where such entities do 
not exist as fundamental quanta. In other words the definition of the $\MSbar$
renormalization constant carries with it no connection of where it is defined.
By contrast the MOM schemes of \cite{16,17} do since they incorporate 
information of the symmetric subtraction point through the inclusion of a 
finite part in the coupling constant renormalization constant. Moreover, as no 
external legs have been nullified there is no issue with the interpretation of 
the fields corresponding to fundamental particles. In another sense a momentum
subtraction scheme would appear to be more natural for a physical coupling as 
the finite part could be regarded as a measure of the associated radiation. 
Therefore, in order to resolve some of these observations it would again seem 
necessary to extend the MOM $\beta$-functions of \cite{16,17} to four loops. 

{\begin{figure}[ht]
\includegraphics[width=7.6cm,height=8cm]{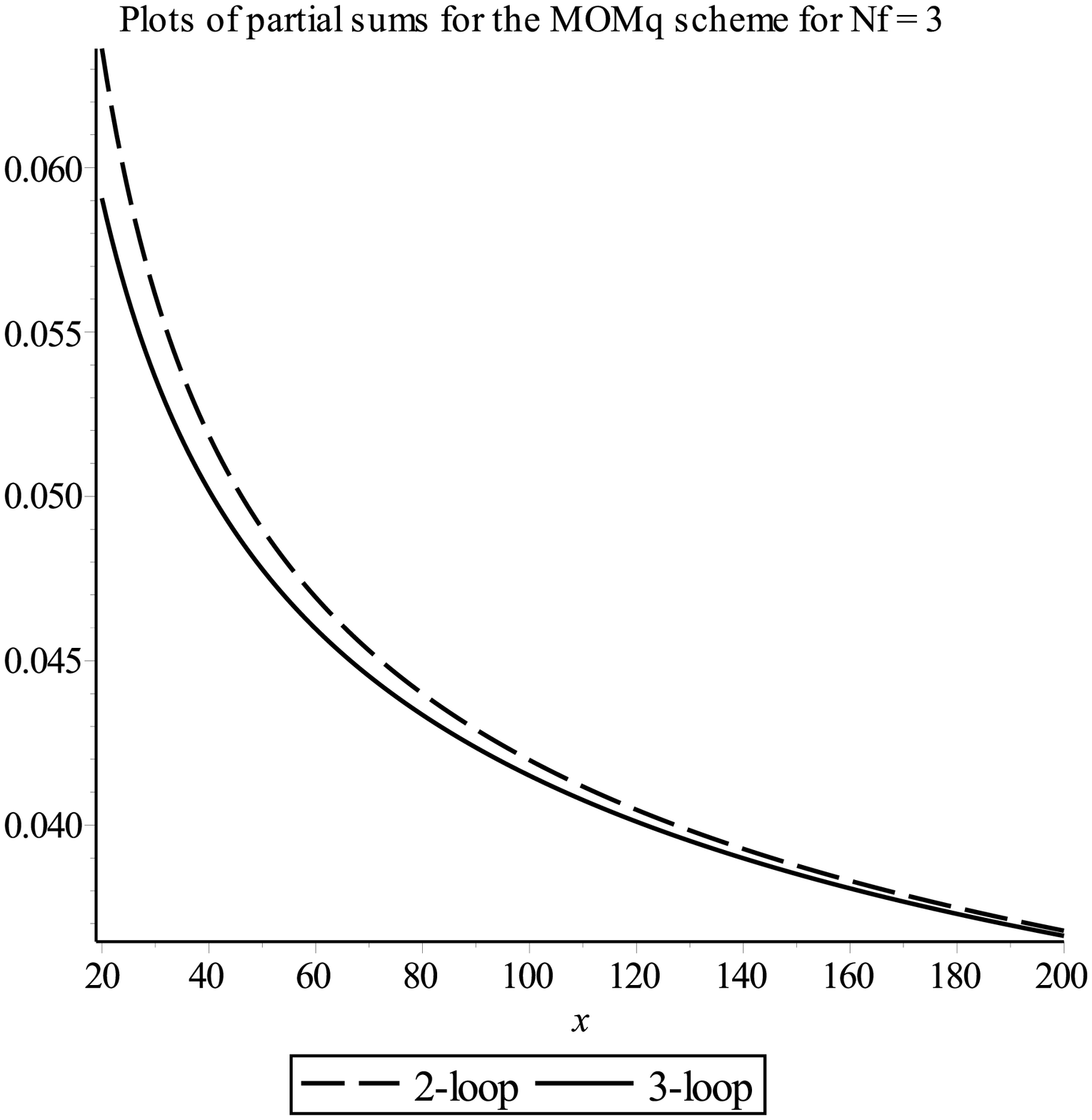}
\quad
\includegraphics[width=7.6cm,height=8cm]{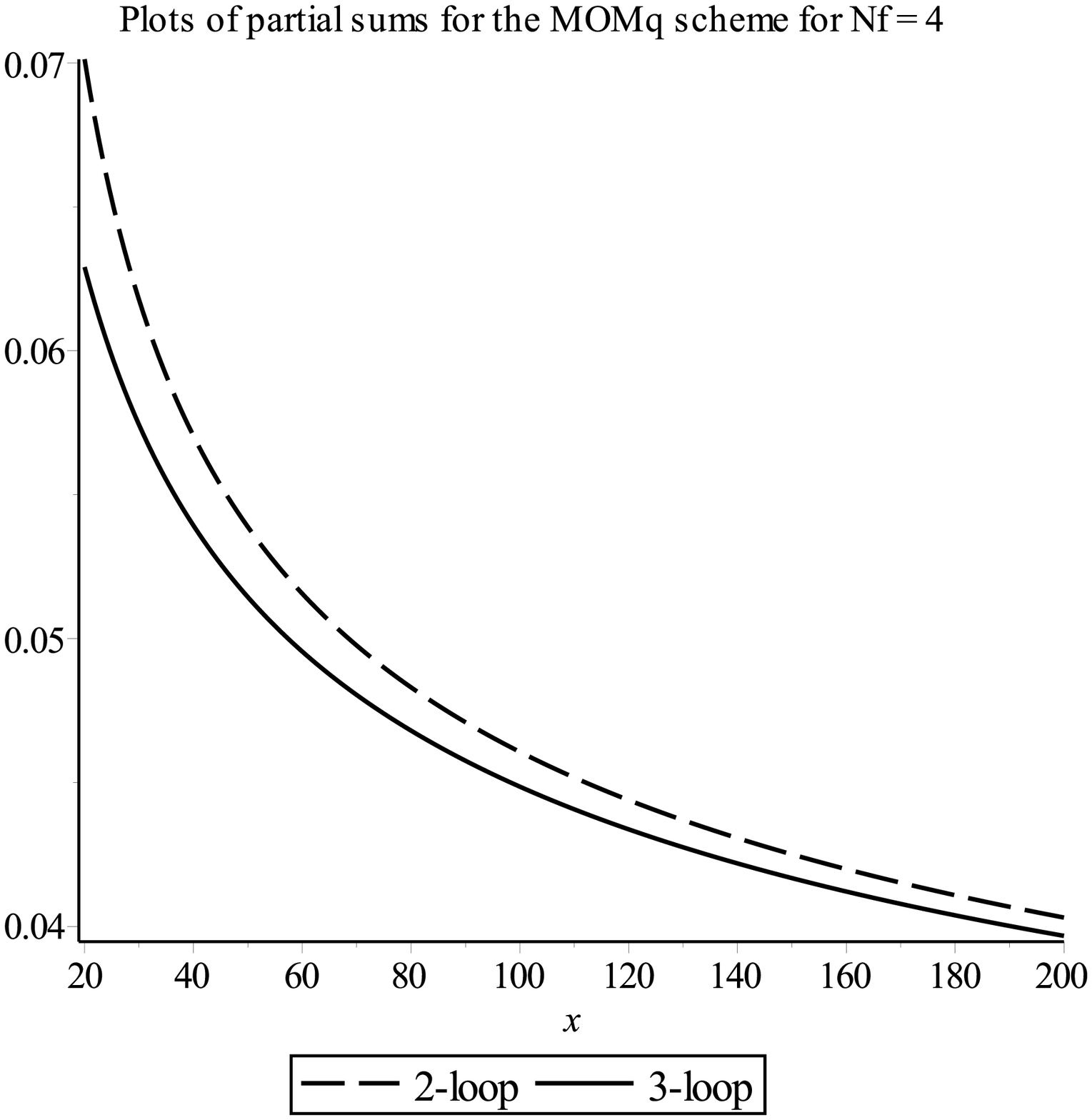}

\vspace{0.8cm}
\includegraphics[width=7.6cm,height=8cm]{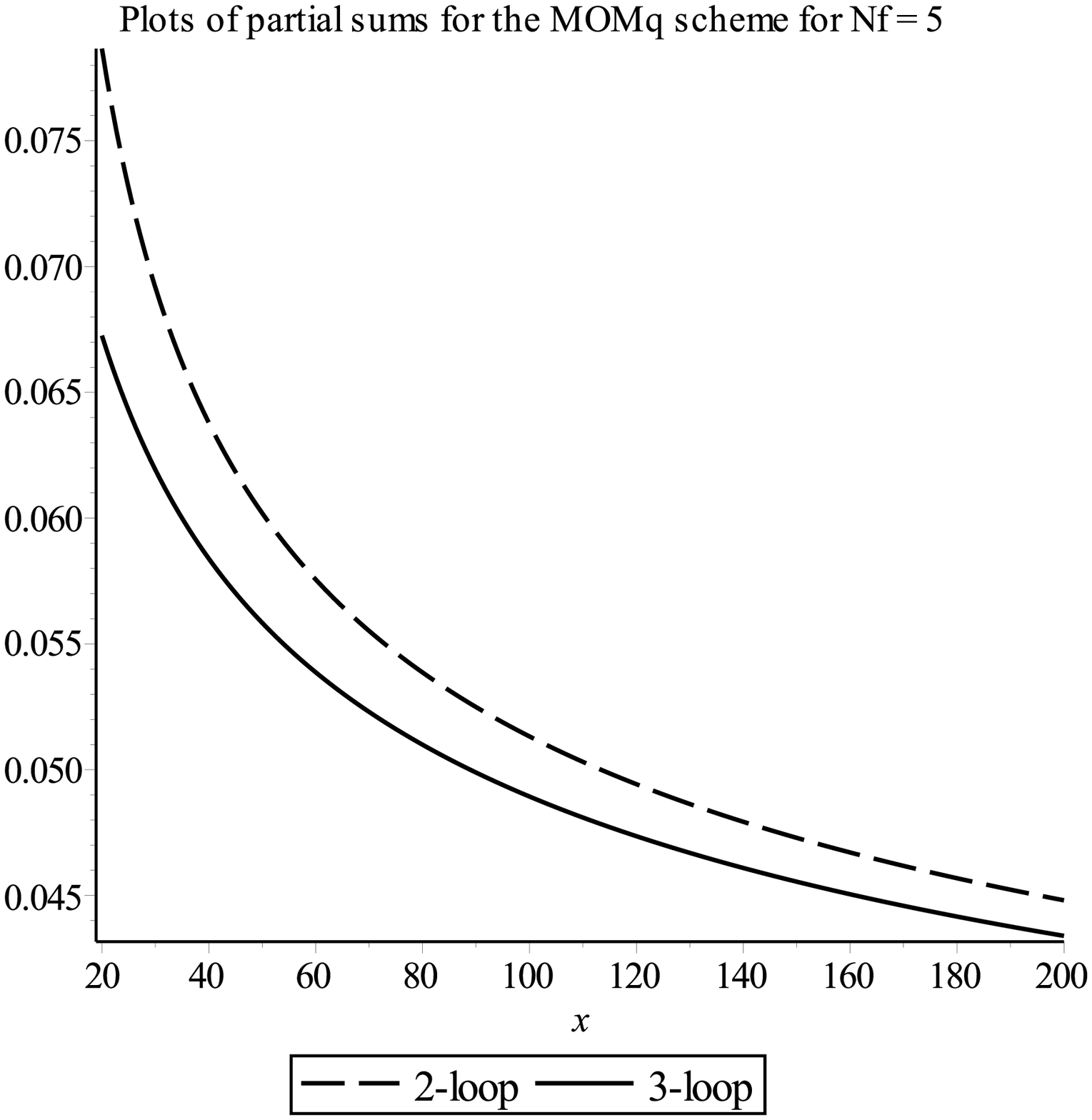}
\quad
\includegraphics[width=7.6cm,height=8cm]{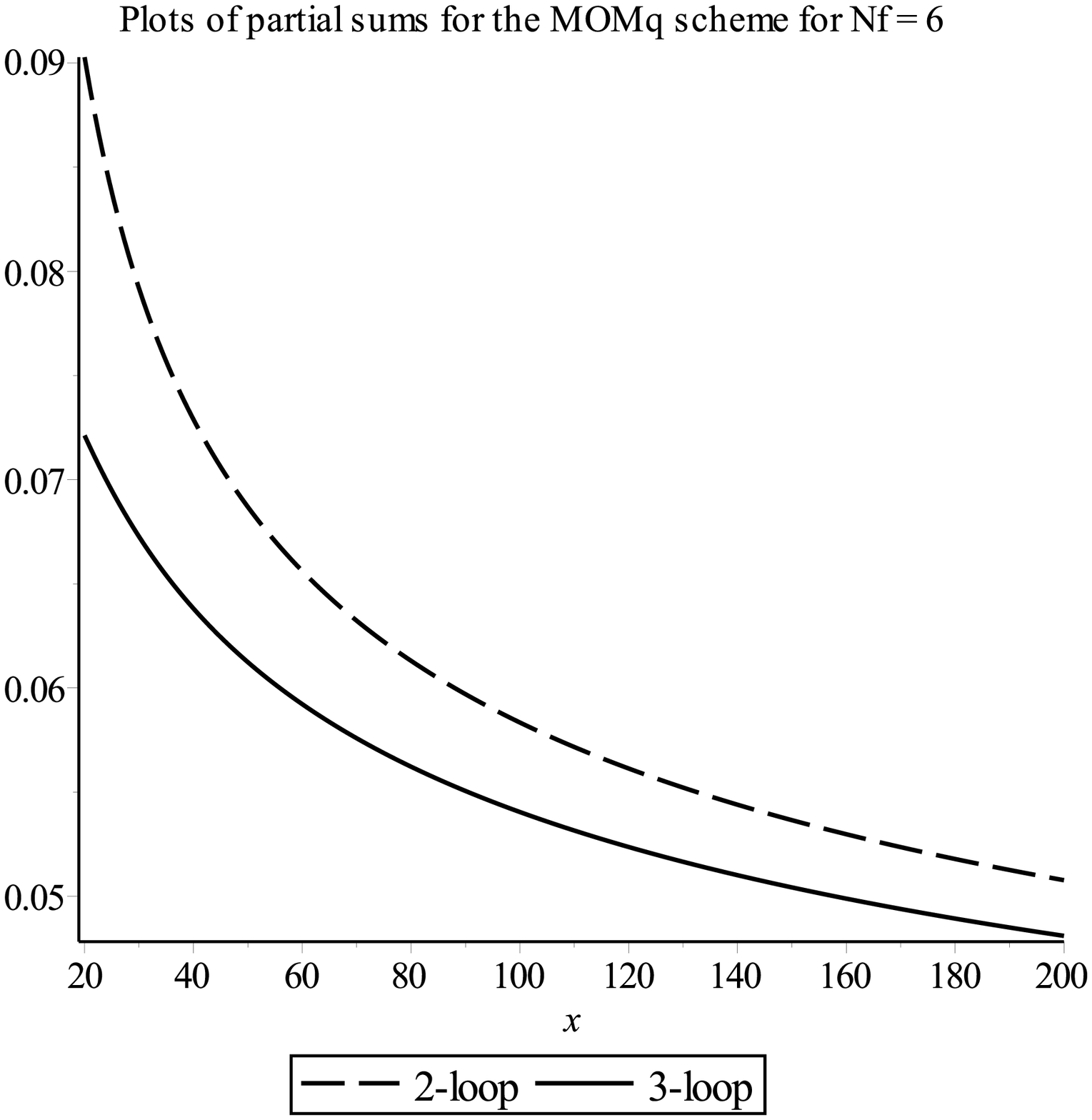}
\caption{Plots of $a_{LL}^{\MOMqs}(x)$ for $L$~$=$~$2$ and $3$.}
\end{figure}}

While we have focused exclusively to this point on the Landau gauge for the MOM
schemes for the reasons we stated earlier, it is worth considering the 
dependence on the gauge parameter as an exercise. For this we have analysed the
$\MOMq$ scheme for various values of $\alpha$ and the same values of $\Nf$ as 
before. For this investigation the same formalism is used where now the 
$\alpha$ dependent coefficients of the three loop $\MOMq$ $\beta$-function must
be included and are available from \cite{16,17,20}. Equally the $\alpha$ 
dependent $\MOMq$ three loop $R$-ratio is constructed from the three loop 
coupling constant map. As the $\MSbar$ $R$-ratio is gauge parameter independent
we do not have to include the mapping of the gauge parameter variable from one 
scheme to another. The gauge parameter dependence arises solely from the 
coupling constant map. 
{\begin{figure}[ht]
\includegraphics[width=7.6cm,height=8cm]{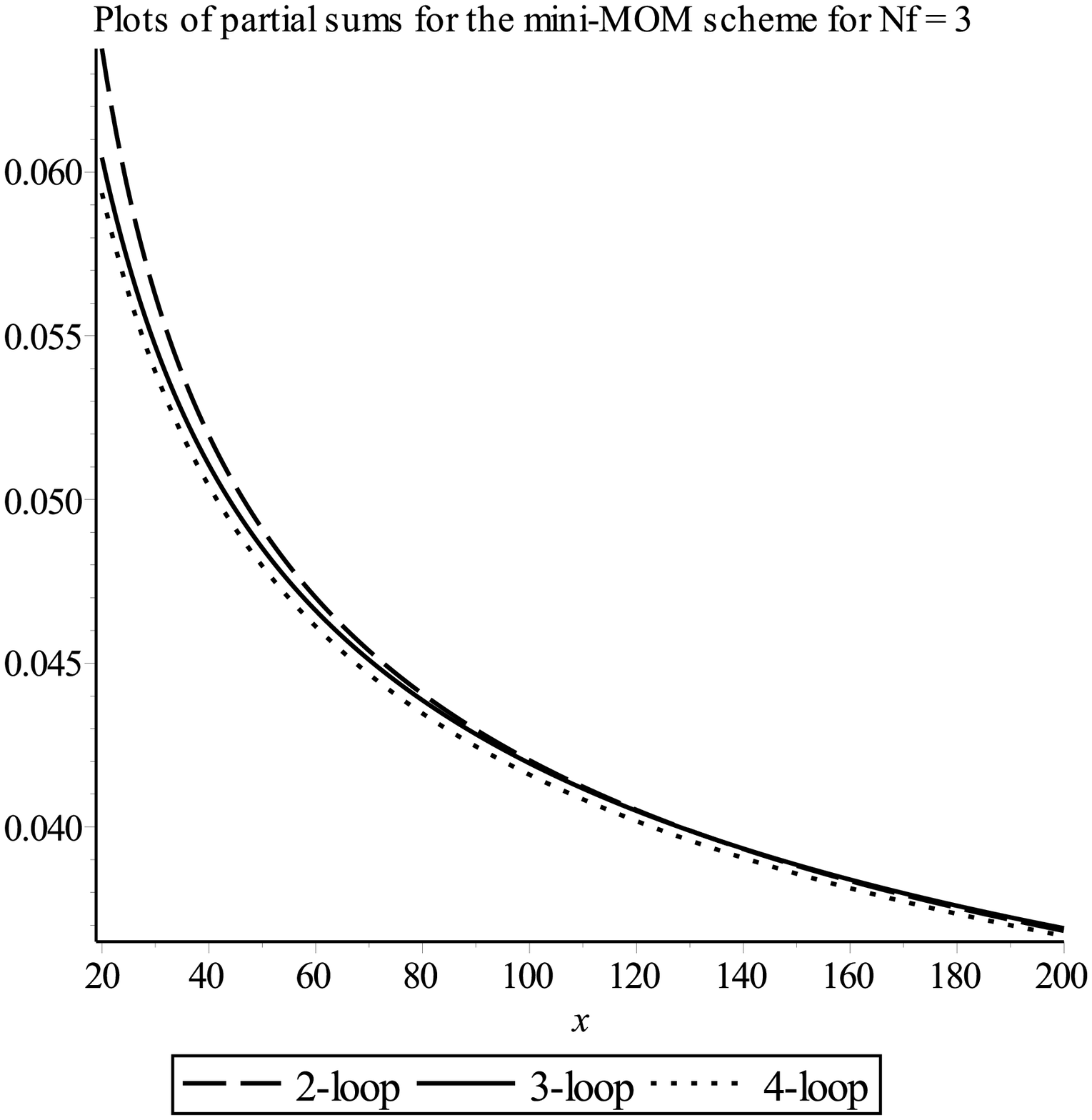}
\quad
\includegraphics[width=7.6cm,height=8cm]{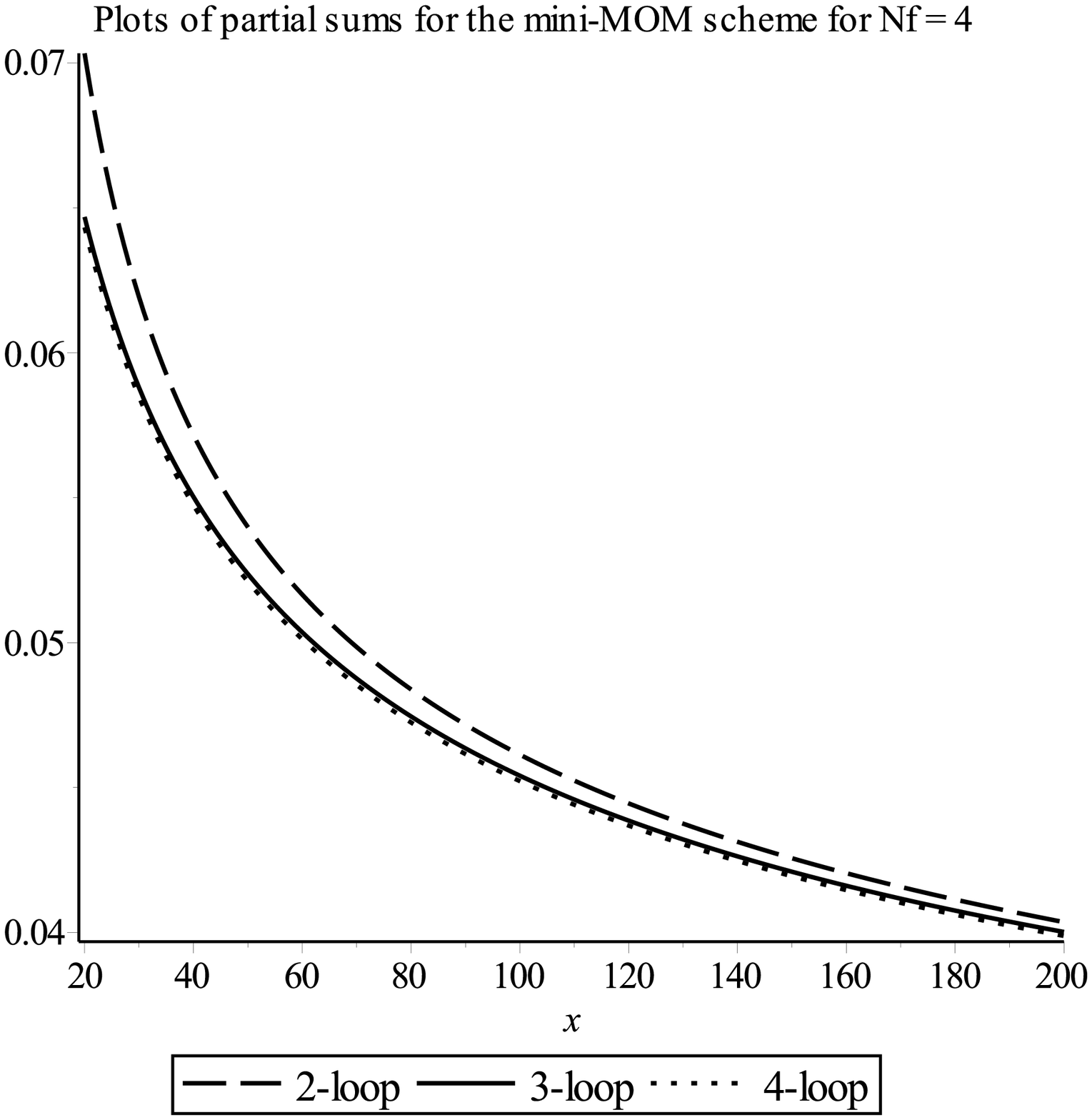}

\vspace{0.8cm}
\includegraphics[width=7.6cm,height=8cm]{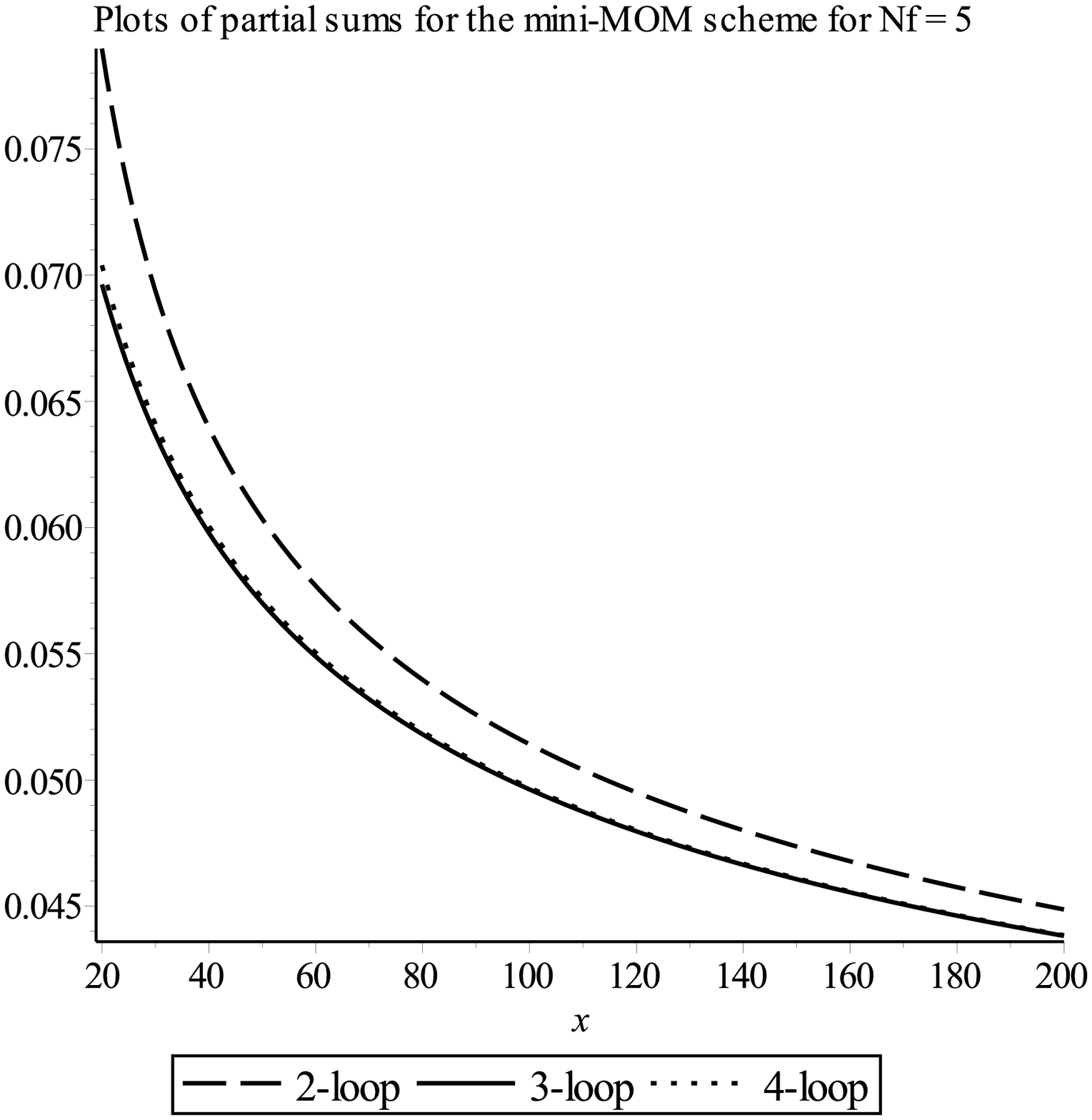}
\quad
\includegraphics[width=7.6cm,height=8cm]{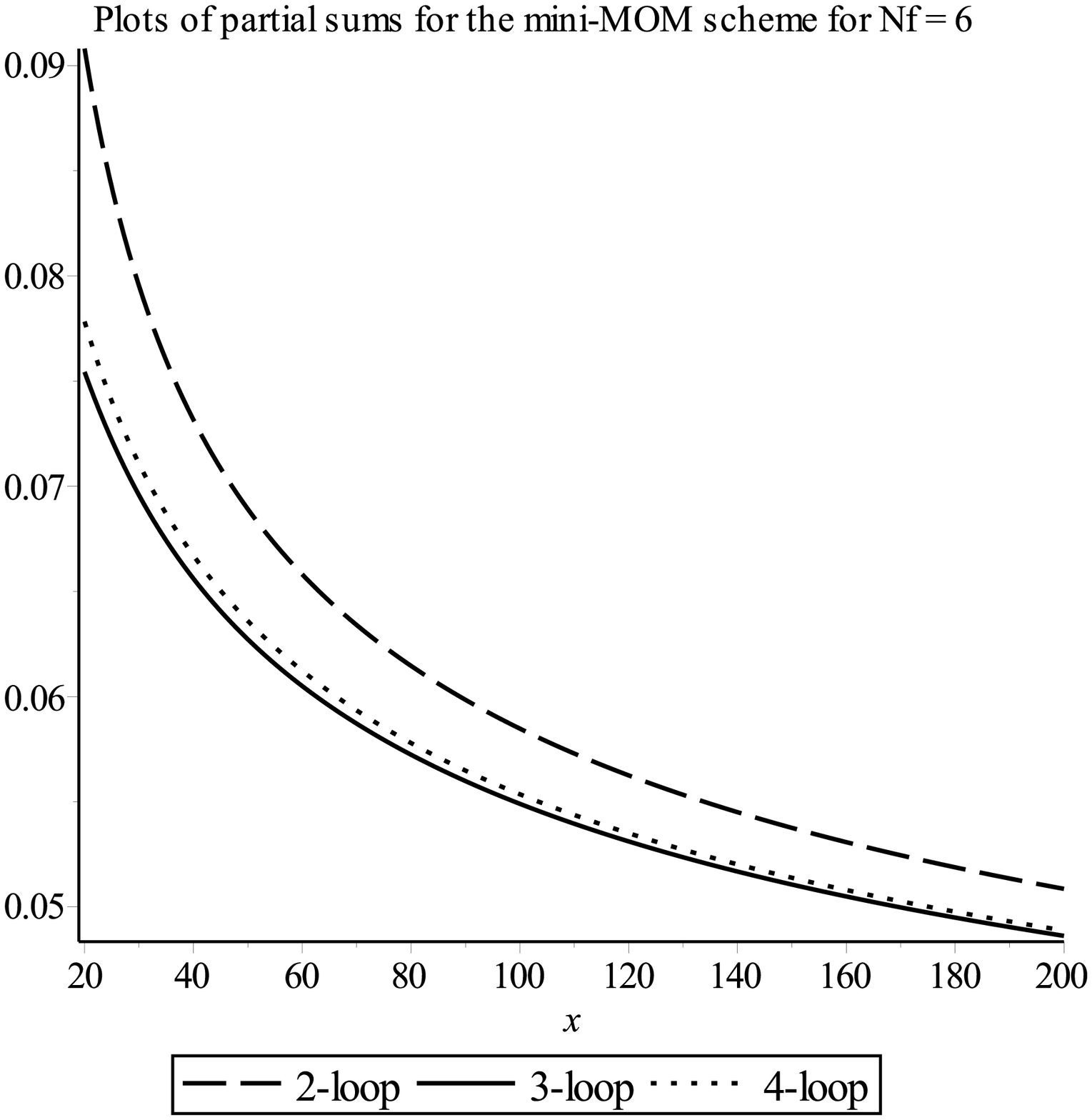}
\caption{Plots of $a_{LL}^{\mMOMs}(x)$ for $L$~$=$~$2$, $3$ and $4$.}
\end{figure}}
In addition we also have to use the $\alpha$ dependent $\Lambda$ parameter 
ratios which are straightforward to extract from the formul{\ae} given in 
\cite{16,17}. In order to assist with the comparison of earlier plots we give 
the same partial sums as before for two loops in Figure $8$ and three loops in 
Figure $9$. In both the quantity {\tt al} corresponds to the canonical gauge 
parameter $\alpha$ with $\alpha$~$=$~$0$ corresponding to the earlier Landau 
gauge results. The plots for that gauge are repeated in both figures as the 
benchmark to compare with. We have not provided comparison plots between loops 
since it is the variation in the behaviour for various $\alpha$ which is of 
interest. The largest value of $\alpha$ we have taken is $10$ as anything 
beyond this has a significantly small $\Lambda$ ratio. From Figure $8$ for each
value of $\Nf$ $a_{22}^{\MOMqs}(x)$ gets progressively smaller as $\alpha$ 
increases. The main point is that as $\alpha$ increases the partial sum 
diverges from not only the Landau gauge $\MOMq$ value but also the $\MSbar$ 
value. This is more evident in the plots of Figure $9$ where the non-zero 
$\alpha$ lines cover a broad range especially for relatively small values of 
$x$. If one recalls the earlier comparisons of $\MOMq$ for the Landau gauge 
with the other schemes while there is clearly not precise agreement there is 
not a broad range suggesting convergence. Such a convergence is difficult to 
perceive from using various fixed values for $\alpha$ as can be seen in Figures
$8$ and $9$ which suggests the Landau gauge is the appropriate choice for a MOM
scheme analysis. 
{\begin{figure}[ht]
\includegraphics[width=7.6cm,height=8cm]{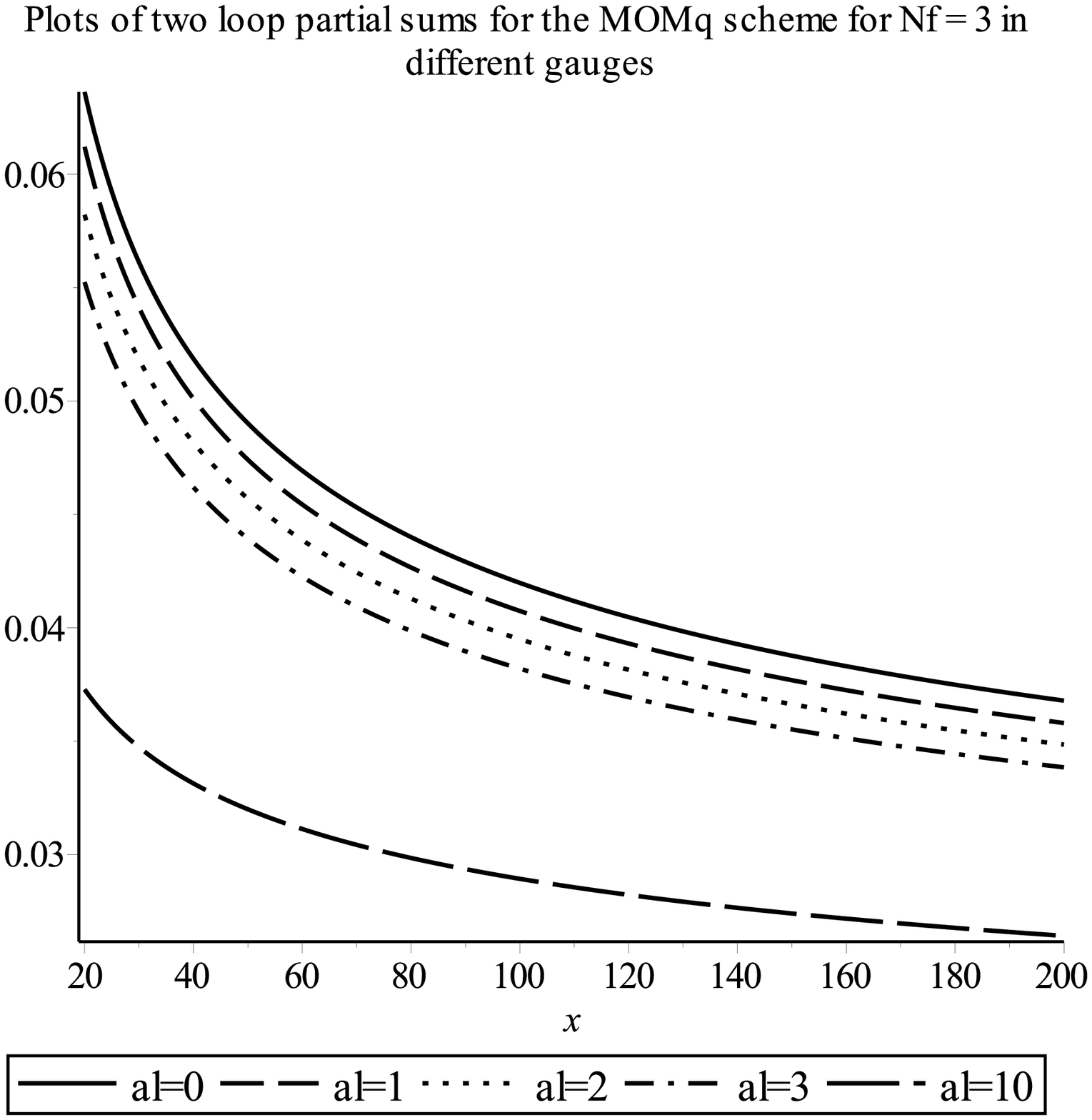}
\quad
\includegraphics[width=7.6cm,height=8cm]{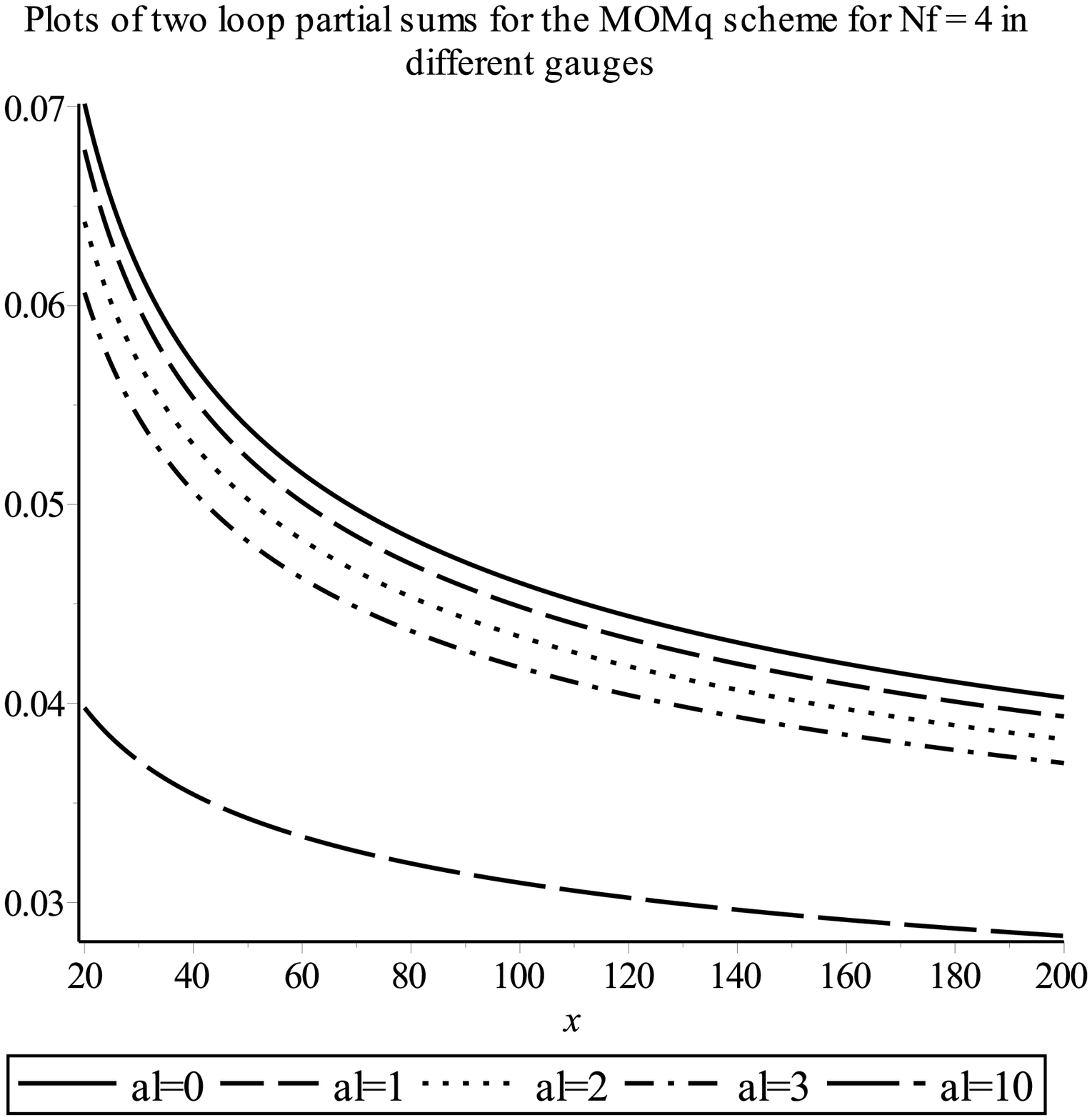}

\vspace{0.8cm}
\includegraphics[width=7.6cm,height=8cm]{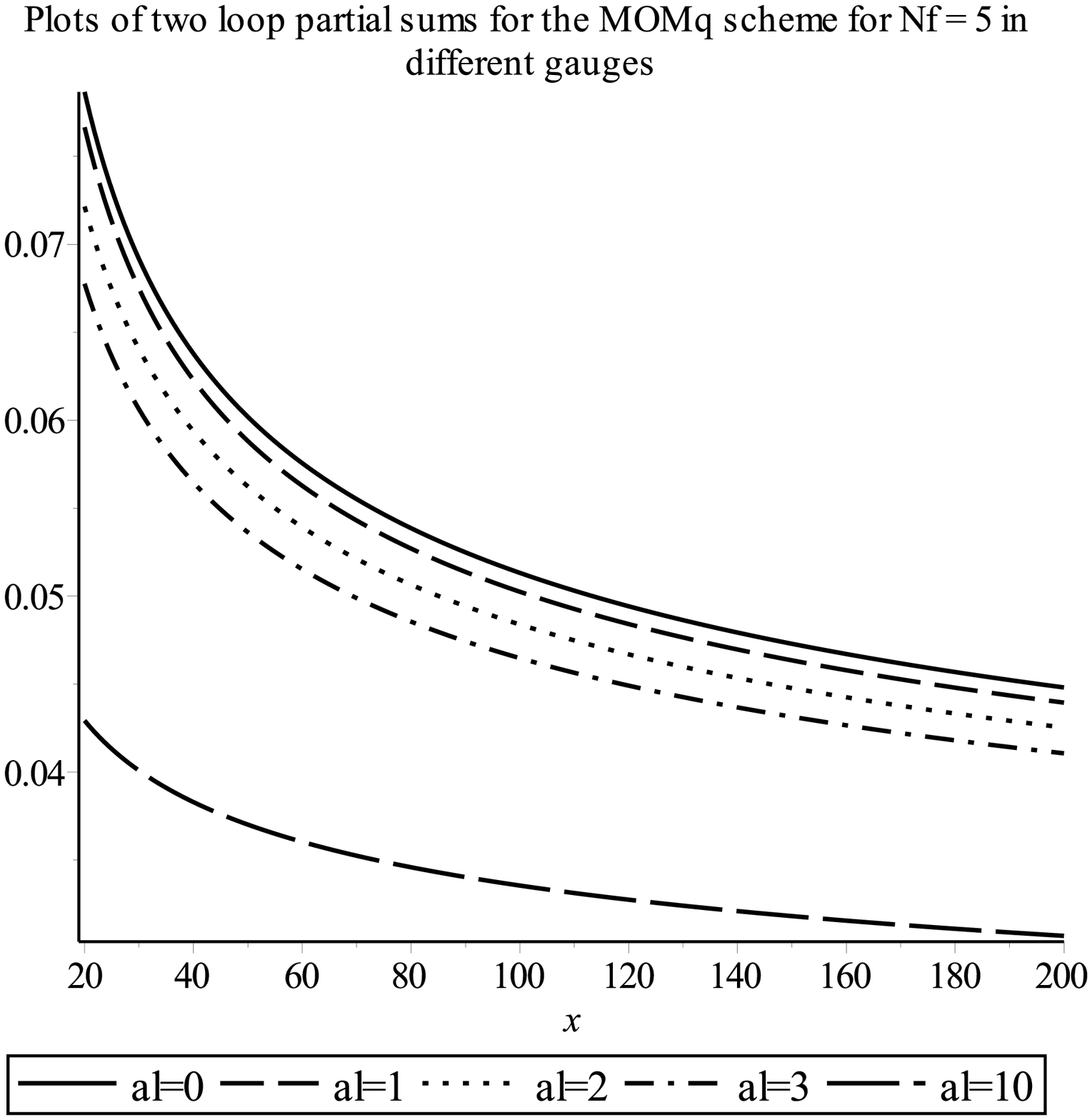}
\quad
\includegraphics[width=7.6cm,height=8cm]{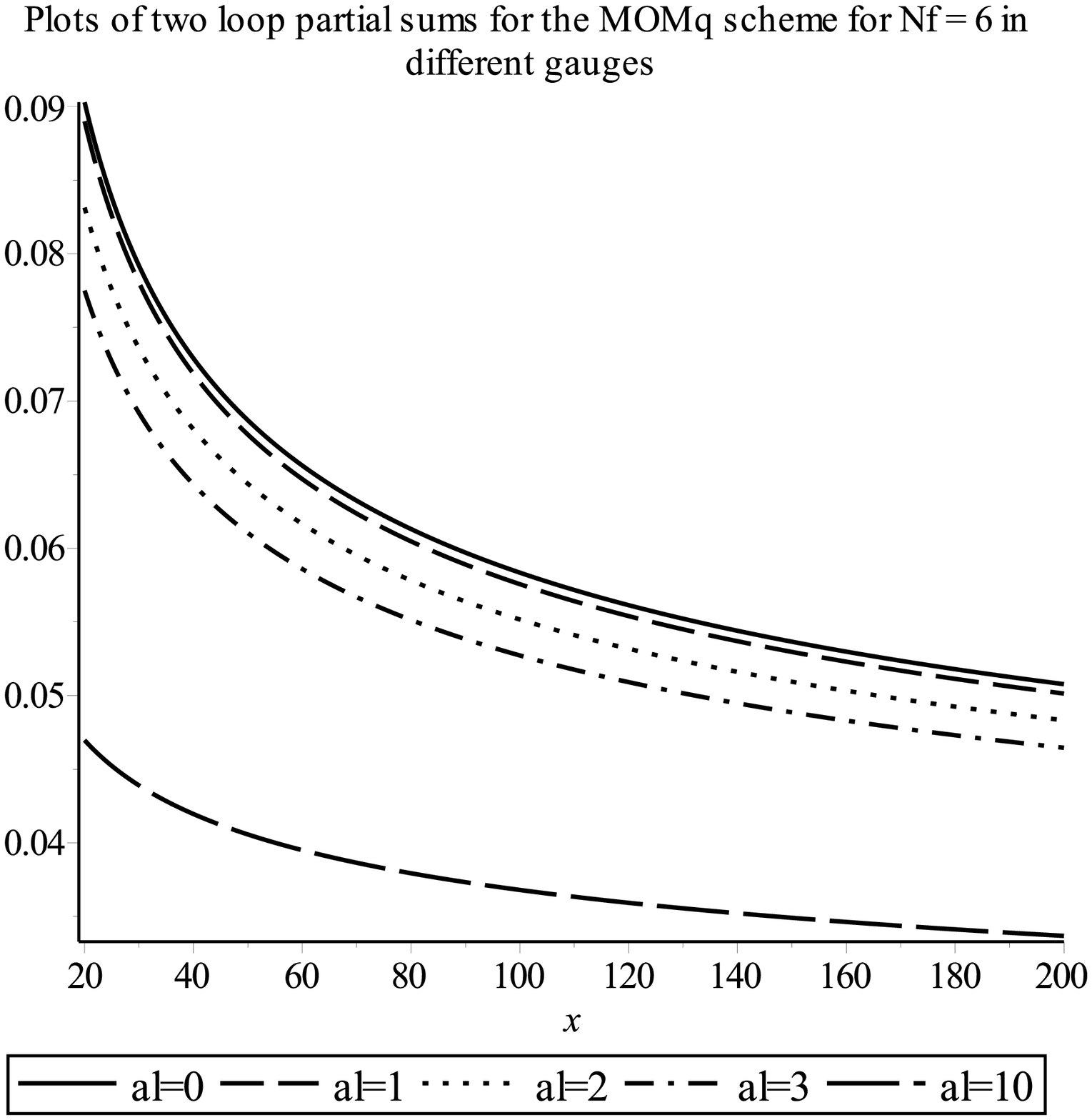}
\caption{Plots of $a_{22}^{\MOMqs}(x)$ for various gauges.}
\end{figure}}

\sect{Discussion.}

We have completed an analysis of the $R$-ratio in various MOM renormalization
schemes. One motivation was to extend the early computations of \cite{5,6} to 
the next loop order since the two loop mappings of the coupling constants
between the $\MSbar$ and MOM schemes of \cite{16,17} as well as the three loop 
$\MOMg$, $\MOMh$ and $\MOMq$ $\beta$-functions are now available, \cite{20}. 
Another aim was to examine whether the observation of \cite{5,6} that the 
$\MOMq$ scheme led to an improved or better convergent series for a physical 
quantity held at next order. 
{\begin{figure}[ht]
\includegraphics[width=7.6cm,height=8cm]{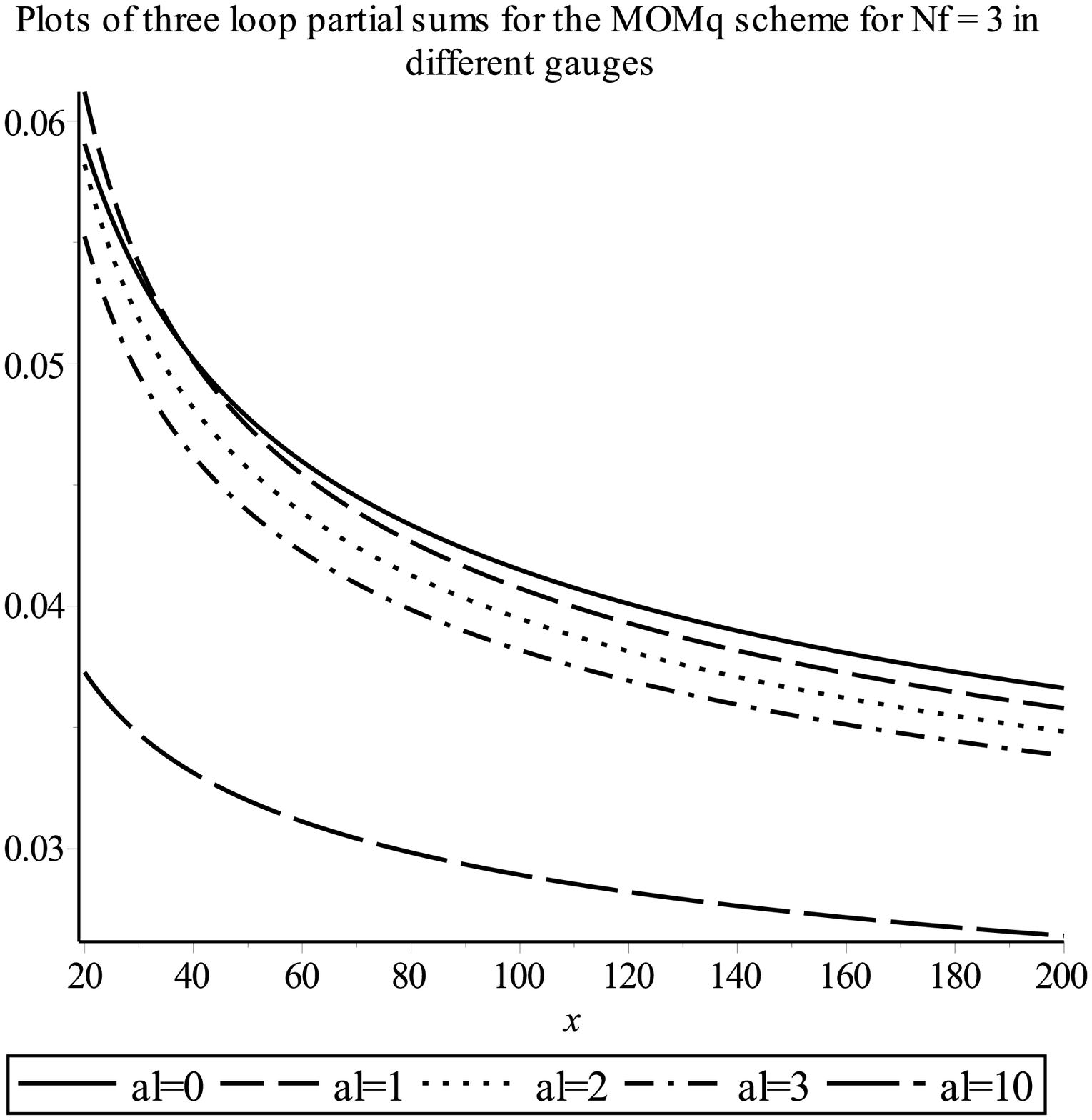}
\quad
\includegraphics[width=7.6cm,height=8cm]{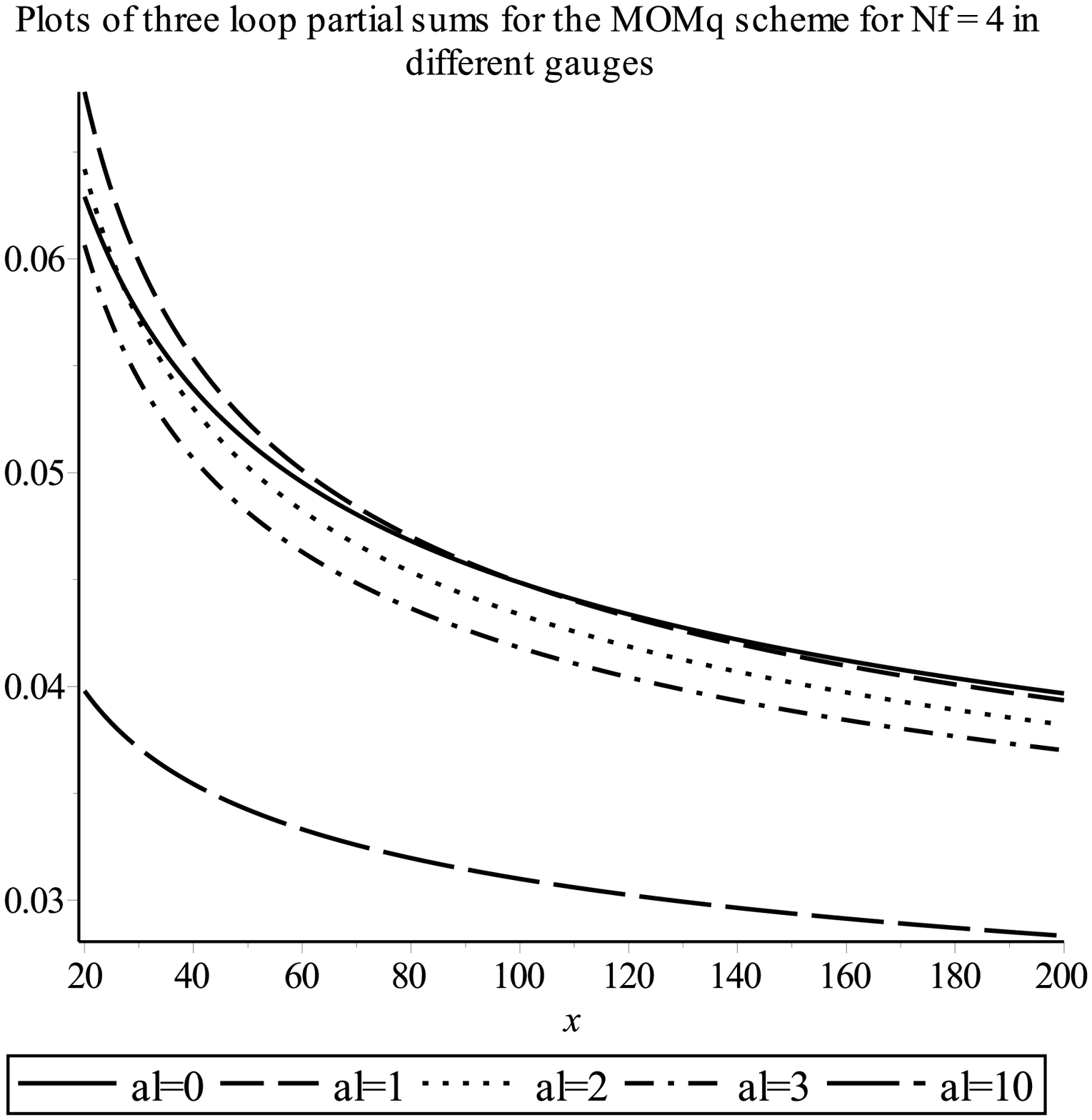}

\vspace{0.8cm}
\includegraphics[width=7.6cm,height=8cm]{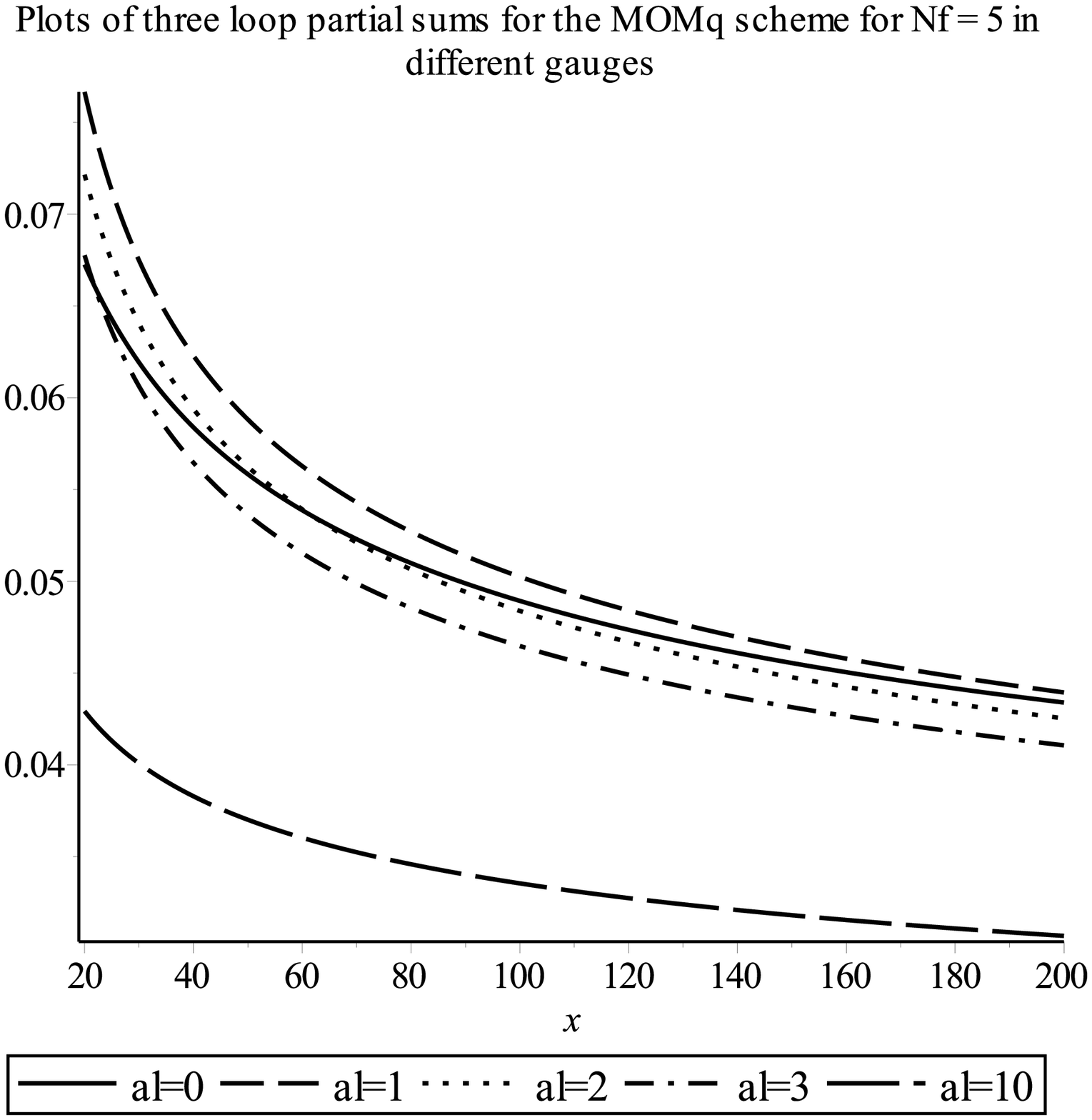}
\quad
\includegraphics[width=7.6cm,height=8cm]{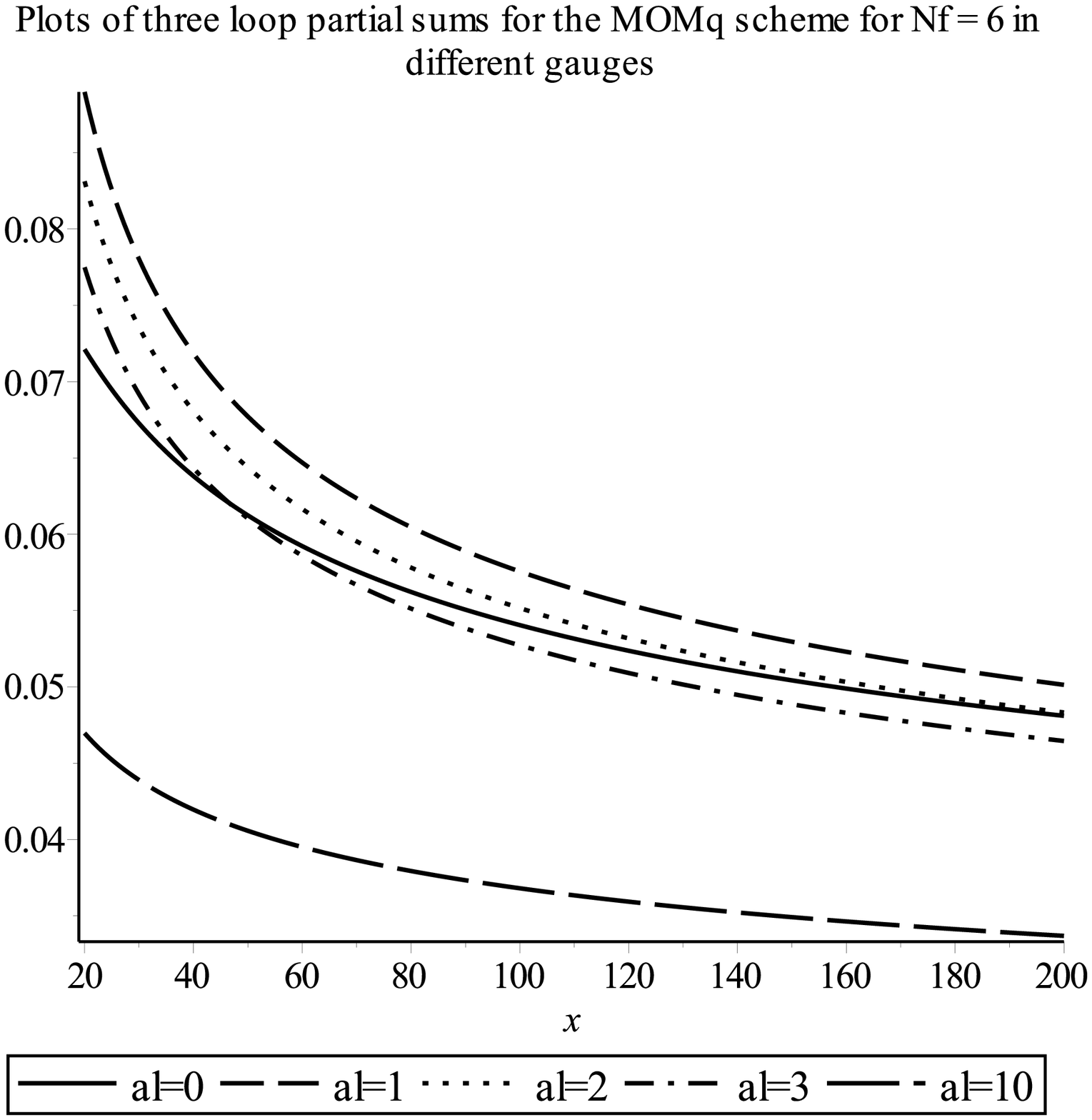}
\caption{Plots of $a_{33}^{\MOMqs}(x)$ for various gauges.}
\end{figure}}
In general terms it does not appear to be the case. By this we mean that from 
the numerical values, (\ref{rnf3})-(\ref{rnf6}), while the magnitude of some of
the $\MOMi$ schemes $L$-loop coefficients may be smaller than their $\MSbar$ 
counterpart, at the next order this position may be reversed. Indeed the sign 
of the corresponding terms is not preserved between schemes. This is not 
unexpected as there is no general reason why this should be. However, it could 
give the impression that convergence of the series in one scheme is better than
another. This is not the case as can be seen from the various plots provided 
here. Once the running coupling constant as a function of the $\Lambda$ 
parameter associated with that scheme is included, as well as its relation to 
$\Lambda^{\MSbars}$ as the reference scale, then the partial sums plotted 
against the momentum scale are generally comparable across schemes. Overall 
there is little difference between the results for large momentum as expected. 
It is at lower values where there are differences. However, the actual 
discrepancy is not significant. Moreover, within schemes there appears to be 
little difference for $\MSbar$ and $\mMOM$ between the three and four loop 
partial sums. At present the four loop $\MOMg$, $\MOMh$ and $\MOMq$ schemes can
not be examined at this level but it would be worth pursuing. This is because 
of the fact that the MOM schemes are in some sense more physical in that they 
are defined at a subtraction point which is completely symmetric and the 
momentum configuration there is non-exceptional. Indeed this would resolve 
whether there is actually a difference between the $\MSbar$ and $\MOMq$ results
as suggested in the $\Nf$~$=$~$6$ plot of Figure $1$. By contrast the $\MSbar$ 
scheme is essentially independent of the momentum subtraction point and to all 
intents and purposes does not carry any information about the structure of the 
vertex which is encoded in the finite parts of the coupling constant 
renormalization in the $\MOMg$, $\MOMh$ and $\MOMq$ schemes. In this context 
and from the plots $\mMOM$ is more akin to $\MSbar$ partly due to the fact that
it is defined at a subtraction point which is exceptional. In carrying 
information about the vertex structure in the associated renormalization 
constant that information differs from the $\MOMg$, $\MOMh$ and $\MOMq$ schemes
because $\mMOM$ has an off-shell leg. Within certain plots where the schemes 
were compared there is a suggestion that the exceptional based schemes are 
slightly different from the other three schemes. However, it is not possible to
draw a definite conclusion on this before the full four loop $\MOMg$, $\MOMh$ 
and $\MOMq$ scheme $\beta$-functions are known. Though one conclusion which 
seems to be assured is that the Landau gauge is the gauge which one should only
consider when applying the $\MOMi$ schemes. While we have focused on the 
$R$-ratio here it would be interesting to examine high loop evaluations of 
other physical quantities in order to see if one can make similar or general 
observations about the MOM schemes of \cite{16,17}. 

\vspace{1cm}
\noindent
{\bf Acknowledgements.} The author thanks J.M. Bell, D.J. Broadhurst, M. 
Gorbahn, D. Kreimer and E. Panzer for valuable discussions as well as the 
organisers of the Summer School on Structures in Local Quantum Field Theory 
held at Les Houches, France where the work was initiated. The Mathematical 
Physics Group at Humboldt University, Berlin, where part of the work was 
carried out, is also thanked for its hospitality.

\end{document}